%% file: main.tex
\documentclass[aps,prx,twocolumn,english,superscriptaddress,floatfix,longbibliography,amssymbol]{revtex4-2}

\usepackage{bbold}
\usepackage{amsmath,amsfonts,amsmath,mathtools,bbm,bm}
\usepackage{graphicx}
\usepackage{braket}
\usepackage{dsfont}
\usepackage[colorlinks,linkcolor=blue,urlcolor=blue,citecolor=blue]{hyperref}
\usepackage[all]{hypcap}

\usepackage{blindtext}
\usepackage{enumitem}
\usepackage{xcolor}
\usepackage[normalem]{ulem}
\usepackage{float}
\setcitestyle{numbers,square}

\usepackage{verbatim}
\usepackage{tensor}
\usepackage{tikz}
\usepackage{tikz-network}
\usetikzlibrary{patterns,decorations.pathreplacing,calc}

\graphicspath{{./Figures/}}

\newcommand{\HIDDEN}[1]{}

\newcommand{\tr}{\mathrm{Tr}}
\newcommand{\bbC}{\mathbb{C}}
\newcommand{\bbE}{\mathbb{E}}
\newcommand{\bbId}{\mathbb{1}}

\newcommand{\gateboundary}{U}
\newcommand{\gatebath}{V}
\newcommand{\evolutionoperator}{\mathcal{U}}
\newcommand{\kOTOC}{C_{ab}^{(k)}}
\newcommand{\kOTOCsingle}{c_{ab}^{(k)}}
\newcommand{\kOTOCsimple}[1][k]{C^{(#1)}}

\newcommand{\channel}[2]{\mathcal{M}_{#1 #2}}

\newcommand{\U}{\mathcal{U}}
\newcommand{\bigO}{\mathcal{O}}
\newcommand{\Hil}{\mathcal{H}}

\newcommand{\smallsquare}{\mathord{\scalebox{0.7}{$\square$}}} 
\newcommand{\sq}{\mathord{\scalebox{0.5}{$\square$}}} 

\newcommand{\figeq}[2][1cm]{%
  \vcenter{\hbox{\includegraphics[width=#1]{#2}}}%
}

\begin{document}

\title{Free Probability in a Minimal Quantum Circuit Model}

\author{Felix Fritzsch}
\email{fritzsch@pks.mpg.de}
\affiliation{Max Planck Institute for the Physics of Complex Systems, 01087 Dresden, Germany}

\author{Pieter~W.~Claeys}
\affiliation{Max Planck Institute for the Physics of Complex Systems, 01087 Dresden, Germany}

\date{\today}

\begin{abstract}
Recent experimental and theoretical developments in many-body quantum systems motivate the study of their out-of-equilibrium properties through multi-time correlation functions.
We consider the dynamics of higher-order out-of-time-order correlators (OTOCs) in a minimal circuit model for quantum dynamics.
This model mimics the dynamics of a structured subsystem locally coupled to a maximally random environment.
We prove the exponential decay of all higher-order OTOCs and fully characterize the relevant time scales, showing how local operators approach free independence at late times.
We show that the effects of the environment on the local subsystem can be captured in a higher-order influence matrix, which allows for a Markovian description of the dynamics provided an auxiliary degree of freedom is introduced.
This degree of freedom directly yields a dynamical picture for the OTOCs in terms of free cumulants from free probability, consistent with recent predictions from the full eigenstate thermalization hypothesis (ETH).
This approach and the relevant influence matrix are expected to be applicable in more general settings and present a first step to characterizing quantum memory in higher-order OTOCs.
\end{abstract}

\maketitle

\section{Introduction}
Recent decades have seen tremendous progress in our understanding of the thermalization of isolated quantum systems under their own unitary dynamics.
The current paradigm through which we understand how local observables relax towards thermal equilibrium is known as the eigenstate thermalization hypothesis (ETH)~\cite{srednickiApproachThermalEquilibrium1999,deutschEigenstateThermalizationHypothesis2018,dalessio_quantum_2016}.
Intuitively, ETH explains the relaxation of local subsystems towards thermal equilibrium by interpreting the complement of this subsystem as a large effective bath. 
This picture can be made precise within the influence-matrix approach to quantum dynamics~\cite{banuls2009matrix,lerose_influence_2021,sonner_influence_2021}.
Inspired by the Feynman-Vernon influence functional~\cite{feynman1963the}, the influence matrix at a given time captures the combined effect of the degrees of freedom in the effective bath on the local subsystem up to that time. 
In the case of maximally ergodic (discrete time) dynamics, the influence matrix represents perfectly dephasing boundary conditions, allowing for a Markovian description of the local subsystem evolution~\cite{lerose_influence_2021,sonner_influence_2021}.
Away from the limit of a perfect Markovian bath, the influence-matrix approach represents a powerful numerical tool for capturing weak memory effects and describing many-body quantum dynamics in a tractable way~\cite{muellerhermes2012tensor,hastings2015connecting,jorgensen_exploiting_2019,cygorek_simulation_2022,friasperez2022light,giudice_temporal_2022,thoenniss_efficient_2023,foligno2023temporal,thoenniss_nonequilibrium_2023,ng_real-time_2023,chen_grassmann_2024,nayak_steady-state_2025,klobas2021entanglement, bertini2022growth, bertini2022entanglement, rampp_entanglement_2023,bertini2023nonequilibrium,yao_temporal_2024, bertini2024dynamics,link_open_2024,sonner_semi-group_2025}.

Nevertheless, the advent of novel experimental techniques, such as quantum simulation and (noisy, intermediate scale) quantum computing platforms provide the opportunity for probing quantum dynamics with unprecedented precision and for asking more refined questions about thermalization.
One example is the concept of deep thermalization, i.e., the dynamical emergence of quantum state designs in local subsystems under projective measurements on their environment~\cite{cotler_emergent_2023,choi_preparing_2023,ippoliti_solvable_2022,ho_exact_2022,claeys_emergent_2022,wilming_high-temperature_2022,ippoliti_dynamical_2023,bhore_deep_2023,liu_deep_2024,chang_deep_2025,lucas_generalized_2023,mark_maximum_2024,manna_projected_2025,chan_projected_2024,varikuti_unraveling_2024,shrotriya_nonlocality_2025,mok_optimal_2025,yu2025mixed,lami2025quantum}.
A complementary approach asks about the relaxation of general multi-time correlation functions, which characterize for instance the scrambling of quantum information in terms of out-of-time-order correlators (OTOCs)~\cite{maldacena2016bound,swingle_unscrambling_2018} and contribute to transport phenomena in hydrodynamic theories beyond the linear-response regime~\cite{fava2021hydrodynamic,doyon2020fluctuations,myers2020transport}.

This move to more refined probes of thermalization was accompanied by corresponding extensions of ETH~\cite{foiniEigenstateThermalizationHypothesis2019a,brenesOutoftimeorderCorrelationsFine2021c,pappalardiEigenstateThermalizationHypothesis2022,jindalGeneralizedFreeCumulants2024,valliniLongtimeFreenessKicked2024,pappalardi_full_2025,fava_designs_2025,alves_probes_2025,fritzschMicrocanonicalFreeCumulants2024}.
While two-point functions, appearing in linear response and dissipation-fluctuation relations, are described by ETH, the latter fails to correctly capture higher-order, multi-time correlation functions.
This failure can be traced back to the ETH ansatz neglecting correlations between eigenstates and matrix elements \cite{chan2019eigenstate,hahn2024eigenstate,jafferis2023matrix,jafferis2023jackiw}.
Reintroducing such correlations recently led to an extended, so-called full ETH ansatz, whose validity has been demonstrated in paradigmatic chaotic many-body quantum systems where it accurately describes the dynamics of OTOCs~\cite{pappalardi_full_2025,fritzschMicrocanonicalFreeCumulants2024}.

The discussion of such OTOCs and more general multi-time correlation functions is greatly simplified by translating the combinatorics of the full ETH ansatz into the combinatorial structure of free probability~\cite{pappalardiEigenstateThermalizationHypothesis2022}.
Free probability is a theory of non-commutative random variables and has found applications in random matrix theory and operator algebras~\cite{voiculescuFreeNoncommutativeRandom1991,speicherFreeProbabilityTheory2003d,mingoFreeProbabilityRandom2017,nica2006lectures,novakThreeLecturesFree2012,speicherFreeProbabilityTheory2017b}. 
In the context of ETH, free probability allows for a decomposition of (multi-time) correlation functions into elementary building blocks known as free cumulants~\cite{speicherFreeProbabilityTheory2003d,pappalardiEigenstateThermalizationHypothesis2022,hruza2023coherent}.
The notion of freeness extends the notion of independence from commuting random variables to noncommuting random variables, and it is expected that in ergodic systems time-evolved observables become asymptotically free from static observables in the limit of large systems, i.e., in the thermodynamic or semiclassical limit.
This emergence of freeness at late times has been linked to the emergence of unitary designs from physical time evolution~\cite{fava_designs_2025} and was proposed as an indicator for quantum chaos~\cite{chen_free_2025,camargo2025quantum,jahnke2025free}.

On the level of correlation functions asymptotic freeness at late times is characterized by the relaxation of generalized, higher-order, OTOCs towards a stationary value described by the correlations between free random variables.
Analytical results on the dynamics of OTOCs, however, remain scarce.
Notable exceptions include holographic and conformally invariant theories~\cite{roberts2015diagnosing,roberts2015localized}, minimal supersymmetric models \cite{das2024late}, and the Sachdev-Ye-Kitaev model~\cite{maldacena2016remarks}, as well as random  quantum circuits \cite{nahum_operator_2018,chan2018solution,yoshimura_operator_2025} and dual-unitary circuits~\cite{claeys_maximum_2020,bertini2020scrambling}. For the latter, partial analytical results on the dynamics of higher-order OTOCs were recently obtained~\cite{chen_free_2025}.
To the best of our knowledge, no other analytical results on the dynamics of higher-order OTOCs exists, while a corresponding generalization of the influence-matrix approach even for perfectly Markovian baths is missing completely. 
Furthermore, it is not clear how the decomposition proposed by full ETH in terms of free cumulants, defined as appropriate summations over matrix elements, can be interpreted in a purely dynamical manner.

We here consider a minimal quantum circuit model for the unitary dynamics of a local system coupled to an environment, motivated by a similar approach for deep thermalization~\cite{ippoliti_solvable_2022}. While the local dynamics is structured, the bath is evolved using random unitary dynamics, resulting in tractable yet nontrivial dynamics for local observables.
This model can be interpreted as the circuit equivalent of Lindblad dynamics, with a separation of relaxation time scales between system and bath.
Through an appropriate averaging over circuit realizations, we completely and rigorously characterize the resulting dynamics of higher-order OTOCs, with the emergent structure reproducing predictions from full ETH.

The main advances of this work are that we
(i) fully characterize the dynamics of higher-order OTOCs in a minimal model of local dynamics and establish asymptotic freeness,
(ii) show how this minimal model returns an influence matrix that extends the Markovian influence matrix for correlation functions to higher-order OTOCs, which can in turn be mapped to a Markovian process on the noncrossing partition lattice,
(iii) in this way bridge full ETH and influence matrices, showing how the influence matrices for higher-order OTOCs are naturally expressed in terms of the noncrossing partitions underlying full ETH and free cumulants.
Our study additionally highlights the different origins of full thermalization and deep thermalization.
We expect this description and explicit structure of the influence matrix to be applicable beyond the specific setup of this work and emerge in more general dynamics in which a Markovian bath appears. The results from this work can be used as a starting point for more involved bath models and characterizations of quantum memory, in the same way that the Markovian limit can be used as the starting point for influence matrices on the level of correlation functions.

\subsection{Main results}

Before going into the details of our approach we provide a synopsis of the main ideas and results in the following, introducing relevant concepts from free probability along the way. 
For the sake of readability we keep this section informal and use simplified notation only.
We refer the reader to the subsequent sections for proper definitions and proofs, and to Appendix~\ref{app:FP} for an introduction to free probability.

\textit{Setting.---} To model physical systems with spatially local interactions we consider a minimal solvable circuit of just three sites, inspired by recent work on deep thermalization~\cite{ippoliti_solvable_2022}.
The left site $A$ models a small local subsystem of a spatially extended many-body system, while the right site $E$ models its environment acting as an effective bath. 
To incorporate the effect of spatial locality the sites $A$ and $E$ are coupled via a small intermediate site $C$, which can be thought of as a local bottleneck preventing the direct flow of quantum information between $A$ and $E$.
At every discrete time step the sites $A$ and $C$ interact via a fixed two-site unitary $\gateboundary$, which can in principle be time-dependent, whereas the sites $C$ and $E$ interact via a Haar random unitary $\gatebath_t$ at time $t$.
More precisely, the evolution operator over a single discrete time step reads
\begin{align}
     \evolutionoperator_t = \left(\bbId_A \otimes \gatebath_t\right)\left(\gateboundary\otimes \bbId_E\right).
     \label{eq:circuit_intro}
\end{align}
The Haar-random unitaries $\gatebath_{t}$ at different time steps are taken independent from each other and we denote the evolution operator up to time $t$ by 
\begin{align}
    \evolutionoperator(t) = \mathcal{U}_t \dots \mathcal{U}_2\, \mathcal{U}_1  = \figeq[0.28\columnwidth]{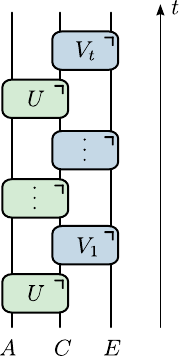}
    \label{eq:evolution_op_intro}
\end{align}
In this model the thermodynamic limit is obtained by taking the bath size to infinity, thereby causing $D \to \infty$ with $D$ the Hilbert space dimension of the full system.

The main focus of this work are correlation functions between local observables $A$ and $B$ acting non-trivially only in the subsystem $A$.
We specifically focus on generalized OTOCs of arbitrary order $k$ ($k$-OTOC) with respect to the infinite temperature state, defined by
\begin{align}
    \kOTOCsimple(t) = \frac{1}{D}\mathrm{Tr}\left([A(t)B]^k\right),
    \label{eq:OTOC_def_intro}
\end{align}
with $A(t)=\evolutionoperator(t) A\, \evolutionoperator(t)^\dagger$ the time-evolved observable. For $k=1$ the above reduces to a two-point function and  for $k=2$ it returns the usual OTOC.

\textit{Dynamics on the noncrossing partition lattice.---} 
As detailed in Sec.~\ref{sec:model}, in the thermodynamic limit and after averaging over the Haar-random unitaries, the $k$-OTOC can be expressed as a summation over different terms indexed by paths on the noncrossing partition lattice, i.e., the partially ordered set of all noncrossing partitions of $k$ elements. 

Noncrossing partitions are the fundamental objects in both free probability and full ETH~\cite{pappalardiEigenstateThermalizationHypothesis2022}. A partition $\sigma$ of $\{1,2, \dots k\}$ with blocks $V_1, V_2 \dots V_{|\sigma|}$ is said to be noncrossing if, when the elements $p_i, q_i$ and $p_j, q_j$ belong to distinct blocks $V_i$ and $V_j$, there are no ``crossings'' in which $p_i < p_j < q_i < q_j$. These noncrossing partitions can be represented graphically by arranging the elements on the circle and connecting elements that are in the same block. For example, the partition $(12)(34)$ is noncrossing and $(13)(24)$ is crossing, as made explicit graphically:
\begin{align}
    (12)(34) = \,\,\figeq[0.16\columnwidth]{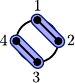}, \quad (13)(24) = \,\,\figeq[0.16\columnwidth]{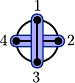} \, .
\end{align}
Two particularly important noncrossing partitions are the partitions consisting of either $k$ blocks or a single block, e.g. for $k=4$,
\begin{align}
        (1)(2)(3)(4) = \,\,\figeq[0.16\columnwidth]{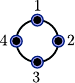}, \quad (1234) = \,\,\figeq[0.16\columnwidth]{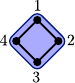} \, ,
\end{align}
which we will refer to as the identity and the cyclic partition respectively.

The set of noncrossing partitions supports a partial ordering, where for two noncrossing partitions $\nu$ and $\sigma$ we say that $\nu \subseteq \sigma$ if and only if all blocks of $\nu$ are contained in the blocks of $\sigma$. This ordering can be illustrated in the noncrossing partition lattice, e.g. for $k=4$,
\begin{align}\label{eq:NC_lattice}
    \figeq[0.65\columnwidth]{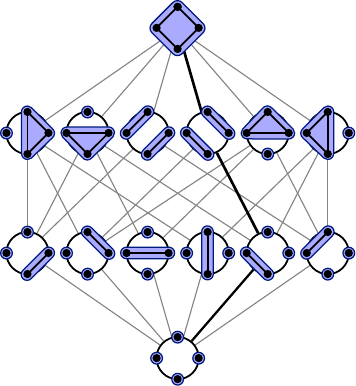}
\end{align}
which illustrates all noncrossing partitions on $k=4$ elements, with the identity partition at the bottom and the cyclic partition at the top. The bold line illustrates a geodesic
\begin{align}
    (1)(2)(3)(4) \subset (1)(2)(34) \subset (12)(34) \subset (1234)\,.
\end{align}
For the expansion of the averaged $k$-OTOC \eqref{eq:OTOC_def_intro}, each sequence of $2t$ nondecreasing partitions on a geodesic defines a sequence of transfer matrices acting on $k$ replicas of the subsystem $A$, with the noncrossing partitions encoding the combinatorics on how replicas can interact. 
We provide a convenient way to track the combinatorial structure of these paths in terms of a Markov process on the noncrossing lattice by introducing an auxiliary degree of freedom corresponding to a noncrossing partition.
This Markovian process corresponds to introducing an influence matrix for $k$ replicas, with a bond dimension set by the number of noncrossing partitions. 

\textit{Time scale for asymptotic freeness.---} Using this approach, we establish late-time asymptotic freeness as characterized for traceless observables by the vanishing of all $k$-OTOCs in the thermodynamic limit $\lim_{t \to \infty} \lim_{D\to \infty} \kOTOCsimple(t) = 0$\,\footnote{Note that the order of limits matters, since for finite systems there is no expected convergence in time due to revivals.}.
In the simplest case $k=1$ the Markovian process reproduces the perfectly dephasing influence matrix indicating Markovianity of the environment $E$ for correlation functions.
In this case all dynamics is described by a unital single-particle quantum channel $\mathcal{M}$ acting on $A$. The subleading eigenvalue $\lambda < 1$ of the quantum channel dominates the late-time dynamics as 
\begin{align}
C^{(k=1)}(t) \propto \lambda^t.
\end{align}
For arbitrary $k>1$ the combinatorics of noncrossing partitions guarantees that the dynamics at each time step is dominated by at least two copies of this single-particle channel. This gives rise to the leading contribution to the $k$-OTOC dynamics 
\begin{align}
    \kOTOCsimple[k \geq 2](t) \propto \lambda^{2t} \, ,  
    \label{eq:intro_kOTOC_decay}
\end{align}
i.e., all higher $k$-OTOCs decay exponentially with the same decay rate.
More precisely, we obtain rigorous bounds on the dynamics of the $k$-OTOCs, which scale as the right hand side of Eq.~\eqref{eq:intro_kOTOC_decay} up to polynomial corrections.
This result demonstrates, in particular, that for $k>2$ the $k$-OTOC cannot be decomposed as a product of two-point functions, as would be expected for free dynamics or Gaussian systems.
A notable exception to the above behavior occurs when the subsystem $A$ hosts a single qubit. 
In this case the assumptions on the local observables imply further algebraic relations causing the $k$-OTOC to decay as $\kOTOCsimple(t) \propto \lambda^{kt}$.

In general, we establish late-time asymptotic freeness between local observables within this solvable model whenever the choice of unitary gates $U$ results in ergodic dynamics for the two-point correlation functions. 
When choosing $U$ to be dual-unitary~\cite{bertini_exactly_2025}, i.e. also unitary when switching the roles of space and time, the dynamics is shown to be maximally ergodic corresponding to $\lambda=0$ and we explicitly characterize the stability of this maximally ergodic point.
 
\textit{Free cumulants.---} Once we drop the requirement on the observables being traceless, $k$-OTOCs decay to a generically non-zero stationary value. From free probability, the stationary value corresponds to
\begin{align}
    \lim_{t \to \infty}\kOTOCsimple(t) =  \sum_{\sigma\, \in \, \textrm{NC}(k)} \!\!\!\kappa_\sigma(A) \,\varphi_{\sigma^*}(B)   \,.
    \label{eq:stationary_state_intro}
\end{align}
Each term in this sum is labeled by a noncrossing partition $\sigma$ and factorizes in a free cumulant $\kappa_\sigma(A)$ and a generalized moment $\varphi_{\sigma^*}(B)$, with $\sigma^*$ the so-called Kreweras complement of $\sigma$ (see Appendix~\ref{app:FP}). Generalized moments  and free cumulants are both indexed by a noncrossing partition and factorize according to the blocks of this partition,
\begin{align}\label{eq:intro:factorization_moments_cumulants}
    \varphi_{\nu}(a) = \prod_{V \in \nu} \varphi_{|V|}(a), \quad\ \kappa_{\sigma}(a) = \prod_{V \in \sigma} \kappa_{|V|}(a),
\end{align}
with $|V|$ the size of each block. In the first expression $\varphi(\bullet)$ is the normalized trace defined as $\varphi_{k}(a) = \mathrm{Tr}(a^k)/\mathrm{Tr}(\mathbb{1})$. For the second expression free cumulants are defined recursively as
\begin{align}\label{eq:intro:phi_from_kap}
\varphi_k(a) =  \!\!\!\!\sum_{\sigma \in \textrm{NC}(k)}\!\!\!\kappa_{\sigma}(a).
\end{align}
The relation between moments and free cumulants can be formally inverted to return
\begin{align}\label{eq:intro:kap_from_phi}
    \kappa_{\sigma}(a) = \sum_{\substack{\nu \in \textrm{NC}(k)\\\nu \subseteq \sigma }}\!\!\!\varphi_{\nu}(a)\,  \mu(\nu,\sigma),
\end{align}
with $\mu(\nu,\sigma)$ the so-called M\"obius function on the noncrossing partition lattice. Both free cumulants and the M\"obius function will be discussed in detail throughout the text.

The stationary value of the $k$-OTOC~\eqref{eq:stationary_state_intro} reflects as much factorization of correlations and consequently as much independence between $A$ and $B$ as is allowed by free probability.
This steady-state follows directly from the Markovian process on the noncrossing lattice, where we can identify the free cumulants and moments in Eq.~\eqref{eq:stationary_state_intro} with leading right and left eigenvectors of the Markovian transfer matrix respectively. 
Subleading and hence decaying eigenvectors can be similarly identified with so-called mixed free cumulants depending on both $A$ and $B$. In this way we recover a decomposition of the OTOC in which each term can be identified with a free cumulant as
\begin{align}
    \kOTOCsimple(t) =  \sum_{\sigma\, \in \, \textrm{NC}(2k)} \!\!\!\kappa_\sigma(A,B,\dots A,B;t)  \,,
    \label{eq:dynamics_state_intro}
\end{align}
where mixed free cumulants are defined analogously to the free cumulants of Eq.~\eqref{eq:intro:phi_from_kap} (see Appendix~\ref{app:FP}). This expansion corresponds exactly to the prediction from full ETH~\cite{foiniEigenstateThermalizationHypothesis2019a,pappalardiEigenstateThermalizationHypothesis2022,fritzschMicrocanonicalFreeCumulants2024,pappalardi_full_2025}. These results are argued to remain valid when adding additional spatial structure to either the subsystem or effective bath.
This stability of the late-time behavior indicates that the minimal circuit model captures the essential dynamics of multi-time correlation functions in strongly ergodic systems.

\section{Solvable Circuit Model and $k$-OTOCs}
\label{sec:model}

We here present a more detailed introduction of the model and higher-order OTOCs, fixing notation along the way.
The replica picture is introduced as a convenient representation for the $k$-OTOCs dynamics, and we show how averaging over the bath dynamics returns an expansion for the $k$-OTOCs in terms of paths on the noncrossing partition lattice.

\subsection{Time evolution}

The minimal circuit model considered here was introduced as a minimal solvable model for deep thermalization in Ref.~\cite{ippoliti_solvable_2022}.
The model is built from a lattice of just three sites, modeling a local subsystem $A$ of a many-body system and its complement $E$, with intermediate site $C$ mediating the local coupling between $A$ and $E$.
The local Hilbert spaces associated to the three sites are $\Hil_A = \bbC^{d_A}$, $\Hil_C = \bbC^{d_C}$ as well as $\Hil_E = \bbC^{d_E}$ with dimensions $d_A, d_C \ll d_E$. The full Hilbert space $\Hil =\Hil_A \otimes \Hil_C \otimes \Hil_E$ has dimension $D=d_Ad_Cd_E$. We keep $d_A$ and $d_C$ fixed, while allowing $d_E\to \infty$ and hence $D\to \infty$ to model the thermodynamic limit.
We denote the canonical (computational) basis in $\Hil_x$ by
$\ket{i}_x$ for $i=1,\ldots,d_x$ and $x=A,C,E$.

Local interactions between $A$ and $C$ are modeled by a fixed unitary two-qudit gate $U$ of dimension $d_Ad_C \times d_Ad_C$ and the interactions between $C$ and $E$ by independently drawn Haar-random $d_Cd_E\times d_Cd_E$ unitaries $V_t$ at each time step $t$. Graphically, we denote the unitary gates as
\begin{align}
     U = \figeq[0.13\columnwidth]{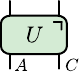}\,\,, \qquad V_t =  \figeq[0.13\columnwidth]{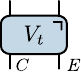}\,\,.
\end{align}
Here the labels indicate the relevant Hilbert space.
Evolution for the $t$-th time step is then given by the minimal two-layer brickwork circuit $\U_t$ from Eq.~\eqref{eq:circuit_intro}.
For a fixed realization of the circuit, corresponding to  a fixed sequence of independent Haar-random unitaries $\{V_1, \dots,V_t\}$, the evolution operator $\U(t)$ is given in terms of Eq.~\eqref{eq:evolution_op_intro} as
\begin{align}
    \mathcal{U}(t) = \figeq[0.18\columnwidth]{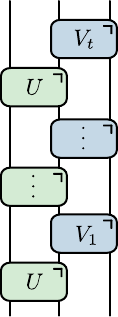}\,\,.
\end{align}
The resulting model for time evolution is not homogeneous in time and should be thought of as a non-periodically driven system.
It is not a Floquet system and hence there is no notion of eigenstates, rendering concepts from ETH and full ETH ill-defined.
The dynamical picture in terms of free cumulants put forward by full ETH, however, persists for the present model.

\subsection{Higher-order OTOCs}

We consider generalized OTOCs of order $k$, defined by Eq.~\eqref{eq:OTOC_def_intro}
for local observables $A$ and $B$ acting non-trivially only on the subsystem $A$.
These observables are determined by Hermitian observables $a$ and $b$ acting on $\Hil_A$ which are embedded into the full system as $A=a \otimes \bbId_C \otimes \bbId_E$ and $B=b \,\otimes \bbId_C \otimes \bbId_E$, respectively.
Unless stated otherwise, we assume the observables to be traceless, $\tr(a)=\tr(b)=0$.
We define evolution in time as
\begin{align}
    A(t)=\evolutionoperator(t) A\, \evolutionoperator(t)^\dagger \, .
\end{align}
Note that this dynamics corresponds to that of density matrices rather than observables in the Heisenberg picture, which will help to simplify the following expressions.

We will also consider more general $k$-OTOCs defined between $k$ time-evolved observables $A_{1}(t),\ldots,A_{k}(t)$ and $k$ static observables $B_{1},\ldots,B_{k}$ built from $k$-tuples of local observables $a=(a_{1},\ldots,a_{k})$ and $(b_{1},\ldots,b_{k})$, respectively.
The corresponding $k$-OTOC reads
\begin{align}
        \kOTOCsingle(t; d_E) = \frac{1}{D}\tr\left[A_{1}(t)B_1A_{2}(t)B_2\cdots A_{k}(t)B_k\right] .
        \label{eq:kOTOC_single_finite_bath}
\end{align}
While we dropped the dependence on the dimension of the environment in $\U(t)$ as well as in $A_i$, $B_i$, we explicitly include it in the single-realization $k$-OTOC.
Originating from a random time evolution, the latter is a random variable itself, and we are generally interested in its expectation value when averaged over the independent Haar-random unitaries $V_t$ in the thermodynamic limit $d_E \to \infty$. We denote this limit as
\begin{align}
     \kOTOC(t) = \lim_{d_E \to \infty} \bbE\, \left[  \kOTOCsingle(t; d_E)\right].
    \label{eq:kOTOC_avg_infinite}
\end{align}
We emphasize that we take the thermodynamic limit at an arbitrary but fixed time and only afterwards study the late-time behavior.
This order of limits, thermodynamic limit first and late time second, is crucial as for finite systems the $k$-OTOCs for both a single realization and on average are expected to exhibit residual finite size fluctuations even after relaxation.
Note also that we take the limit of the finite-size OTOCs, i.e., a limit of ordinary real or complex numbers, rather than explicitly constructing observables for the infinite system and computing OTOCs between them. 
This former type of limit is the usual limit considered in free probability theory and gives rise to a convergence in distribution.
In Appendix~\ref{app:averaging} we additionally prove the stronger almost sure convergence, ensuring that the above average describes typical circuit realizations as well.

\subsection{Replica picture}

A convenient choice for representing the $k$-OTOCs diagrammatically and for analytical considerations is the so-called folded or replica picture.
Graphically the folded picture is obtained by folding all copies (replicas) of $\U(t)$ and its inverse on top of each other in an alternating fashion.
This folding gives rise to a single evolution operator
\begin{align}
    \evolutionoperator^{(k)}(t)=\left(\evolutionoperator(t) \otimes \evolutionoperator(t)^*\right)^{\otimes k} \, ,
\end{align}
with $*$ denoting complex conjugation, acting on $k$ replicas, each of which consists of a forward and a backward time sheet. 
This $k$-replica is built from the $k$-folded gate
\begin{align}
    U^{(k)}= \frac{1}{d_C^k}\,\left(U \otimes U^*\right)^{\otimes k}
\end{align}
or, graphically,
\begin{align}
     \figeq[0.12\columnwidth]{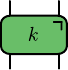}\,\, =  \frac{1}{d_C^k}\,\, 
     \figeq[0.18\columnwidth]{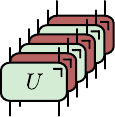}\,\,,
\end{align}
here illustrated for $k=3$. Note that we include an unconventional prefactor $d_C^{-k}$ in the folded gate in order to simplify all following graphical expressions. The green (red) gates denote $U$ ($U^*$). A folded version of $V_t$ can be similarly defined.

In this representation, the $k$-OTOC can be obtained as a matrix element of $\evolutionoperator^{(k)}(t)$ with respect to an initial and a final state on $\Hil^{\otimes 2k}$, which encode both the observables $A$ and $B$ as well as how the replicas are connected. 
Graphically
\footnote{Due to unitarity of $V_t$ or, equivalently, the strict causality in the brickwork circuit, the above $k$-OTOC does not depend on $V_t$.},
\begin{align}\label{eq:kOTOC_noavg}
    \kOTOCsingle(t; d_E) \propto \,\,\figeq[0.21\columnwidth]{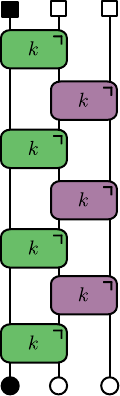}\,\,.
\end{align}
The boundary conditions, here represented by squares and circles, can be expressed in terms of permutation states.
We denote the product basis on $\Hil_x^{\otimes 2k}$ by $|i_1,i'_1,\ldots,i_k,i'_k)_x$ for $x=A,C,E$ and introduce the unnormalized permutation state $|\sigma)$ for a permutation $\sigma \in S_k$ of $k$ elements via 
\begin{align}\label{eq:permutation_states}
    (i_1, i_1', \dots i_k,i_k'| \sigma ) = \delta_{i_{\sigma(1)},i_1'} \dots \delta_{i_{\sigma(k)},i_k'}\, .
\end{align}
Graphically such a permutation state connects the backward sheet of the $i$-th replica with the forward sheet of the $\sigma(i)$-th replica, where e.g. for $\sigma=(12)(3)$
\begin{align}
|\sigma=(12)(3))\,\, =\,\, \figeq[0.04\columnwidth]{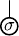} \,\,=\,\, \figeq[0.10\columnwidth]{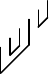}\,\,.
\end{align}
Note that we denote permutations using cycle notation.
We write the identity permutation as $\circ=(1)(2)\cdots(k)$ and the cyclic permutation as $\smallsquare=(1,2,3,\ldots,k)$, with corresponding states of the form
\begin{align}\label{eq:circ_sq}
|\circ)\,\, =\,\, \figeq[0.028\columnwidth]{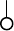} \,\,=\,\, \figeq[0.10\columnwidth]{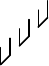}\,\,,\qquad 
|\smallsquare) \,\,=\,\, \figeq[0.028\columnwidth]{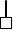} \,\,=\,\, \figeq[0.10\columnwidth]{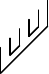}\,\,,
\end{align}
again illustrated for $k=3$ for concreteness. 
Permutation states are neither normalized nor orthogonal, since for $\sigma,\nu \in S_k$ one has
\begin{align}
    _x(\sigma | \nu )_x = d_x^{|\sigma^{-1}\nu|}\,\,,
    \label{eq:overlaps_permutations}
\end{align}
with $|\pi|$ the number of cycles of a permutation $\pi \in S_k$.

The relevant boundary vectors for the OTOC in Eq.~\eqref{eq:kOTOC_noavg} are obtained by dressing these permutation states with operators as
\begin{align}\label{eq:fullcirc}
    &\figeq[0.028\columnwidth]{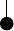} \,\, = |a_\sigma)_A = \left(a_1 \otimes \bbId_A \otimes \cdots \otimes a_k \otimes \bbId_a\right)|\sigma)_A  \, , \\
    & \figeq[0.028\columnwidth]{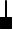} \,\, = _A\!\!(b_{\sigma}| = _A\!\!(\smallsquare| \left(\bbId_A \otimes b_1 \otimes \cdots \otimes \bbId_A \otimes b_k\right) 
    \label{eq:fullsq}
\end{align}
and similarly for arbitrary permutations $\sigma$.
In terms of matrix elements, these read
\begin{align}
    _A(i_1, i_1', \dots i_k,i_k'|a_\circ )_A = (a_1)_{i_1 i_1'} (a_2)_{i_2 i_2'} \dots (a_k)_{i_k i_k'}, \\
    _A(b_{\sq} | i_1, i_1', \dots i_k,i_k')_A= (b_1)_{i_1' i_2} (b_2)_{i_2' i_3}\dots (b_k)_{i_k' i_1}\, ,
\end{align}
with $a_{ij} = _A\!\!\langle i|a|j\rangle_A$ and $b_{ij} = _A\!\!\langle i|b|j\rangle_A$.

\subsection{Averaged $k$-OTOC}

The averaging over unitary gates $V_t$ to obtain the averaged $k$-OTOC in the thermodynamic limit [Eq.~\eqref{eq:kOTOC_avg_infinite}] can now be performed using the Weingarten calculus. This calculation is detailed in Appendix~\ref{app:averaging}. While this averaging involves standard methods, care needs to be taken due to the nontrivial interplay between the overlaps between the permutations themselves with the boundary conditions corresponding to the identity and the cyclic permutation [cf. Eq.~\eqref{eq:overlaps_permutations}]. The final expression is given by
\begin{widetext}
\begin{align}\label{eq:OTOC_summation}
    C_{ab}^{(k)}(t) = d_A^{-1} d_C^{k-1} \sum_{\substack{\sigma_i,\nu_i \in \textrm{NC}(k) \\\ldots \subseteq \sigma_i \subseteq \nu_i \subseteq \sigma_{i+1} \subseteq \ldots } }\left(\prod_{i=1}^{t-1}W_{\nu_i\sigma_{i+1}}\right) (b_{\sq}|\mathcal{M}_{\nu_t \sigma_t} \dots \, \mathcal{M}_{\nu_2\sigma_2} \mathcal{M}_{\nu_1 \sigma_1}|a_\circ).
\end{align}
\end{widetext}
Each term in this summation is indexed by a set of permutations $\sigma_1, \nu_1, \dots, \sigma_t, \nu_t$. These permutations are constrained to be noncrossing. 
We can identify permutations with partitions by choosing the blocks of the partition as the cycles of the permutation, and we will freely switch between these interpretations. 
As mentioned above, noncrossing partitions support a partial ordering, where $\nu \subseteq \sigma$ if all cycles of $\nu$ are contained in the cycles of $\sigma$. 
In Eq.~\eqref{eq:OTOC_summation} the permutations are further restricted to so-called multichains on the noncrossing partition lattice,  i.e. geodesics connecting the identity with the cyclic permutation as
\begin{align}
    \circ = \sigma_1 \subseteq \nu_1 \subseteq \sigma_2 \subseteq \nu_2 \dots \subseteq \sigma_t \subseteq \nu_t = \smallsquare \,.
\end{align}
The factors $W_{\nu \sigma}$ originate from the Weingarten calculus as
\begin{align}
    W_{\nu\sigma} &= d_C^{-k+|\nu^{-1}\sigma|} \mu(\nu,\sigma),
    \label{eq:Weingarten_C}
\end{align}
where $\mu(\nu, \sigma)$ is shorthand notation for
\begin{align}
     \mu(\nu,\sigma) = \prod_{V \in  \nu^{-1}{\sigma}}(-1)^{|V|-1}C_{|V|-1},
     \label{eq:mobius_intro}
\end{align}
with $V$ the different cycles of $\nu^{-1}\sigma$.
Crucially, this function is a M\"obius function for the noncrossing partition lattice, satisfying
\begin{align}
    \sum_{\sigma \subseteq \rho \subseteq \nu}\mu(\nu,\rho)  = \delta_{\nu\sigma}\, ,
    \label{eq:sumrule_moebius_intro}
\end{align}
which can also be taken as its defining property. From the above definition it also directly follows that the M\"obius function encodes the transformation from moments to free cumulants, cf. Eq.~\eqref{eq:intro:kap_from_phi}. This connection will be directly responsible for the appearance of free cumulants in the OTOC dynamics.

The remaining terms $\mathcal{M}_{\nu\sigma}$ are non-expanding linear maps acting on the space of $k$ replicas [as in Eqs.~\eqref{eq:fullcirc} and \eqref{eq:fullsq}] defined as
\begin{align}
\mathcal{M}_{\nu\sigma} = \,\, \figeq[0.17\columnwidth]{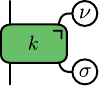} \,\,.
\end{align}
Written out explicitly, these maps are given by 
\begin{align}
\mathcal{M}_{\nu \sigma} =  d_C^{-k}\left( \bbId_A^{\otimes 2k} \otimes {_C}( \nu| \right) \left(U \otimes U^*\right)^{\otimes k} \left( \bbId_A^{\otimes 2k} \otimes |\sigma )_C \right) \, .
\end{align}

The remainder of this work analyzes Eq.~\eqref{eq:OTOC_summation} and its implications. 
In Sec.~\ref{sec:free_independence} we establish the late-time behavior of the $k$-OTOCs on the level of individual multichains, identifying the slowest-decaying contributions and their combinatorial interpretation. In this way we present a physical picture of the decay channels of the $k$-OTOC and the effect of increasing $k$.
In Sec.~\ref{sec:influence_matrix} we show how the full summation can be recast in an influence matrix formalism, resulting in a Markovian picture for the $k$-OTOC. In this section we explicitly recover the structure of free probability: The eigenmodes of the Markovian transfer matrix can be identified with (mixed) free cumulants, returning the dynamical behavior postulated by full ETH. 

Before doing so it is instructive to compare the structure of Eq.~\eqref{eq:OTOC_summation} with the results of Ref.~\cite{ippoliti_solvable_2022} on deep thermalization in this model, which helps to make clear the differences between deep and full thermalization.
In the context of deep thermalization projective measurements are performed on the environment $E$, where the resulting projected ensemble of states on $A$ is expected to form a maximally Haar-random ensemble at late times~\cite{choi_preparing_2023}. 
In order to form a $k$-design, the first $k$ moments of this projected ensemble need to agree with the moments of the Haar ensemble, expressed in terms of permutations of the form~\eqref{eq:permutation_states}. Within deep thermalization, there is no preferred `ordering' between the $k$ replicas, such that the relevant symmetry group is $S_k$. 
Within full thermalization, however, the different replicas have a preferred ordering and are only invariant under shifts: the choice of $k$-OTOC as observable introduces boundary conditions corresponding to the identity and the cyclic permutation. The relevant permutations are restricted to be noncrossing $\textrm{NC}(k)$, since noncrossing permutations can be connected to the identity and cyclic permutation through a minimal number of transpositions, minimizing the (Cayley) distance to these boundaries [i.e., satisfying a geodesic condition, also made apparent in the noncrossing partition lattice~\eqref{eq:NC_lattice}].
The noncrossing partitions are additionally required to change between different time steps, starting at the identity and ending at the cyclic permutation, thereby giving rise to the summation over multichains in Eq.~\eqref{eq:OTOC_summation} and a nontrivial interaction between the different replicas.
In contrast, in deep thermalization, the permutations do not change in time and for a fixed permutation the replicas do not interact.

We also note that a similar expression involving paths on the noncrossing partition lattice appeared in Ref.~\cite{chen_free_2025} for $k$-OTOCs in dual-unitary circuits, showing how these OTOCs spread with a maximum velocity, using a similar approach as in Ref.~\cite{claeys_maximum_2020} for $2$-OTOCs. Eigenmodes of a light-cone transfer matrix were identified and labeled by paths on the noncrossing partition lattice, but the complicated overlaps between these eigenmodes required taking the limit of an infinite-dimensional Hilbert space for each site and no connection with full ETH was made.

Remarkably, Eq.~\eqref{eq:OTOC_summation} remains valid when allowing for additional structure in the environment: As described in Appendix~\ref{app:extended_bath}, replacing the Haar-random unitaries by a random brickwork circuit with large local dimension leaves the $k$-OTOCs unchanged, hinting towards a more general applicability of the presented structure.

\section{Free Independence at Late Times}
\label{sec:free_independence}

We now consider the dynamical behavior of the averaged $k$-OTOC at late times, which will also help to make clear the role of the different terms in Eq.~\eqref{eq:OTOC_summation}.
In particular, we show that all $k$-OTOCs between traceless observables exponentially decay to zero at late times with the same rate if $k\geq 2$.
We derive these results from the combinatorial structure of multichains in the noncrossing lattice and their interplay with the quantum channels $\mathcal{M}_{\nu \sigma}$ and their spectra.
We moreover characterize the steady state of the $k$-OTOC dynamics for arbitrary observables.
In the free probability language the decay of the $k$-OTOCs gives rise to free independence between time evolved and static observables.
Correspondingly, the steady state is described by the correlations between freely independent observables.

\subsection{Example: $k=1$}

We first consider the simplest example, $k=1$, for which the $k$-OTOC simplifies to a standard (time-ordered) correlation function, 
\begin{align}
    C^{(k=1)}_{ab}(t) = \lim_{d_E\to\infty} \bbE \, \frac{1}{D} \mathrm{Tr}\left[A(t)B\right].
\end{align}
In this limit the underlying combinatorial structure simplifies considerably as there is only a single (noncrossing) partition, $\circ=\smallsquare = (1)$, and a single corresponding quantum channel $\mathcal{M}=\channel{\circ}{\circ}$.
There is hence only a single multichain and all terms from the Weingarten function yield a factor of $W_{\circ\circ}=1$.
Consequently, Eq.~\eqref{eq:OTOC_summation} reduces to 
\begin{align}
  \label{eq:OTOC_k1}
  C_{ab}^{(k=1)}(t) = \frac{1}{d_A} \mathrm{Tr}[b\, \mathcal{M}^t(a)]= \frac{1}{d_A}\,\, \figeq[0.223\columnwidth]{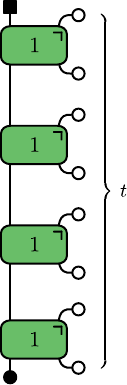} \,.
\end{align}
All dynamics follows from the action of $\mathcal{M}$ as defined by
\begin{align}\label{eq:def_channel}
    \mathcal{M} = \,\,\figeq[0.15\columnwidth]{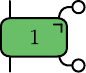} = \frac{1}{d_C}\,\,\figeq[0.13\columnwidth]{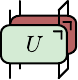}\,\,.
\end{align}
Written out explicitly, $\mathcal{M}$ acts as
\begin{align}
    \mathcal{M}(a) = \frac{1}{d_C} \mathrm{Tr}_C \left[U(a\otimes \bbId_C)U^{\dagger}\right],
\end{align}
corresponding to a unital quantum channel, i.e., a trace-preserving completely positive map which preserves the identity~\cite{nielsen2010quantum}.
The long-time behavior of the correlation function depends on the spectral properties of this quantum channel. From unitality we have that $\mathcal{M}(\mathbb{1}_A) = \mathbb{1}_A$, leading to a guaranteed eigenoperator with eigenvalue one. 
All other eigenvalues are bounded in magnitude by one.

If the identity is the only eigenoperator with eigenvalue with modulus one, at late times we can replace $\mathcal{M}$ with a projector on this operator, leading to thermalization to the infinite temperature state $\mathbb{1}_A$.  
If the operators are traceless, this projector vanishes since $\tr(\mathbb{1}_A \cdot a) = \tr(a)=0$. If the operators are not traceless, the steady-state follows from this projector as
\begin{align}
  \lim_{t\to \infty}C_{ab}^{(k=1)}(t) = \frac{1}{d_A^2}\tr(a)\tr(b) = \varphi(a)\varphi(b)\, ,
\end{align}
with $\varphi(\bullet)=\tr(\bullet)/d_A$ the normalized trace.

The dynamics of eigenoperators of $\mathcal{M}$ is particularly straightforward. For a right eigenoperator $a_\lambda$ with eigenvalue $\lambda$, i.e. $\mathcal{M}(a_\lambda) = \lambda a_\lambda$, we can choose $a=a_\lambda$ and $b=b_\lambda$ the corresponding left eigenoperator to obtain the correlation function as
\begin{align}
C_{ab}^{(k=1)}(t) = \frac{1}{d_A} \mathrm{Tr}[b_\lambda \mathcal{M}^t(a_\lambda)] = \frac{1}{d_A} \lambda^t\, \mathrm{Tr}[b_\lambda a_\lambda]= \lambda^t\,,
\end{align}
where we have taken this eigenoperator to be Hilbert-Schmidt normalized as $\tr(b_\lambda a_\lambda)/d_A=1$. If $|\lambda|<1$, this autocorrelation function decays to the zero ergodic value as is illustrated in Fig.~\ref{fig:kOTOC}~(a) below.
General traceless local observables $a$ and $b$ can be expanded in eigenoperators of $\mathcal{M}$ that are necessarily orthogonal to $\mathbb{1}_A$, such that the largest (in modulus) of these eigenvalues dominates the late-time behavior. 
In the following we will always denote this eigenvalue by $\lambda$.
Alternatively, the above can be formulated in terms of the operator norm of $\mathcal{M}$ restricted to the space orthogonal to $\bbId_A$, i.e., the space of traceless operators.
For normal $\mathcal{M}$ and real $\lambda\geq 0$ both coincide and give rise to an upper bound for traceless $a$ and $b$:
\begin{align}
\left|C_{ab}^{(k=1)}(t)\right| \leq \lambda^t \|a\|_2 \|b\|_2 \, ,
\end{align}
with $\|\cdot \|_2$ the Hilbert-Schmidt (Frobenius) norm of the local observables $a$ and $b$.
For nonnormal $\mathcal{M}$, the eigenvalue $\lambda$ needs to be replaced by the above norm.
For simplicity we will formulate all of the following discussion in terms of the subleading eigenvalue, which we also assume to be unique, real, and nonnegative.
We note, however, that our results hold in full generality upon replacing $\lambda$ by the appropriate norm.

The dynamics is ergodic and mixing if $\lambda$ is strictly smaller than one. 
While for concreteness we have taken all gates $U$ to be identical, this conclusion directly extends to the case where $U$ is time-dependent, as long as a nonvanishing fraction of the gates are sufficiently ergodic.
A similar quantum channel construction appears in the calculation of light-cone correlation functions in dual-unitary circuits, for which the inside of the causal light cone acts as a random environment, and the ergodicity of dual-unitary circuits can be classified in terms of the eigenspectrum of a similar quantum channel~\cite{bertini_exact_2019}. The same classification applies in this model and we repeat it here for convenience: 

\begin{itemize}
\itemsep-0.1em 
\item[(i)] \emph{Non-interacting}: All $d_A^2$ eigenvalues equal 1. All dynamical correlations remain constant.
\item[(ii)] \emph{Non-ergodic}: More than one but less than $d_A^2$ eigenvalues are equal to 1, some dynamical correlations decay to a non-zero constant.
\item[(iii)] \emph{Ergodic and non-mixing}: All nontrivial eigenvalues are different from 1, but there exists at least one nontrivial eigenvalue with unit magnitude. All correlations oscillate around a time-averaged value corresponding to the ergodic value.
\item[(iv)] \emph{Ergodic and mixing}: All nontrivial eigenvalues lie within the unit disc and all dynamical correlations decay (exponentially) to their ergodic value.
\item[(v)] \emph{Maximally ergodic}: All $d_A^2-1$ nontrivial eigenvalues equal zero. Dynamical correlations decay instantaneously to the ergodic value.
\end{itemize}

As an aside, we note that it is possible to obtain a condition on the operator entanglement of the unitary $U$ that guarantees mixing and, when maximized, maximally ergodic dynamics. 
Choosing $d_A=d_C=d$ for convenience, the operator Schmidt decomposition of $U$ is given by
\begin{align}
U = \sum_{j=1}^{d^2} \sqrt{\gamma_j}\, X_j \otimes Y_j,
\end{align}
where $X_j$ and $Y_j$ are single-site operators acting on $A$ and $C$, respectively, orthonormalized such that $\mathrm{Tr}(X_j^{\dagger}X_k) = \mathrm{Tr}(Y_j^{\dagger}Y_k) = \delta_{jk}$. The linear entropy of the gate $U$, or operator entropy in short, is defined as~\cite{rather_creating_2020,aravinda_dual-unitary_2021,rather_construction_2022,aravinda_ergodic_2024}
\begin{align}
E(U) = 1 - \frac{1}{d^4}\sum_{j=1}^{d^2} \gamma_j^2= 1-\frac{1}{d^2}\,\,\figeq[0.12\columnwidth]{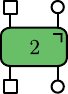}\,\,,
\end{align}
where the folded gate acts on two replicas, such that $\circ=(1)(2)$ and $\smallsquare=(12)$.

This operator entropy is bounded as $0 \leq E(U) \leq 1-1/d^2$. As shown in Appendix~\ref{app:channel} in general and in Ref.~\cite{aravinda_ergodic_2024} for the case of equal Hilbert space dimensions, there exists a threshold value $E_* \equiv 1-2/d^2$ such that the channel is guaranteed to be mixing if $E(U) > E_*$. Intuitively, if the operator is sufficiently entangled, at each time step quantum information is guaranteed to `leak' into the environment, resulting in correlations that decay to the ergodic value. The proof is based on a similar proof from Ref.~\cite{aravinda_dual-unitary_2021} bounding the ergodicity of dual-unitary gates in terms of their entangling power. 

Furthermore, for the maximal value $E(U)=1-1/d^2$ the corresponding quantum channel is guaranteed to be maximally ergodic. In this limit $U$ is dual-unitary, which enforces \emph{all} information to leak into the environment after a single time step, consistent with the use of dual-unitary circuits as minimal models of maximal quantum chaos (for a recent review, see Ref.~\cite{bertini_exactly_2025}). Dual-unitary gates are unitary along the horizontal (spatial) direction~\cite{gopalakrishnan_unitary_2019,bertini_exact_2019}, which can be graphically represented as
\begin{align}\label{eq:channel:du}
    \figeq[0.12\columnwidth]{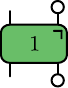} \,\, \propto \,\,
    \figeq[0.025\columnwidth]{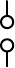}
\end{align}
This condition directly implies that $\mathcal{M}$ is a projector on the identity. 
Interestingly, the above bound implies that the ergodic behavior of dual-unitary gates is stable for gates `close to' dual-unitarity, where the proximity to dual-unitarity is quantified by the operator entanglement. The operator entanglement previously appeared as a measure for proximity to dual-unitarity in the study of $2$-OTOCs~\cite{rampp_dual_2023}, similar to the appearance of the closely related entanglement velocity in the study of entanglement dynamics~\cite{zhou_maximal_2022}.
As discussed in the remainder of this paper, ergodicity on the level of correlation functions guarantees freeness at late times, such that ergodicity on the level of $k$-OTOCs is structurally stable near dual-unitarity.
The structural stability of dual-unitarity is more generally an open question~\cite{kos_correlations_2021,riddell2024structural}.

In the remainder of this paper we focus exclusively on the ergodic and mixing case, corresponding to a subleading eigenvalue $\lambda<1$.

\subsection{Example: $k=2$}

We now consider $k=2$, returning the usual OTOC
\begin{align}
 C^{(k=2)}_{ab}(t) = \lim_{d_E\to\infty} \bbE \, \frac{1}{D}\mathrm{Tr}\left[A(t)BA(t)B\right],
\end{align}
and show how the underlying combinatorial structure first emerges.
In this case there are two partitions, both noncrossing, $\circ = (1)(2)$ and $\smallsquare = (12)$, and multichains are characterized by domain walls between $\circ$ and $\smallsquare$.
The resulting $2$-OTOC follows as a sum over such domain-wall configurations
\begin{align}\label{eq:2OTOC_summation}
  C_{ab}^{(k=2)}(t) = \frac{d_C}{d_A}\sum_{s=1}^t\,\, \figeq[0.27\columnwidth]{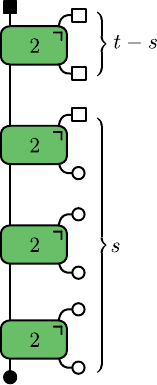}\!\!\!\!\!\!\!\!\!\!\!\!\!-\frac{1}{d_A}\sum_{s=1}^{t-1}\,\, \figeq[0.27\columnwidth]{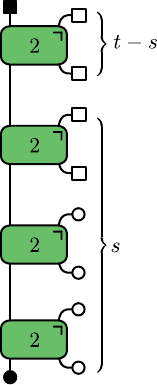}\, .
\end{align}
The first sum runs over domain walls at half-integer time steps characterized by $\sigma_i=\circ$ and $\nu_i=\smallsquare$, whereas the second sum runs over domain walls at integer time steps corresponding to $\nu_i=\circ$ and $\sigma_{i+1}=\smallsquare$. The latter are accompanied by a non-trivial factor of $W_{\circ\sq}=-1/d_C$. 
Such domain wall configurations are ubiquitous in the study of $2$-OTOCs in quantum circuit models, where for random quantum circuits the dynamics of the $2$-OTOC was originally understood through an effective model describing the dynamics of such domain walls as a biased random walk~\cite{nahum_operator_2018,von_keyserlingk_operator_2018,fisher_random_2023}.

Similar as in the previous section, the $2$-OTOC decays to zero in the long-time limit provided the quantum channel $\mathcal{M}$ is ergodic and the operators are traceless. 
This is again best seen when taking $a$ and $b$ to be right/left eigenoperators of the single-particle channel $\mathcal{M}$ with eigenvalue $\lambda$. It follows from the factorization of $\mathcal{M}_{\circ\circ}$ and $\mathcal{M}_{\sq \sq}$ that
\begin{align}
\mathcal{M}_{\circ\circ} |a_{\circ}) = \lambda^{2} |a_{\circ}) ,\quad (b_{\sq}|\mathcal{M}_{\sq \sq} = \lambda^{2} (b_{\sq}|. \label{eq:propagation_of_boundaries}
\end{align}
These identities can be used to `propagate' the boundaries through the sequences of transfer matrices to return
\begin{align}\label{eq:2OTOC:eigen_decay}
C_{ab}^{(k=2)}(t) &= \lambda^{2t-2} t  \frac{d_C (b_{\sq}|\mathcal{M}_{\sq \circ}|a_{\circ})}{d_A} \nonumber\\
&\qquad \qquad-\lambda^{2t} (t-1)\frac{(b_{\sq}|a_{\circ})}{d_A} \, , 
\end{align}
indicating exponential decay of OTOCs, as illustrated in Fig.~\ref{fig:kOTOC}~(a) below.

The decay of the OTOC as $C_{AB}^{(k=2)}(t)  \propto t \lambda^{2t}$ can be argued to hold for generic traceless operators.
Graphically, the factorization of $\mathcal{M}_{\circ\circ}|a_{\circ})$ can be made apparent by noting that both $\mathcal{M}_{\circ\circ}$ and $|a_{\circ})$ factorize in the same way,
\begin{align}
  \mathcal{M}_{\circ \circ}|a_{\circ})  = \,\,  \figeq[0.15\columnwidth]{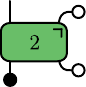}= \frac{1}{d_C^2}\,\,  \figeq[0.21\columnwidth]{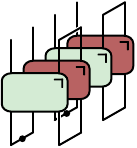}\,\,,
\end{align}
where the first and second layers are decoupled from the third and fourth layers, returning two copies of $\mathcal{M}(a)$:
\begin{align}
     \figeq[0.21\columnwidth]{fig_2OTOC_bottom_1} = 
     \figeq[0.14\columnwidth]{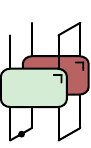} \otimes
     \figeq[0.14\columnwidth]{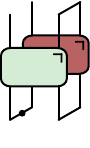} \, .
\end{align}
A similar factorization holds for $(b_{\sq}|$ and $\mathcal{M}_{\sq \sq}$,
\begin{align}
 (b_{\sq}| \mathcal{M}_{\sq \sq} = \,\,  \figeq[0.15\columnwidth]{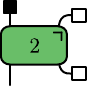}= \frac{1}{d_C^2}\,\,  \figeq[0.21\columnwidth]{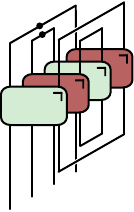}\,\,,
\end{align}
where the first and fourth layers are now decoupled from the second and third layers, such that the right-hand side factorizes as
\begin{align}
     \figeq[0.21\columnwidth]{fig_2OTOC_top_1} = 
     \figeq[0.21\columnwidth]{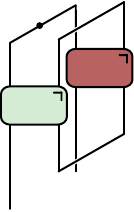} \otimes\,\,
     \figeq[0.14\columnwidth]{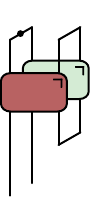} \,\,.
\end{align}
The latter factorization returns the former when shifting by `half a replica', i.e., by regrouping the backward sheet of the original replica $i$ with the forward sheet of replica $i+1$.
This picture has a precise formulation on the level of the noncrossing partitions as it corresponds to mapping a nonrossing partition $\sigma$ to its Kreweras complement $\sigma^*=\sigma^{-1}\smallsquare$, see Appendix~\ref{app:FP}. For $k=2$ this yields $\circ^*=\smallsquare$ and $\smallsquare^*=\circ$.

This factorization fails when acting with $\mathcal{M}_{\sq \circ}$. However, for a total number of $t$ time steps there can only be a single domain wall and corresponding `off-diagonal' map $\mathcal{M}_{\sq \circ}$, such that the remaining $t-1$ other maps are `diagonal' maps $\mathcal{M}_{\circ \circ}$ or $\mathcal{M}_{\sq \sq}$. Using the same arguments as in the previous section, we can bound
\begin{align}
    &|| \mathcal{M}_{\circ \circ}^s |a_{\circ}) ||_2 \leq \lambda^{2s}\, ||a||_2^2 , \\
    &|| (b_{\sq}|\mathcal{M}^{t-s}_{\sq\sq} ||_2 \leq \lambda^{2(t-s)}\,||b||_2^2\,.
\end{align}
By combining these two bounds, it directly follows that every term in Eq.~\eqref{eq:2OTOC_summation} decays as $\propto \lambda^{2t}$. Since the total number of terms scales as $t$, we expect that the $2$-OTOC decays to the ergodic zero value $\propto t \lambda^{2t}$, consistent with Eq.~\eqref{eq:2OTOC:eigen_decay}.

Compared to the $k=1$ case the decay is now twice as fast due to the two copies of the channel $\mathcal{M}$ in $\channel{\circ}{\circ}$ and $\channel{\sq}{\sq}$ at each timestep, respectively.
Additionally, due to the increasing number of multichains, the exponential decay is accompanied by a linear in $t$ prefactor.
This linear corrections are a first hint towards an underlying Jordan structure, which we discuss in later sections.

If the observables are not traceless, the steady-state value can be calculated by repeatedly subtracting the appropriate traces, i.e., by computing the OTOC for the traceless observables $a_i - \varphi(a_i)$ and $b_i-\varphi(b_i)$, to return
\begin{align}
\label{eq:2OTOC_steady_state}
  &\lim_{t\to \infty}C_{ab}^{(k=2)}(t) = - \varphi(a_1)\varphi(a_2)\varphi(b_1)\varphi(b_2) \nonumber\\
  &\quad+ \varphi(b_1 b_2) \varphi(a_1)\varphi(a_2)+\varphi(a_1 a_2) \varphi(b_1)\varphi(b_2) \,.
\end{align}
We will return to this result in later sections. 

\subsection{General $k$}

We now turn our attention to the decay of general $k$-OTOCs. For traceless observables we will argue that they exhibit the same exponential decay $\propto \lambda^{2t}$ as in the $k=2$ case, albeit with a modified polynomial prefactor due to the growing number of multichains.

For $k=2$ there are only two permutations, i.e. the identity of the cyclic permutation, guaranteeing a `matching' between the permutations in the linear maps $\mathcal{M}_{\nu \sigma}$ and one of the two boundaries, such that the decay directly follows from the product of two quantum channels $\mathcal{M}$ acting on the operators $a$ ($b$) from the right (left). For $k >2$ the summation can however involve permutations $\sigma$ that are distinct from both the identity and cyclic permutation, and no such factorization can be performed. This lack of factorization results in a slower decay compared to the scaling $\propto \lambda^{kt}$  which would naively be expected. 

For concreteness, we first illustrate this argument for $k=3$ and $\sigma = (12)(3)$ and consider a `diagonal' path of the form $(b_{\sq}|\mathcal{M}_{\sigma \sigma}^t|a_{}\circ)$. Wile not technically part of the summation in Eq.~\eqref{eq:OTOC_summation} due to the different boundary conditions, this example helps to illustrate the general argument. Graphically, 
\begin{align}\label{eq:kOTOC:path_illustration}
     (b_{\sq}|\mathcal{M}_{\sigma \sigma}^t|a_{\circ}) \propto \,\,\figeq[0.28\columnwidth]{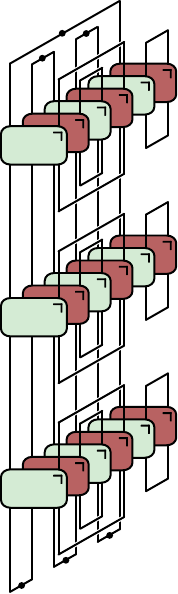}\,\,,
\end{align}
where for concreteness we have fixed $t=3$. Neither the action on $a$ nor $b$ factorizes fully, but we can identify a sequence of the form $\mathcal{M}^t(a)$ in the third replica (starting from the bottom). We can isolate this term to factorize Eq.~\eqref{eq:kOTOC:path_illustration} as
\begin{align}
     \figeq[0.28\columnwidth]{fig_kotoc_fac_0}\,\,=\,\,\figeq[0.21\columnwidth]{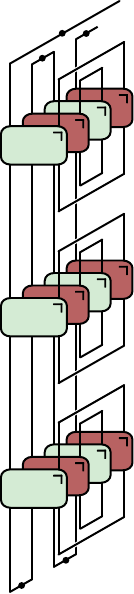}\,\, \times \,\,
     \figeq[0.15\columnwidth]{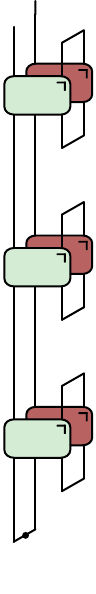}  \, .
\end{align}
Here the multiplication should be interpreted as performing the appropriate contractions at the top. 
The absolute value of this expression can be bounded as the product of the operator norms of both terms.
Since $\mathcal{M}^t(a)$ decays exponentially with $t$, the norm of the second term can again be bounded by $\lambda^t ||a||_2$ (when reintroducing the appropriate prefactors of $d_C$).

The first term on the right-hand side can now be bounded by identifying a sequence where the quantum channels act to the left on $b$ (starting from the top), which can be factorized as
\begin{align}
     \figeq[0.21\columnwidth]{fig_kotoc_fac_1}\,\,=\,\,\figeq[0.21\columnwidth]{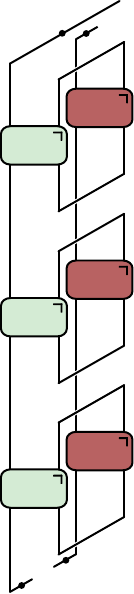}\,\, \times \,\,
     \figeq[0.14\columnwidth]{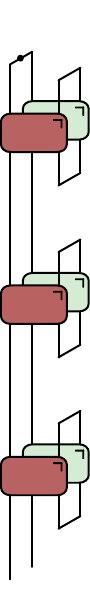} \,\,.
\end{align}
Multiplication should here be interpreted as performing the appropriate contractions at the bottom.
The norm of the second term on the right-hand side, i.e. the sequence of quantum channels acting on $b$ from the left, can again be bounded as $\lambda^t ||b||_2$. Combining this bound with the previous bound, we recover an exponential decay as $\lambda^{2t}$ from the two sequences of quantum channels.

No further simplifications can be made, since at no other point can we identify sequences of the quantum channel acting on a traceless operator. Indeed, the remaining term generally does not vanish at late times but rather decays to a constant. If $a^2$ and $b^2$ are not traceless, then contracting this term with the identity at the top and the bottom returns
\begin{align}
 \lim_{t\to \infty}\,\, \figeq[0.27\columnwidth]{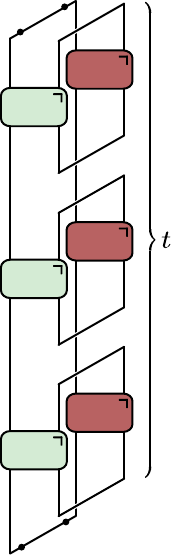} \,\, = \mathrm{Tr}(a^2)\mathrm{Tr}(b^2) = \mathcal{O}(1)\,,
\end{align}
again when reintroducing the appropriate factors of $d_C$.

Taking these results together, we find that we can bound Eq.~\eqref{eq:kOTOC:path_illustration} as
\begin{align}
    |(b_{\sq}|\mathcal{M}_{\sigma \sigma}^t|a_{}\circ)| \leq \lambda^{2t} ||a||_2^k ||b||_2^k\,,
\end{align}
for $\sigma=(12)(3)$.

This argument can now be extended to generic noncrossing partitions and the $k$-OTOC [Eq.~\eqref{eq:OTOC_summation}]. For sequences of `diagonal' quantum channels we have the bound
\begin{align}
    ||(b_{\sq}|\mathcal{M}_{\sigma \sigma}^t|a_{\circ}) ||_2 \leq \lambda^{(n(\sigma)+n(\sigma^*))t}\, ||a||_2^k \, ||b||_2^k \,,
\end{align}
where $n(\sigma)$ is the number of singletons (cycles of length one) in $\sigma$, and $n(\sigma^*)$ is the number of singletons in its Kreweras complement. The former describes the number of quantum channels acting from the right (bottom) on $a$, i.e. the number of sequences $\mathcal{M}^t(a)$ that can be identified, whereas the latter describes the number of quantum channels acting from the left (top) on $b$. For $\sigma=(12)(3)$ we have $n(\sigma)=n(\sigma^*)=1$, returning the previous result for Eq.~\eqref{eq:kOTOC:path_illustration} establishing a decay $\propto \lambda^{2t}$.
The diagonal channels corresponding to the identity and the cyclic permutation always decay as $\lambda^{kt}$, since $n(\circ)=n(\smallsquare^*)=k$ and $n(\smallsquare) = n(\circ^*)=0$. 
The number of singletons in a permutation $\sigma$ satisfies $n(\sigma)+n^*(\sigma) \geq 2$, where the inequality is satisfied for the permutations in the middle of the noncrossing partition lattice. 
For any $k >2$ there are permutations [see for instance the two partitions in the middle of Fig.~\ref{fig:NC_factorizing} below] in which the combined number of singletons equals 2, independent of $k$, such that the slowest decay is given by $\propto \lambda^{2t}$. 

These arguments can be directly extended to bound the contribution $(b_{\sq}|\mathcal{M}_{\Sigma} |a_\circ)$ for a generic nondiagonal multichain 
$\Sigma = (\sigma_1 \subseteq\nu_1\subseteq \ldots \subseteq \nu_t)$ of length $2t$ and $\mathcal{M}_{\Sigma}$ the corresponding product of channels. 
As there are at most $k$ distinct partitions along the multichain at most $k-1$ of the channels $\mathcal{M}_{\nu_i \sigma_i}$ are offdiagonal, i.e., $\nu_i \neq\sigma_i$.
For $t$ sufficiently large, the majority of channels are hence diagonal.
By definition of the order of the permutations the above argument can be directly repeated, such that we can upper bound the contribution from the whole multichain 
\begin{align}
    |(b_{\sq}|\mathcal{M}_{\Sigma}|a_\circ)|\leq \lambda^{m(t-k+1)}\|a\|^k_2\|b\|^k_2 \, ,
    \label{eq:bound_multichain}
\end{align}
where $t-k+1$ and $m=\min\{n(\sigma) + n^*(\sigma), \sigma \in \textrm{NC}(k)\}$ are the minimal number of diagonal channels along the chain and the minimal possible number of singletons within this chain, respectively.

Each multichain additionally comes with a prefactor $\prod_{i=1}^{t-1}W_{\nu_i\sigma_{i+1}}$ from the Weingarten functions~\eqref{eq:Weingarten_C}.
Again, at most $k-1$ terms in this product are different from unity and those terms are bounded in modulus by the maximum value the Möbius function takes. This prefactor can hence be bounded by a constant that only depends on $k$ and does not scale with $t$.

The above bound applies for every multichain, and we can get an estimate for the decay by combining the total number of multichains, scaling as $t^{k-1}$, with the slowest decay, scaling as $\lambda^{2t}$. Putting all of the above together it follows that there exist a constant $K_k$ such that
\begin{align}
    \label{eq:kOTOC_bound}
    \big|\kOTOC(t)\big|\leq K_kt^{k-1}\lambda^{2t}\|a\|^k_2\|b\|^k_2.
\end{align}
At late times all $k$-OTOCs hence decay at least exponentially up to polynomial corrections.
At sufficiently late times the above bound returns an exponential decay $\propto \lambda^{2t}$ and by considering eigenoperators $a_\lambda$ of the channel $\mathcal{M}$ this asymptotic decay is saturated (at least for $d_A\geq 3$; see below for $d_A=2$). 

\begin{figure}[t]
	\centering
	\includegraphics[width=\linewidth]{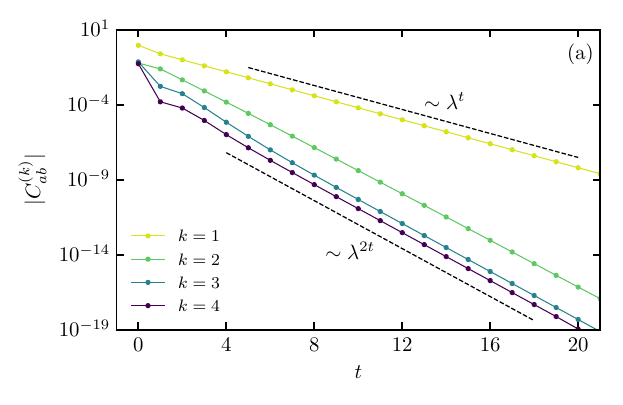}
	\includegraphics[width=\linewidth]{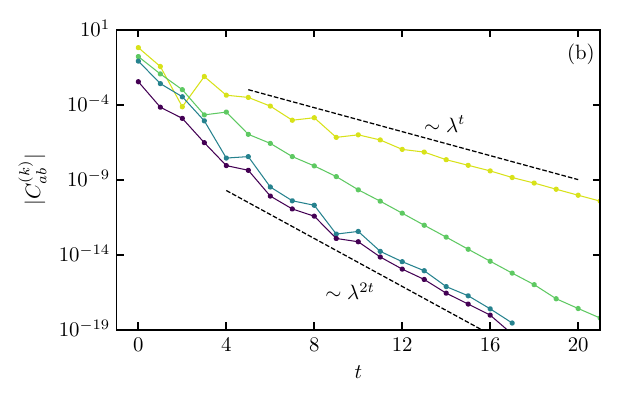}
	\caption{$k$-OTOC for $d_A=d_C=3$ for (a) left and right eigenoperators $a_\lambda$ and $b_{\lambda}$ and (b) generic traceless $a$ and $b$ for different $k$ and a gate $\gateboundary$ with $\lambda=0.398$. The asymptotic decay $\propto \lambda^{2t}$ ($k=1$) and $\propto \lambda^{2t}$ ($k\geq 2$), respectively, are depicted as dashed lines.
    \label{fig:kOTOC}}
\end{figure}

This scaling is compared with numerical simulations in Fig.~\ref{fig:kOTOC} for a generic operator and for an eigenoperator of the quantum channel, showing excellent agreement.
Both from the theoretical bound and the numerical data, it follows that the decay of the $k$-OTOC is the same for all $k\geq 2$, implying that higher-order OTOCs with $k>2$ can not be decomposed into two-point functions, highlighting the chaotic nature of our model. This result can be seen as the dynamical manifestation of higher-order correlations in the full ETH ansatz.

\begin{figure}[t]
	\centering
	\includegraphics[width=\linewidth]{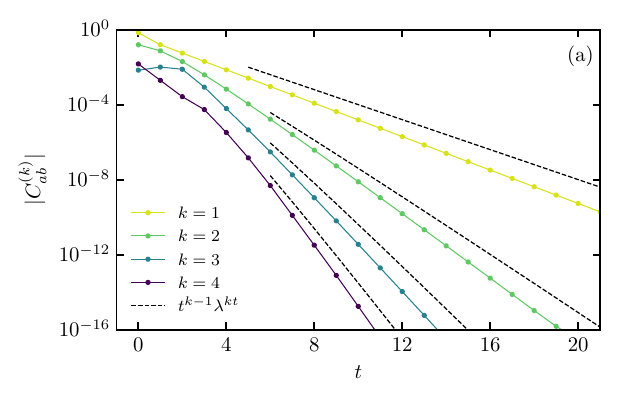}
	\includegraphics[width=\linewidth]{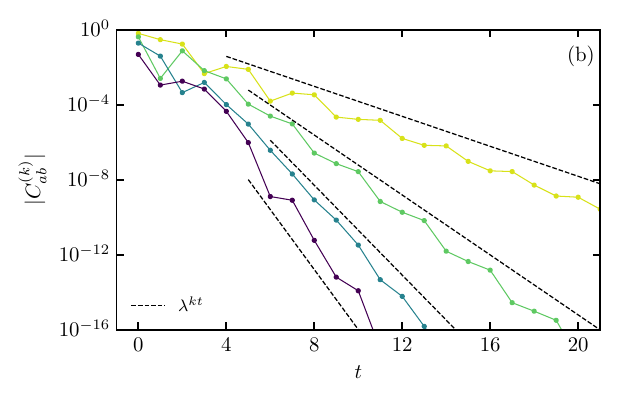}
	\caption{$k$-OTOCs for $d_A=d_C=2$ for (a) left and right eigenoperators $a_\lambda$ and $b_{\lambda}$ and (b) generic traceless $a$ and $b$ for different $k$ and a gate $\gateboundary$ with $\lambda=0.358$. The asymptotic decay (a) $\propto t^{k-1}\lambda^{kt}$ and (b) $\propto \lambda^{kt}$ is depicted as dashed lines.
    \label{fig:kOTOC_qubits} }
\end{figure}

This bound can be further sharpened when the subsystem $A$ hosts a single qubit, i.e. $d_A=2$, and all the local observables $a_1=\ldots = a_k=a$ and $b_1 = \ldots = b_k = b$ are the same. 
In this case being Hermitian and traceless directly implies $a^2=b^2 \propto \bbId_A$. 
Consider again the example of Eq.~\eqref{eq:kOTOC:path_illustration}: Fixing $a = a_{\lambda}$ and $b = b_{\lambda}$ to be decaying left and right eigenoperators of the quantum channel and propagating these through the multichain, the total expression is proportional to $\tr(b_{\lambda} a_\lambda b_{\lambda}) \propto \tr(a_\lambda)=0$, using that the subleading eigenoperator $a_\lambda$ is orthogonal to the leading one $\bbId_A$ and therefore traceless. 
This relation gives rise to additional cancellations and leads to an asymptotic decay of $k$-OTOC of the asymptotic form
\begin{align}
    \kOTOC(t)\propto\lambda^{kt},
    \label{eq:kOTOCs_qubits}
\end{align}
showing faster decay if $k\geq3$. For a general sequence of diagonal quantum channels $\mathcal{M}_{\sigma \sigma}$ the decay is determined by the unique eigenvector of the linear map $|(a_{\lambda})_\sigma)$ with eigenvalue $\lambda^k$, i.e. the permutation operator $|\sigma)$ where every replica is dressed with an eigenoperator $a_{\lambda}$ [as in Eqs.~\eqref{eq:fullcirc} and \eqref{eq:fullsq}]. This behavior is illustrated in Fig.~\ref{fig:kOTOC_qubits}. 

\subsection{Steady State}
\label{sec:steady_state_multichains}

Having established that the $k$-OTOCs between traceless observables decays to zero at late times , we now turn our attention to the steady-state value of the $k$-OTOCs once the constraint on having zero trace is lifted. An example of which is given in Eq.~\eqref{eq:2OTOC_steady_state} for $k=2$, whereas we provide a systematic derivation and interpretation for arbitrary $k$ in the following.
More precisely, we characterize the limit $\lim_{t \to \infty} \kOTOC(t)$ in terms of generalized moments $\varphi_{\sigma^*}$ and free cumulants $\kappa_\nu$ between the observables  [cf. Eqs.~\eqref{eq:intro:factorization_moments_cumulants} and \eqref{eq:intro:phi_from_kap}].
While this expression can be directly obtained from free probability or by repeatedly subtracting the non-traceless part of the operator, it is instructive to see how these expressions appear here.

To obtain the steady state, we observe that allowing for $\varphi(a_i)\neq0\neq\varphi(b_i)$ leads to nonvanishing overlaps with the leading eigenoperator $\bbId_A$ of the channel $\mathcal{M}$. The steady state is governed by the leading eigenoperators of the channels $\mathcal{M}_{\nu\sigma}$. By unitarity of the local gates $U$, it follows that the leading left and right eigenoperators are permutation states, obeying
\begin{align}
    \mathcal{M}_{\nu\sigma}|\sigma) = d_C^{-k + |\nu^{-1}\sigma|}|\sigma),
\end{align}
and 
\begin{align}
   (\nu| \mathcal{M}_{\nu\sigma} = d_C^{-k + |\nu^{-1}\sigma|}(\nu| .
\end{align}
For the diagonal channels $\nu=\sigma$, such that the leading left and right eigenoperators coincide and the leading eigenvalue is 1.
Replacing the channels $\mathcal{M}_{\nu\sigma}$ along each multichain by the projector onto the leading eigenoperators, which after proper biorthonormalization reads
\begin{align}
    \channel{\nu}{\sigma} = d_C^{-k + |\nu^{-1}\sigma|} d_A^{- |\nu^{-1}\sigma|}|\sigma)(\nu| + \ldots
    \label{eq:channels_leading}
\end{align}
returns the stationary state.
Here the dots represent subleading terms (bi)orthogonal to the leading eigenoperators, which give rise to decaying contributions to the $k$-OTOC.

The steady state can be obtained by replacing the channels $\mathcal{M}_{\nu\sigma}$ by the above projection, simplifying the resulting overlaps between permutation states, and using the combinatorial properties of the Möbius function $\mu(\nu, \sigma)$. Inserting the definition of the Weingarten functions~\eqref{eq:Weingarten_C} and repeatedly using $|\sigma^{-1}\rho|+|\rho^{-1}\nu| = k + |\sigma^{-1}\nu|$ for $\sigma \subseteq\rho \subseteq \nu$, for each multichain $\Sigma = \circ \subseteq \nu_1 \subseteq \ldots \subseteq \sigma_{t-1} \subseteq \smallsquare$ the contribution to Eq.~\eqref{eq:OTOC_summation} at late time reads
\begin{align}
C_{ab, \Sigma}^{(k)}(t) = &\left(\prod_{i=1}^{t-1}\mu(\sigma_i, \nu_i)\right) \nonumber\\
&\qquad \times\varphi_{\nu_1}(a_1,\ldots,a_k)\varphi_{\sigma_{t-1}^*}(b_1,\ldots,b_k),
\end{align}
up to subleading terms which vanish as $t \to \infty$. 
Here we additionally used, that the overlaps of the permutation states with the boundary conditions  yield the moments, e.g., 
$d_A^{-|\nu|}(\nu|a_\circ)=\varphi_\nu(a_1,\dots,a_k)$.
To perform the summation over all multichains we first perform the sum over all multichains with fixed $\nu_1=\nu$ and $\sigma_{t-1}=\sigma$, since
\begin{align}
    \sum_{\substack{\dots \subseteq \sigma_i \subseteq \nu_i \subseteq \dots \\\nu_1 = \nu,\sigma_{t-1}=\sigma}}\prod_{i=1}^{t-1}\mu(\sigma_i, \nu_i) = \mu(\sigma,\nu),
\end{align}
where we repeatedly use the defining property of the Möbius function [Eq.~\eqref{eq:sumrule_moebius_intro}], here repeated for convenience:
\begin{align}
    \sum_{\sigma \subseteq \rho \subseteq \nu}\mu(\nu,\rho)  = \delta_{\nu,\sigma}\, .
\end{align}
The steady state for the $k$-OTOC follows as
\begin{align}
\label{eq:steadystate_multichain}
    \lim_{t \to \infty}\kOTOC(t) &= \sum_{\nu \subseteq \sigma}
    \mu(\nu, \sigma)\varphi_\nu(a_1,\ldots,a_k)\varphi_{\sigma^*}(b_1,\ldots,b_k)  \nonumber\\
    &=\sum_{\sigma\, \in \, \textrm{NC}(k)}
    \varphi_{\sigma^*}(b_1,\ldots,b_k)\kappa_\sigma(a_1,\ldots,a_k) \, ,
\end{align}
where we have used that the free cumulants are related to the moments through the M\"obius function [see Eq.~\eqref{eq:intro:kap_from_phi}].

This expression describes the correlations between freely independent observables and corresponds to the correlations between time evolved observables and static observables if we were to replace the evolution operator $\evolutionoperator(t)$ by a Haar random unitary and subsequently taking the thermodynamic limit \cite{nica2006lectures}.
In other words, within the minimal model presented here we recover the random matrix result in the thermodynamic limit at late times.

\subsection{Approach to Equilibrium}
\label{sec:approach_to_equilibrium_multichains}

\begin{figure}[]
	\centering
	\includegraphics[width=\linewidth]{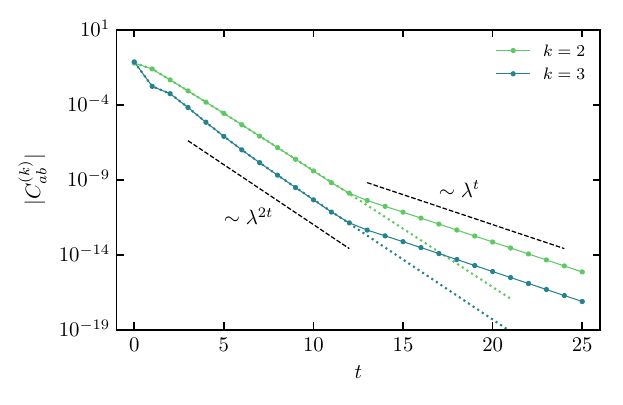}
	\caption{$k$-OTOCs for $d_A=d_C=3$ for $k=2,3$ and a gate $U$ with $\lambda=0.398$. 
    The observables are left and right eigenoperators $a_\lambda$ and $b_{\lambda}$, respectively, except for $a_0=a_\lambda + \epsilon\, \bbId_A$ with $\epsilon=10^{-4}$ ($k=2$) and $\epsilon=10^{-5}$ ($k=3$).
    The decay at initial times $\propto \lambda^{2t}$ and at late times $\propto \lambda^t$ is represented by dashed lines, while the $k$-OTOCs for $\epsilon=0$ are shown as dotted lines.
    }
    \label{fig:kOTOC_2step}
\end{figure}

Having established the decay of $k$-OTOCs between traceless observables as well as the emergence of the steady state for observables with nonzero trace, we combine both aspects in the following to obtain yet a different decaying behavior $\propto \lambda^t$ for arbitrary $k$. Provided the observables are no longer traceless, there will be nontrivial contributions from single sequences of quantum channels leading to a decay $\propto \lambda^t$, corresponding to a two-time correlation function being factored out.
This decay is illustrated in Fig.~\ref{fig:kOTOC_2step} and will be discussed in detail in Sec.~\ref{sec:approach_IM}.
Interestingly, the presence of a single operator that is not traceless can lead to a two-step decay, a recent topic of interest in OTOC dynamics~\cite{bensa_two-step_2022,znidaric_two-step_2023,jonay_physical_2024,jonay_two-stage_2025}. Choosing a single operator to be traceless, the steady-state value of the $k$-OTOC is still zero, but the decay to this steady state is modified from $\propto \lambda^{2t}$ to $\propto \lambda^t$. If this non-traceless part is sufficiently small, the term decaying $\propto \lambda^t$ is perturbative, leading to a two-stage relaxation of the OTOC, as illustrated in Fig.~\ref{fig:kOTOC_2step}.

To conclude this section, we consider the maximally ergodic case when $U$ is dual-unitary~\cite{bertini_exactly_2025}. In this case the quantum channel~\eqref{eq:def_channel} is a projector on the identity and all correlation functions for $k=1$ reach their steady-state ergodic value after a single time step [see Eq.~\eqref{eq:channel:du}]. 
For $k \geq 2$ the OTOC is nonvanishing after a single time step, since $(b_{\sq}|\mathcal{M}_{\sq \circ}|a_{\circ}) \neq 0$, but, in accordance with the previous discussion, for traceless observables the $k$-OTOC with $k \geq 2$ vanish identically after two time steps. Consider e.g. again $k=3$ and $\sigma=(12)(3)$, then
\begin{align}
    \mathcal{M}_{\sigma \circ}|a_{\circ}) \propto \figeq[0.28\columnwidth]{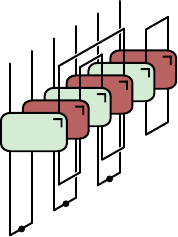} = \figeq[0.21\columnwidth]{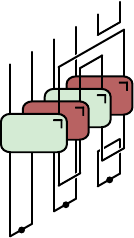} =0,
\end{align}
where in the first equality we have used dual-unitarity for the two gates in the third replica, corresponding to the singleton in $\sigma$, and in the second equality we have used the tracelessness. More generally, $\mathcal{M}_{\sigma \circ}|a_{\circ})$ vanishes whenever $\sigma$ has a singleton, whereas $(b_{\sq}|\mathcal{M}_{\sq \nu }$ vanishes whenever $\nu^*$ has a singleton. Combined with the constraint that $\sigma \subseteq \nu$, which implies that every singleton in $\nu$ is a singleton in $\sigma$ and that every singleton in $\sigma^*$ is a singleton in $\nu^*$, it immediately follows that either $\sigma$ has a singleton or $\nu^*$ has a singleton, resulting in an identically vanishing $k$-OTOC after two time steps.
For traceless observables and dual-unitary dynamics, it follows that
\begin{align}
    C_{ab}^{(k=1)}(t \geq 1)  = 0, \qquad C_{ab}^{(k=2)}(t \geq 2) = 0.
\end{align}
For operators that are not traceless, the non-traceless part can again be factored out, such that the steady-state value of Eq.~\eqref{eq:steadystate_multichain} is reached after two discrete time steps. These results further cement dual-unitary circuits as maximally ergodic/chaotic models of many-body quantum dynamics.

\section{Markovian Influence Matrix Dynamics}
\label{sec:influence_matrix}

In this section we show how the obtained expression for the $k$-OTOC [Eq.~\eqref{eq:OTOC_summation}] can be recast in an influence matrix formalism, which captures the effect of the random matrix environment on the local subsystem. This influence matrix corresponds to a matrix product state acting on the temporal lattice, where the number of sites corresponds to the number of time steps. 
As opposed to influence matrices for correlation functions, where this influence matrix is unentangled, the minimal structure imposed by the random environment results in a low but nonvanishing (temporal) entanglement for $k$-OTOCs with $k \geq 2$. This entanglement can be bound by the necessary bond dimension, which for the $k$-OTOC corresponds to the number of noncrossing partitions of $k$ elements, i.e. the Catalan numbers $C_k$.
The auxiliary noncrossing partition degree of freedom effectively tracks the system’s position in the noncrossing partition lattice, encoding how different replicas have interacted over time, i.e., which replicas have become dynamically correlated due to system-bath interactions.

In this way the influence-matrix representation can be used to recast the $k$-OTOC dynamics as a Markovian process, provided we keep track of this additional noncrossing partition degree of freedom. This approach allows for numerical simplifications when expressing the summation in Eq.~\eqref{eq:OTOC_summation} and has the advantage that the leading eigenstates of this Markovian process can be constructed analytically. These eigenstates can be labeled by noncrossing partitions, which directly returns the late-time dynamics in terms of free cumulants indexed by noncrossing partitions. In this way we recover the expansion of the $k$-OTOC predicted by full ETH, with the different free cumulants satisfying the expected factorization properties from free probability.

\subsection{Influence matrix representation}
The summation over different paths in the expression for the $k$-OTOC [Eq.~\eqref{eq:OTOC_summation}] can be recast in a more compact form by treating the noncrossing partition degrees of freedom as states in an auxiliary Hilbert space. 
We now represent the $k$-OTOC of Eq.~\eqref{eq:OTOC_summation} as
\begin{align}\label{eq:IM}
  C_{ab}^{(k)}(t) = \,\, \figeq[0.24\columnwidth]{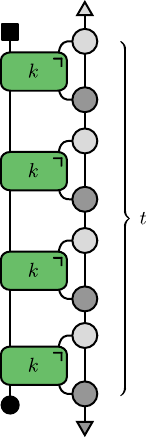}
\end{align}
The relevant influence matrix $\mathcal{I}_k$ corresponds to a matrix product state acting on the temporal Hilbert space of $2t$ sites as
\begin{align}
  \mathcal{I}_k\,\, =\,\,  \figeq[0.6\columnwidth]{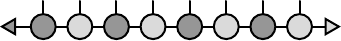}\,,
  \label{eq:influence_matrix}
\end{align}
here rotated by $90^{\circ}$ for convenience.
This matrix product state is defined in terms of tensors 
\begin{align}
    \figeq[0.077\columnwidth]{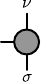} = \, \figeq[0.084\columnwidth]{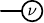}\,\,  W_{\nu\sigma} \,  \zeta_{\sigma \nu}\,, \qquad 
    \figeq[0.077\columnwidth]{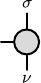} = \, \figeq[0.084\columnwidth]{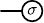}\,\, \zeta_{\nu  \sigma}\,,
\end{align}
where the states in this auxiliary space are labelled by noncrossing partitions such that the bond dimension equals $C_k$. The `physical' space is $\Hil_C^{\otimes 2k}$, i.e., the $(d_C)^{2k}$-dimensional space of the replicas of $C$. These tensors enforce the noncrossing partitions to be non-decreasing in time by introducing the zeta function of the noncrossing lattice: $\zeta_{\sigma \nu} = 1$ if $\sigma \subseteq \nu$ and zero otherwise, see Appendix~\ref{app:FP}.
Moreover, the tensors introduce the M\"obius functions $\mu(\nu, \sigma)$ encoded in the Weingarten functions $W_{\nu \sigma}$ at alternating time steps.

The boundary vectors are defined as
\begin{align}
    \figeq[0.036\columnwidth]{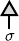} \,\,=\,\, (d_A d_C)^{-1}\delta_{\sigma {\sq}}\,, \qquad  \figeq[0.036\columnwidth]{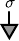} \,\,=\,\, d_C^{|\sigma|}\,. 
\end{align}
These boundary vectors are chosen such that the partitions at the top and bottom correspond to the cyclic permutation and the identity permutation, respectively, and the prefactor from Eq.~\eqref{eq:OTOC_summation} has been distributed over these terms to simplify later expressions.
 
For the contraction at the bottom we have that
\begin{align}
    \figeq[0.077\columnwidth]{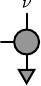} \,\,=\,\, \figeq[0.07\columnwidth]{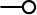} \,\,d_C^k\,\delta_{\nu\circ}\,,
\end{align}
using the properties of the geodesics and the M\"obius function,
\begin{align}
    \sum_{\sigma \subseteq \nu}& \mu(\nu,\sigma)\,d_C^{-k+|\nu^{-1}\sigma|+|\sigma|} \nonumber\\
    &=\sum_{\sigma \subseteq \nu} \mu(\nu,\sigma)\,d_C^{|\nu|} = d_C^{k}\,\delta_{\nu\circ}\,.
\end{align}
For the contraction at the top we directly find that
\begin{align}
    \figeq[0.077\columnwidth]{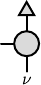} \,\,=\,\, \figeq[0.07\columnwidth]{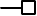}\,\,(d_A d_C)^{-1}   \,\,.
\end{align}

\subsection{Example: $k=1$}

For $k=1$ there is only a single partition, $\circ = (1)$, such that this matrix product state reduces to a product wave function on the temporal lattice. We can write
\begin{align}
    \mathcal{I}_{k = 1} \,\,\propto \,\,\figeq[0.3\columnwidth]{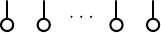}\,\,,
\end{align}
where the number of sites corresponds to $2t$.
Consequently, the diagrammatic representation in Eq.~\eqref{eq:influence_matrix} reduces to Eq.~\eqref{eq:OTOC_k1}.
This influence matrix was previously obtained in Ref.~\cite{sonner_influence_2021}, where it was termed the `perfect dephaser' limit. For such influence matrices the environment cancels out all interference effects and acts as a perfectly Markovian bath. This perfect dephaser limit was analytically shown to be the influence matrix for dual-unitary dynamics, minimal models of ergodic many-body dynamics for which the correlation functions of local operators vanish identically after a single time step. In this limit the temporal entanglement vanishes identically, and it was argued in Ref.~\cite{sonner_influence_2021} that temporal entanglement is expected to stay small even away from the perfect dephaser limit. This observation motivated the more general applicability of influence matrices away from this perfectly Markovian limit. For a detailed discussion of temporal entanglement in chaotic circuits, we refer the reader to Ref.~\cite{foligno2023temporal}.

\subsection{Example: $k=2$}

For $k=2$ there are two partitions, both noncrossing, $\circ = (1)(2)$ and $\smallsquare = (12)$. The influence matrix can be expanded in its components, where the only nontrivial term from the Weingarten functions is $W_{\circ \smallsquare} = -1/d_C$. The influence matrix describes a sum of domain wall states, with weight depending on the parity of the position of the domain wall, i.e.
\begin{align}
    \mathcal{I}_{k = 2} \,\,\propto&\,\, \sum_{s=1}^{2t-1} I_s\,\,\figeq[0.5\columnwidth]{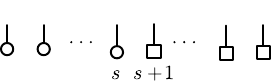}\,,
\end{align}
where $I_s = 1$ if $s$ is odd and $I_s = -1/d_C$ if $s$ is even.
Consequently, the tensor network representation of the influeence matrix in Eq.~\eqref{eq:influence_matrix} can be written as a sum over domain wall configurations as it is done in Eq.~\eqref{eq:2OTOC_summation}.
In structured dual-unitary circuits these influence matrices directly appear in the calculation of the OTOC~\cite{claeys_maximum_2020,bertini2020scrambling,bertini_operator_i_2020,rampp_dual_2023}. An influence matrix of this form also appeared for the random phase model in Ref.~\cite{yoshimura_operator_2025} (with the identity and cyclic permutation corresponding to Gaussian and non-Gaussian diagrams in the language of Ref.~\cite{yoshimura_operator_2025}). A similar structure, away from solvable points, was observed in Ref.~\cite{huang_out--time-order_2023} and argued to underpin the generic dynamics of the $2$-OTOC, again motivating the more general applicability of the developed framework.

\subsection{Markovian dynamics}

Markovian dynamics can be obtained by truncating the summation in Eq.~\eqref{eq:OTOC_summation} after a fixed number of time steps $t$ and keeping track of the final noncrossing partition in the summations and the state returned by the action of the corresponding transfer matrix. 
This picture gives rise to a dynamical semigroup which generalizes the semigroup influence matrix developed for $k=1$ \cite{sonner_semi-group_2025}.
Alternatively, this Markovian dynamics can be interpreted as truncating the matrix product state representation of Eq.~\eqref{eq:IM} and performing a cut along the horizontal direction after a fixed number of time steps.

We define a state 
\begin{align}
    |\varphi_t)) = \,\, \figeq[0.24\columnwidth]{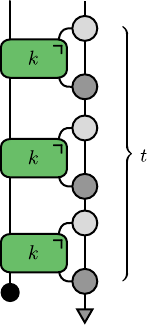}
\end{align}
where the total number of folded gates equals $t$, which satisfies the dynamical rule
\begin{align}\label{eq:Markov_update}
    |\varphi_{t})) = \mathcal{T}\, |\varphi_{t-1})), 
\end{align}
with
\begin{align}
\mathcal{T}\,\,=\,\,\figeq[0.17\columnwidth]{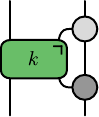}
\end{align}
Note that we use double brackets to emphasize that these states act on both the Hilbert space of $k$ replicas and the auxiliary space. 
The transfer matrix $\mathcal{T}$ corresponds to the generator of the dynamical semigroup.

Fixing the index of the auxiliary space as e.g. $\nu$, we can write
\begin{align}
    |\varphi_t)) = \sum_{\nu\, \in \, \textrm{NC}(k)} |\varphi_t)_{\nu}
\end{align}
Written out explicitly, the dynamics of Eq.~\eqref{eq:Markov_update} for $|\varphi_t)_{\nu}$ is of the form
\begin{align}\label{eq:Markov_T}
    |\varphi_t)_{\nu} = \sum_{\rho\,\in\,\textrm{NC}(k) } \mathcal{T}_{\nu \rho}\,|\varphi_{t-1})_{\rho}\,,
\end{align}
where 
\begin{align}\label{eq:Markov_T_nurho}
    \mathcal{T}_{\nu \rho} =\,\,\figeq[0.17\columnwidth]{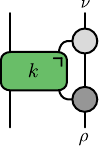} \,\,= 
    \begin{cases}
         \sum_{\rho \subseteq \sigma \subseteq \nu} \mathcal{M}_{\nu \sigma} W_{\sigma\rho} \quad &\textrm{if}\,\,  \rho \subseteq \nu \\
        0 \quad &\textrm{if} \,\, \rho \not\subseteq \nu
    \end{cases}
\end{align}
The full $k$-OTOC can be written as
\begin{align}
    C_{ab}^{(k)}(t) =((\psi_b| \mathcal{T}^t |\psi_a))\,,
\end{align}
with boundary vectors
\begin{align}\label{eq:Markov_boundaries}
    ((\psi_b| &= (d_A d_C)^{-1}\,(b_{\sq}|_{\sq}\,, \\
    |\psi_a)) &= \sum_{\rho \in \textrm{NC}(k)}d_C^{|\rho|}\,|a_{\circ})_{\rho}\,.
\end{align}

For small values of $k=1,2$, these results again allow for a simple representation and connection with known results.
For $k=1$, we again only have a single noncrossing partition, we can identify $|\varphi_t)) = |\varphi_t)_{\circ} = |\varphi_t)$ and the transfer matrix $\mathcal{T}$ reduces to the quantum channel $\mathcal{M}$ acting on the vectorized operator. The dynamics of Eq.~\eqref{eq:Markov_update} returns the previously discussed quantum channel construction for correlation functions as
\begin{align}
    |\varphi_t) = \mathcal{M}_{\circ\circ} |\varphi_{t-1}) = \mathcal{M}|\varphi_{t-1})\,.
\end{align}
For $k=2$, there are two noncrossing partitions, which are coupled through their dynamics as
\begin{align}
    &|\varphi_t)_{\sq} =  \mathcal{M}_{{\sq}{\sq}}|\varphi_{t-1})_{\sq} + \left(\mathcal{M}_{{\sq} \circ} - \frac{1}{d_C} \mathcal{M}_{\circ \circ}\right)|\varphi_{t-1})_{\circ}\, \\
    &|\varphi_t)_{\circ} = \mathcal{M}_{\circ \circ}|\varphi_{t-1})_{\circ}\,
\end{align}
It is a direct check that this expression is equivalent to Eq.~\eqref{eq:2OTOC_summation}.

This example illustrates a general property of the Markov process: The corresponding transfer matrix is naturally written in an upper-diagonal form, since $\mathcal{T}_{\nu \rho} = 0$ if $\rho \not\subseteq \nu$. The noncrossing partition degree of freedom can never decrease. 
In the same way that the eigenspectrum of an upper-diagonal matrix follows directly from its diagonal elements, the spectrum of the full matrix $\mathcal{T}$ follows from the spectrum of the diagonal elements. These diagonal elements here correspond to the `diagonal' transfer matrices $\mathcal{T}_{\nu\nu} = \mathcal{M}_{\nu \nu}$. 
Since these operators factorize in terms of the quantum channel $\mathcal{M}$ describing the correlation functions ($k=1$), their spectrum is directly known and the spectrum of the full Markovian dynamics directly follows as the union of these spectra. In this way the time scales for the decay of the $k$-OTOC directly relate to the decay rate of the correlation functions. Note however that Jordan blocks can and generally will appear, as can be directly checked for $k=2$, leading to the polynomial prefactors already observed in Eq.~\eqref{eq:2OTOC_summation}.

Before presenting a detailed analysis of the eigenspectrum of $\mathcal{T}$, we note that this Markovian dynamics can be written in an alternative way that makes the connection with free probability more explicit.
Due to the alternation of transfer matrices and M\"obius functions, the dynamics can also be written as a two-step process 
\begin{align}\label{eq:Markov_update_double_0}
    &|\kappa_t)_{\sigma} = \sum_{\nu \subseteq \sigma} W_{\sigma\nu}|\varphi_t)_{\nu}\,, \\
    &|\varphi_t)_{\nu} = \sum_{\sigma \subseteq \nu} \mathcal{M}_{\nu \sigma} | \kappa_{t-1})_{\sigma}\,,
    \label{eq:Markov_update_double_1}
\end{align}
with initial condition corresponding to
\begin{align}\label{eq:Markov_boundaries_double}
    |\kappa_{t=0})_{\sigma} = d_C^k\, |a_\circ)\,.
\end{align}
The important observation is that Eqs.~\eqref{eq:Markov_update_double_0} and~\eqref{eq:Markov_update_double_1} strongly resemble the self-consistent relations between moments and free cumulants from Eqs.~\eqref{eq:intro:phi_from_kap} and~\eqref{eq:intro:kap_from_phi}, up to unimportant factors of $d_C$. In the long-time limit where both states become time-independent, this relation enforces that these states indeed satisfy the expected relations from free probability (see discussion below).
 
We note that the move from two sets of variables and an update rule that depends on a single time step to a single set of variables with a two-step update rule is reminiscent of the move from Hamiltonian to Lagrangian dynamics. This analogy is strengthened by the observation that the M\"obius function implements a discrete derivative on the noncrossing partition lattice, effectively returning the states $|\kappa_t)_{\sigma}$ as the conjugate variables of the states $|\varphi_t)_{\nu}$.

\subsection{Steady state} 

We already argued that, for traceless operators $a$ and $b$, the steady-state value of the $k$-OTOC vanishes for all $k$, indicating asymptotic freeness between $a(t)$ and $b$. The dynamics of Eq.~\eqref{eq:Markov_update} can additionally be used to make explicit the appearance of free probability and free cumulants under ergodic many-body dynamics. 

For ergodic correlation functions, the quantum channel \eqref{eq:def_channel} has a unique leading eigenvalue $1$ from unitality, with the corresponding eigenvector being the vectorized identity matrix. From the block structure of Eq.~\eqref{eq:Markov_T_nurho} and the surrounding discussion, it follows that the Markovian process for the $k$-OTOC has $C_k$ leading eigenvalues 1, corresponding to the total number of diagonal blocks indexed by a noncrossing partition $\sigma$ with corresponding operators $\mathcal{M}_{\sigma \sigma}$. The corresponding left and right eigenvectors can be directly obtained, which will in turn allow for obtaining the steady-state value of the $k$-OTOC for general operators through a projection on this eigenspace.

\emph{Eigenstates.---}
The full derivation of the leading eigenstates is detailed in Appendix~\ref{app:eigenstates} and we here only cite the final results. Each left and right eigenstate is labelled by a noncrossing partition, and we denote the right eigenstates as $|\varphi_{\sigma}))$ and the left eigenstates as $((\kappa_{\nu}|$. These eigenstates are given by
\begin{align}\label{eq:leading_eigenstates}
    |\varphi_{\sigma})) &= d_A^{|\sigma|-k}\,\sum_{\sigma \subseteq \rho} \, d_C^{|\rho|}\, |\sigma)_{\rho}\,,\\
    ((\kappa_{\nu}| &= \sum_{\rho \subseteq \nu} (d_A d_C)^{-|\rho|}\,\mu(\nu,\rho) (\rho|_{\rho}\,.
\end{align}
Note that the subscript $\rho$ denotes the auxiliary (bond dimension) degree of freedom and $|\sigma)$ and $(\rho|$ present the vectorized permutation operators $\sigma$ and $\rho$ respectively. The notation of $|\varphi_{\sigma}))$ and $((\kappa_{\nu}|$ is again motivated by the similarity of these definitions to the relation between moments and free cumulants, Eqs.~\eqref{eq:intro:phi_from_kap} and~\eqref{eq:intro:kap_from_phi}, which will be made more explicit shortly. Overlaps between such states can be directly evaluated, where it's useful to observe that $|\varphi_{\sigma}))$ only has components $\rho$ that lie `above' $\sigma$, whereas $((\kappa_{\nu}|$ only has components $\rho$ that lie `below' $\nu$. The overlap hence only depends on the noncrossing partitions $\rho$ that lie on a path from $\sigma$ to $\nu$, i.e. $\sigma \subseteq \rho \subseteq \nu$, such that we can e.g. evaluate
\begin{align}
    ((\kappa_{\nu}|\varphi_{\sigma})) &= \sum_{\sigma \subseteq \rho \subseteq \nu} d_A^{|\sigma|-|\rho|-k} \mu(\nu,\rho)(\sigma|\rho) \nonumber\\
    &= \sum_{\sigma \subseteq \rho \subseteq \nu}d_A^{|\sigma|+|\sigma^{-1}\rho|-|\rho|-k} \mu(\nu,\rho) \nonumber\\
    &= \sum_{\sigma \subseteq \rho \subseteq \nu}\mu(\nu,\rho)  = \delta_{\nu\sigma}\,.
\end{align}
In the second equality we used that $(\sigma|\rho)=d_A^{|\sigma^{-1}\rho|}$, in the third equality we used that a triangle inequality \eqref{eq:triangle} is saturated for $\sigma \subset \rho \subset \nu$, and in the final equality we used the defining property of the M\"obius function~\eqref{eq:sumrule_moebius_intro}.

These states hence present a biorthogonal eigenbasis satisfying
\begin{align}
    ((\kappa_{\nu}| \varphi_{\sigma})) = \delta_{\nu\sigma}\,.
\end{align}

\emph{Steady-state value.---}
For ergodic dynamics these states exhaust the full eigenspace of nondecaying operators and at late times we can write
\begin{align}
    \lim_{t \to \infty} \mathcal{T}^t = \sum_{\sigma \in \textrm{NC}(k)} |\varphi_{\sigma}))((\kappa_{\sigma}|\,.
\end{align}
In order to calculate the steady-state value of the $k$-OTOC \eqref{eq:OTOC_summation}, we note that the overlap of these eigenstates with the boundary states encoding the operators $a$ and operators $b$ [Eq.~\eqref{eq:Markov_boundaries}] corresponds to
\begin{align}
     ((\psi_b|\varphi_{\sigma})) &= \varphi_{\sigma^*}(b_1,\ldots,b_k), \\
     ((\kappa_{\sigma}|\psi_a)) &= \sum_{\nu \subseteq \sigma} \mu(\sigma,\nu) \varphi_{\nu}(a_1,\ldots,a_k)  \nonumber \\
     & = \kappa_{\nu}(a_1,\ldots,a_k)\,.
\end{align}
Here we have used that the moments of $a$ follow from the overlaps with the permutation operators, $\varphi_{\nu}(a) = d_A^{-|\nu|} (\nu|a_{\circ})$, and that the M\"obius functions express the free cumulants in terms of the moments.

\begin{figure*}[t]
	\centering
	\includegraphics[width=\linewidth]{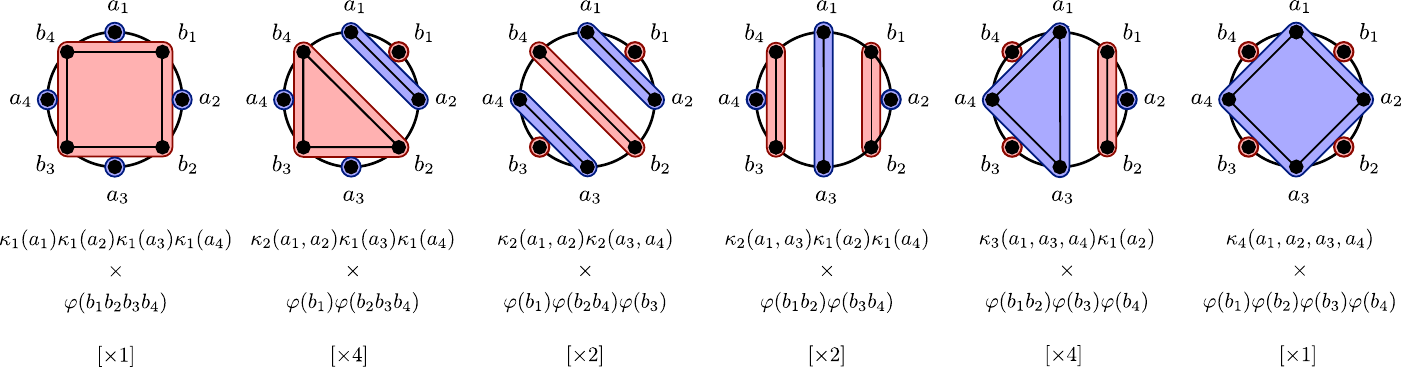}
	\caption{Graphical illustration of the terms determining the steady-state value of the $k$-OTOC for $k=4$. The shaded blue areas denote the free cumulants labeled by a noncrossing partition $\sigma$ on the set $\{a_1,a_2,a_3,a_4\}$, the shaded red areas denote the moment labeled by its complement $\sigma^*$ on the set $\{b_1,b_2,b_3,b_4\}$. The $[\times n]$ indicates that there are $n$ equivalent arrangements of this diagram. Based on a similar figure in Ref.~\cite{fava_designs_2025}.}
    \label{fig:NC_factorizing}
\end{figure*}

Taking the above results together, it immediately follows that
\begin{align}\label{eq:steadystate}
   &\lim_{t\to \infty}  C_{ab}^{(k)}(t) = \lim_{t\to \infty} ((\psi_b|\mathcal{T}^t|\psi_b)) \nonumber\\
   &\qquad =   \sum_{\sigma\, \in \, \textrm{NC}(k)} \varphi_{\sigma^*}(b_1,\ldots,b_k) \,\kappa_{\sigma}(a_1, \ldots, a_k)\, ,
\end{align}
in accordance with Eq.~\eqref{eq:steadystate_multichain}.

\emph{Asymptotic freeness.---} 
This expression is generally expected when $a_i(t)$ and $b_j$ are asymptotically free~\cite{nica2006lectures}. 
The different contributions can be graphically illustrated, where e.g. choosing $\sigma=(12)(3)(4)$ for $k=4$ returns
\begin{align}
    \figeq[0.35\columnwidth]{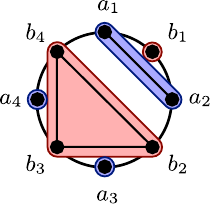} \nonumber
\end{align}
\begin{align}
= \varphi(b_1) \varphi(b_2b_3b_4)\kappa_2(a_1,a_2)\kappa_1(a_3)\kappa_1(a_4) 
\end{align}
Here we illustrate $\sigma$ in blue and $\sigma^*$ in red, where both factorize in terms of free cumulants and moments respectively.
All such terms are illustrated in Fig.~\ref{fig:NC_factorizing} for $k=4$, which also makes apparent the Kreweras complement of each noncrossing partition. The interpretation of Eq.~\eqref{eq:steadystate} as asymptotic freeness can be made explicit by noting that operators are free if their mixed free cumulants vanish and by expanding the $k$-OTOC in free cumulants. Assuming $a_i(t)$ and $b_j$ to be free, we have, with a slight abuse of notation, that $\kappa_n(\dots a_i(t) \dots b_j \dots)=0$ whenever a free cumulant contains both an $a$ and a $b$. We can use this in the expansion of the $k$-OTOC to write
\begin{align}
    &\lim_{t \to \infty}\varphi\left( a_1(t)b_1a_2(t)b_2 \dots a_k(t) b_k\right) \nonumber\\
    &= \sum_{\pi \,\in\,\textrm{NC}(2k)} \kappa_{\pi}(a_1(t),b_1,a_2(t),b_2, \dots ,a_k(t), b_k)\nonumber\\
    & = \sum_{\sigma \in \textrm{NC}(k)} \sum_{\nu \subseteq \sigma^*} \kappa_{\sigma} (a_1, \dots, a_k) \kappa_{\nu}(b_1, \dots b_k) \nonumber \\
    &=\sum_{\sigma \in \textrm{NC}(k)} \kappa_{\sigma} (a_1 \dots a_k) \varphi_{\sigma^*}(b_1, \dots, b_k).
\end{align}
In the second equality we have used that all noncrossing partitions on $2k$ elements in which $a$ and $b$ are not connected can be written as the form $\pi = \sigma \cup \nu$, with both $\sigma$ and $\nu$ noncrossing partitions on $k$ elements and where $\nu \subseteq \sigma^*$, factorized the free cumulant accordingly and used $\kappa_\sigma(a_1(t),\dots,a_k(t))=\kappa_\sigma(a_1,\dots,a_k)$ following from $\varphi(a(t))=\varphi(a)$.
In the final equality we have used the expansion of the moments in terms of free cumulants. 
Returning to Eq.~\eqref{eq:steadystate}, we find that we have a decomposition of all nondecaying terms in terms of free cumulants, which satisfy the expected factorization properties from full ETH and free probability.

\subsection{Approach to Equilibrium}
\label{sec:approach_IM}

Full ETH and free probability more generally predict that all terms in the $k$-OTOC can be given an interpretation as free cumulants -- not just the static terms.
As discussed in the previous section, the leading eigenstates of $\mathcal{T}$ [Eq.~\eqref{eq:leading_eigenstates}] are in one-to-one correspondence with the nonvanishing mixed cumulants determining the steady-state value of the OTOC.
We here show that subleading eigenstates, which contribute terms that decay in time, can be given a similar interpretation and the resulting terms in the dynamics correspond to free cumulants in which $a$ and $b$ are in the same block. We illustrate this correspondence by constructing the exact eigenstates of $\mathcal{T}$ corresponding to the leading nontrivial eigenvalue.
As indicated in Sec.~\ref{sec:approach_to_equilibrium_multichains}, these eigenstates with eigenvalue $\lambda$ govern the approach towards equilibrium $\propto \lambda^t$. While for traceless observables these eigenstates will have vanishing overlaps with the boundary conditions, leading to a decay $\propto \lambda^{2t}$, these terms are nonvanishing for non-traceless operator and simplify the following analysis. 
The construction of these subleading eigenstates and the correspondence with mixed cumulants can be directly extended to general eigenstates, albeit in a slightly more involved way.

\emph{Eigenstates.---} For an ergodic quantum channel $\mathcal{M}$ with leading nontrivial eigenvalue $\lambda$, such that the correlation functions decay as $\lambda^t$, the transfer matrix $\mathcal{T}$ has the same leading nontrivial eigenvalue $\lambda$. 
The corresponding right and left eigenstates $a_\lambda$ and $b_\lambda$ can be used to construct appropriately dressed left and right eigenstates of $\mathcal{T}$ with (not necessary the subleading) eigenvalue $\lambda$.
In the previously defined leading eigenstates, one of the contractions in the permutation operators is replaced by the right (left) eigenoperator $a_\lambda$ ($b_\lambda$)
\begin{align}
    |\varphi^m_{\sigma})) = a_{\lambda,m}|\varphi_{\sigma})), \qquad
    ((\kappa^n_{\nu}| = ((\kappa_{\nu}| b_{\lambda,n},
\end{align}
where $m,n = 1 \dots k$ denote the contraction to be dressed and $a_{\lambda,m}$ acts on all permutations in the expansions of these states as by dressing the $m$'th contraction with this operator. 
These operators remain left and right eigenstates (see Appendix~\ref{app:eigenstates}) of $\mathcal{T}$ and present the complete set of $C_k \times k$ eigenoperators with eigenvalue $\lambda$. However, they are no longer biorthogonal. 

Again fixing the right states as $|\varphi_{\sigma}^m))$, states in which the same contraction is dressed again satisfy the expected orthogonality relations
\begin{align}
    ((\kappa_{\nu}^n|\varphi_{\sigma}^n)) = \delta_{\nu\sigma}\,.
\end{align}
However, as detailed in Appendix~\ref{app:eigenstates}, for $m \neq n$ the state $((\kappa_{\nu}^n|$ additionally has a unit overlap with states $|\varphi_{\sigma}^m))$ where (i) $m$ and $n$ are in the same cycle of $\nu$ and (ii) all cycles of $\sigma$ equal the cycles of $\nu$ except for the cycle that contains $m$ and $n$, which in $\sigma$ is split in two in such a way that the Kreweras complement of this cycle consists solely of singletons and a single transposition $(m,n)$. This $\sigma$ is uniquely defined and will be denoted as $\tilde{\nu}(m,n)$. This permutation is constructed in such a way that $m$ and $n$ are in the same cycle in $\tilde{\nu}(m,n)^{-1}\nu$ and in different cycles in all $\tilde{\nu}(m,n)^{-1}\rho$ with $\tilde{\nu}(m,n) \subseteq \rho \subset \nu$. In all other cases these states are orthogonal. 
Additionally, left and right eigenstates corresponding to different eigenvalues are automatically orthogonal.

\emph{Biorthogonalization.---} From the above relation a biorthogonal basis can be defined iteratively. The free cumulant states in which $n$ is a singleton within $\nu$ are already properly biorthogonalized w.r.t. the moment states, and we define these as
\begin{align}
    ((\tilde{\kappa}^n_{\nu}| = ((\kappa^n_{\nu}| \qquad \textrm{if $n$ is a singleton in $\nu$},
\end{align}
and 
\begin{align}
    ((\tilde{\kappa}^n_{\nu}| = ((\kappa^n_{\nu}| - \sum_{m \neq n} ((\tilde{\kappa}^m_{\tilde{\nu}(m,n)}|  \qquad \textrm{otherwise},
\end{align}
where the summation $m \neq n$ runs over all elements that are in the same cycle of $\nu$ as $n$. This relation can be evaluated recursively: The states are defined when $n$ is a singleton in $\nu$, and when $n$ is not part of a singleton in $\nu$, then $n$ is necessarily part of a smaller cycle of $\tilde{\nu}(m,n)$. In this way states where $n$ is part of a transposition in $\nu$ can first be defined, after which states in which $n$ is part of a cycle of length three can be defined, and so on.

These states satisfy the expected biorthogonality, where
\begin{align}
    ((\tilde{\kappa}^n_{\nu}| \varphi^m_{\sigma})) = \delta_{mn}\delta_{\nu\sigma}\,.
\end{align}
Abstractly, the above construction of eigenstates is the same as for the leading ones if we replace the noncrossing lattice $\textrm{NC}(k)$ by the partially ordered set $\textrm{NC}(k)\times \{1,2,\ldots,k\}$, where the second component keeps track of the dressing.
The partial order is defined by declaring $(\mu, n)\subseteq (\nu, m)$ if $\mu \subseteq \nu$ and $m$ and $n$ are in the same cycle of $\mu^{-1}\nu$, and the recursive construction above corresponds to computing the Möbius function for this extended partially ordered set.

\emph{Projectors as mixed free cumulants.---} 
The corresponding projectors again have a direct interpretation as mixed free cumulants. Considering for concreteness $\sigma = (12)(3)(4)$ and fixing $n=3$ such that the singleton construction applies, the relevant overlaps directly return
\begin{align}
    ((\psi_b|\varphi_{(12)(3)(4)}^{n=3}))((\tilde{\kappa}_{(12)(3)(4)}^{n=3}&|\psi_a)) \nonumber\\
    = \,\varphi(b_1)\varphi(b_2 a_\lambda b_4b_1)&\kappa_2(a_1,a_2)\kappa_1(a_4)\varphi(b_\lambda a_3)\, .
\end{align}
The operator $b_{\lambda}$ effectively `annihilates' the part of the operator $a_n$ that has an overlap with $a_\lambda$ from the free cumulants and inserts $a_{\lambda}$ in between $b_{n-1}$ and $b_{n}$ in the moments.
This result can be made explicit by choosing $a_3 = a_\lambda$, in which case the resulting contribution to the dynamics can be graphically represented as
\begin{align}
    \figeq[0.35\columnwidth]{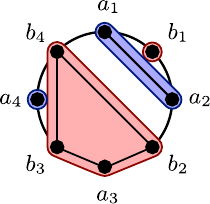} \nonumber
\end{align}
\begin{align}
=\varphi(b_1)\varphi(b_2a_3b_3b_4)\kappa_2(a_1,a_2)\kappa_1(a_4) 
\end{align}
This argument directly extends to eigenstates in which $n$ is not part of a singleton. Consider again $\sigma = (12)(3)(4)$ but $n=2$.
The corresponding projector evaluates to
\begin{align}
((\psi_b&|\varphi_{(12)(3)(4)}^{n=2}))((\tilde{\kappa}_{(12)(3)(4)}^{n=2}|\psi_a)) \nonumber\\
= &\,\varphi(a_\lambda b_1)  \varphi(b_2 b_3 b_4) \kappa_1(a_3)\kappa_1(a_4)\nonumber\\
&\times[\varphi(a_1 b_\lambda  a_2 ) - \varphi(b_\lambda a_1) \varphi(a_2) - \varphi(a_1) \varphi(b_\lambda a_2)]
\end{align}
The operator $b_{\lambda}$ now annihilates part of the combined operator $a_2 a_1$, i.e. the product of all operators in the same cycle, and reinserts these in the moments, merging this term with the cycle containing $b_1$.
Note that the additional terms in the last line, which arise from the biorthogonalization, here appear to avoid double counting contributions from the singletons.
If we choose $a_2 a_1 = a_\lambda$ and assume that $a_1$ and $a_2$ have a vanishing overlap with $b_\lambda$, this contribution can be graphically represented as
\begin{align}
    \figeq[0.35\columnwidth]{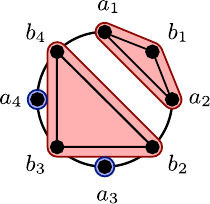} \nonumber
\end{align}
\begin{align}
    = \varphi(a_1b_1a_2)  \varphi(b_2b_3b_4) \kappa_1(a_3)\kappa_1(a_4)
\end{align}
More generally, these eigenstates annihilate a full cycle of $a$'s and insert it in a cycle of $b$'s.

We find that the slowest decaying modes again correspond to mixed cumulants, where the decaying mode of the quantum channels introduces free mixed cumulants that contains a single block in which both $a$'s and $b$'s appear. The contribution from these mixed cumulants decays exponentially in time as $\lambda^t$, as necessary for the emergence of free independence. These results are consistent with free cumulants $\kappa_n$ in which $a$ and $b$ do not mix not decaying, and free cumulants $\kappa_n$ in which a single cycle of $a$'s is merged with a noncrossing partition of $b$'s decaying $\propto \lambda^t$.

The construction of eigenstates directly extends to eigenvalues $\lambda^m$ via dressing by multiple operators, where now however no two replicas within a single cycle can be dressed (see Appendix~\ref{app:eigenstates}). Due to this restriction the eigenstates no longer form a complete basis, consistent with the appearance of Jordan blocks. The corresponding projectors again allow for an interpretation in terms of mixed free cumulants, where additional blocks mixing $a$ and $b$ appear.

\section{Concluding Remarks}
\label{sec:conclusion}

We have presented a full characterization of the dynamics of higher-order out-of-time-order correlation functions (OTOCs) in a minimal quantum circuit model for local unitary dynamics. 
In doing so we uncovered a dynamical picture in terms of free cumulants from free probability.
Free probability was recently shown to underpin higher-order extensions of the eigenstate thermalization hypothesis (ETH) and, even though there is no meaningful notion of eigenstates in our discussed model, we reproduce its predicted decomposition of higher-order OTOCs in free cumulants and establish its predicted applicability of free probability beyond Hamiltonian or Floquet dynamics.
In recent years quantum circuits have gained attention as minimal models for many-body quantum dynamics and were fundamental in uncovering the dynamics of scrambling and entanglement, but results on higher-order OTOCs are limited and no connection with full ETH was made until now.
In this way this work constitutes the first analytical result on full ETH in ergodic many-body dynamics, as well as recasting the eigenstate-based picture in a dynamical picture.

For the OTOC dynamics, all decay time scales were exactly characterized, and it was shown how ergodicity on the level of correlation functions here implies ergodicity on the level of the OTOCs, i.e. the emergence of freeness. We showed how in this model all $k$-OTOCs decay at the same rate $\propto \lambda^{2t}$, twice as fast as the correlation functions $\propto \lambda^{t}$ and significantly slower than the naive expectation $\propto \lambda^{kt}$. As a special limit, it was shown how dual-unitary gates return maximally ergodic dynamics, where both the correlation functions and $k$-OTOCs reach their ergodic steady-state value (almost) immediately. Furthermore, ergodicity was shown to be structurally stable away from this limit -- more generally an open problem in many-body quantum chaos.
These results are relevant for general studies of quantum ergodicity, chaos, and operator scrambling, and can be readily probed in current gate-based quantum computing platforms.

A fundamental result in this connection is the identification of an influence matrix for higher-order OTOCs, capturing the effects of the ergodic environment.
This influence matrix reproduces the `perfect dephaser' limit of a perfectly Markovian bath in the limit of time-ordered correlation functions, extending the perfect Markovianity to higher-order OTOCs.
The general appearance of Markovian baths throughout statistical physics and many-body dynamics, combined with the combinatorial structure of noncrossing partitions encoded in this Markovian influence matrix, suggests the more general applicability of full ETH and free probability in many-body quantum dynamics. Eigenmodes of the Markovian process, labeled by noncrossing partitions ---the fundamental objects in both free probability and full ETH---, could be identified with free cumulants. Through these eigenmodes we explicitly obtained steady-state values for the OTOCs consistent with time-evolved observables becoming free w.r.t. static observables, with an explicit identification of the decaying eigenmodes as mixed free cumulants.

While the model studied in this work is particularly simple, we expect the presented influence matrix to be relevant for more structured dynamics.
The influence matrix approach has proven successful for studying correlation functions, with the perfect dephaser limit as a natural limit, and our results provide a starting point for extending such methods to higher-order OTOCs.
Still, an open question remains whether it is possible to obtain this influence matrix in a fully structured Floquet system with a finite number of local degrees of freedom, e.g., spin chains. 
As such models would exhibit ergodic and (higher order) mixing dynamics they can act as a starting point for a perturbative expansion to extend the presented framework towards increasingly realistic chaotic many-body systems.
Ref.~\cite{chen_free_2025} previously established the role of the noncrossing partition lattice in structured dual-unitary circuit dynamics, but exact results required the limit of a diverging Hilbert-space dimension, limiting its applicability.

The minimal model can also be considered as a circuit equivalent of Lindblad dynamics, where a structured subsystem is coupled to a bath with a separation in timescales, and it would be interesting to obtain a continuum-time limit of the presented dynamics and study the interplay with external noise and dissipation. Another direction of future work is the extension of these results to multi-point correlations functions $\langle A(t_1) A(t_2)\dots A(t_k)\rangle$.
In both cases, the general applicability of full ETH in these contexts suggests the existence of a similar dynamical picture.

\begin{acknowledgements}
We thank Gabriel O. Alves, John Chalker, Silvia Pappalardi and Xhek Turkeshi for useful discussions.
F.F. acknowledges support from the European Union's Horizon Europe program under the Marie Sk{\l}odowska Curie Action GETQuantum (Grant No. 101146632).
P.W.C. acknowledges support from the Max Planck Society.

\end{acknowledgements}

\section*{Data Availability}

The numerical code for computing $k$-OTOCs using the Markovian process on the noncrossing lattice as well as the data and scripts for generating Figs.~\ref{fig:kOTOC}-\ref{fig:NC_factorizing} are provided in Ref.~\cite{zenodo}.

\appendix

\section{A primer on free probability}
\label{app:FP}
In this Appendix we briefly review relevant aspects of free probability. 

\emph{Noncommutative probability spaces.---} The main idea of free probability is to replace the commutative algebra formed by random variables on a classical probabilitity space by elements of a possibly noncommutative unital $*$-algebra $\mathcal{A}$.
A prime example is the algebra of bounded linear operators on some Hilbert space, i.e. in the finite-dimensional case the algebra of $D\times D$ complex matrices.
The notion of computing expectation values is replaced by a unital and positive $*$-linear functional $\varphi$ on $\mathcal{A}$ obeying (i) $\varphi(\mathbb{1})=1$, (ii) $\varphi(a^\dagger a)\geq 0$, and (iii) $\varphi(a^\dagger)=\varphi(a)^*$.
Such a functional is called a state on $\mathcal{A}$.
The canonical example for the above matrix algebra is $\varphi(\bullet) = \mathrm{Tr}(\bullet)/D$, which further exhibits the property of being tracial, i.e., $\varphi(ab)=\varphi(ba)$.
The algebra $\mathcal{A}$ equipped with the state $\varphi$ is called a noncommutative probability space.

\emph{Freeness.---} The notion of free independence extends the notion of independence to noncommuting variables. 
Two subalgebras $A$ and $B$ of $\mathcal{A}$ are said to be freely independent if, for all $a_i \in A$ and $b_j \in B$,
\begin{align}
    \varphi(a_1 b_1 a_2 b_2 \dots a_k b_k)=0\,,
\end{align}
provided $\varphi(a_i) = \varphi(b_j)=0, \forall i,j$. 
Informally, two operators being freely independent means that their eigenbases appear maximally random w.r.t. each other. 
From this definition it follows e.g. that for non-traceless variables,
\begin{align}
    \varphi(ab) = \varphi(a)\varphi(b)\,,
\end{align}
as also expected for classically independent variables. The noncommutativity however prevents a similar factorization of $\varphi(abab) \neq \varphi(a^2)\varphi(b^2)$, as would be the case for commuting independent variables, but rather leads to 
\begin{align}
    \varphi(abab) = \varphi(a^2)\varphi(b)^2 + \varphi(a)^2\varphi(b^2)-\varphi(a)^2\varphi(b)^2\,.
\end{align}

\emph{Noncrossing partitions.---} In order to describe the combinatorics of free probability it is useful to introduce noncrossing partitions~\cite{kreweras1972partitions}. 
The noncrossing partitions of $k$ elements are a subset of the partitions of $k$ elements. Writing a partition as $\sigma = \{ V_1, V_2, ..., V_{| \sigma|}\}$, a partition is said to be noncrossing if, when the elements $p_i, q_i$ and $p_j, q_j$ belong to distinct blocks $V_i$ and $V_j$, there are no ``crossings'' in which $p_i < p_j < q_i < q_j$. These noncrossing partitions are often represented graphically by arranging the elements on the circle and connecting elements that are in the same block. Focusing on $k=4$ for concreteness, the partition $(12)(34)$ (read as $\{\{1,2\},\{3,4\}\}$) is noncrossing and $(13)(24)$ is crossing, as made explicit in their graphic representation:
\begin{align}
    (12)(34) = \,\,\figeq[0.16\columnwidth]{fig_NC_ex_3}, \quad (13)(24) = \,\,\figeq[0.16\columnwidth]{fig_NC_ex_4} \, .
\end{align}
A special role is played by the partition with $k$ blocks, such that all blocks are singletons, and the partition with a single block:
\begin{align}
\label{eq:identity_shift_partition}
        (1)(2)(3)(4) = \,\,\figeq[0.16\columnwidth]{fig_NC_ex_1}, \quad (1234) = \,\,\figeq[0.16\columnwidth]{fig_NC_ex_2} \, .
\end{align}
These correspond to the identity $\circ = (1)(2)\dots (k)$ and the cyclic shift permutation $\smallsquare = (12 \dots k)$ respectively.

\emph{Permutations and partial order.---}
We can identify partitions with permutations by interpreting the blocks of a partition $\sigma$, with the elements of the blocks arranged in ascending order, as the cycles of a permutation $\sigma \in S_k$ of $k$ elements.
We usually don't distinguish between both interpretations.

To characterize the relevant permutations we consider the rank or Cayley weight of a permutation $\sigma \in S_k$, defined as $C(\sigma) = k - |\sigma|$, where $|\sigma|$ is the number of cycles in the cycle decomposition of $\sigma$.
The rank corresponds to the smallest number of transpositions it takes to realize the permutation $\sigma$. The identity $\circ$ minimizes the rank as $C(\circ) = 0$, whereas the cyclic permutation $\smallsquare$ maximizes it as $C(\smallsquare) = k-1$.
The Cayley weight gives rise to a notion of distance in the symmetric group $S_k$ or, more precisely, in its Cayley graph.
The vertices of the latter are the permutation $\sigma \in S_k$ and $\sigma$ and $\nu$ are connected by an edge if $\sigma^{-1}\nu$ is a transposition.
The length of the shortest paths, i.e. the geodesics, connecting $\sigma$ and $\nu \in S_k$ is determined by the Cayley weight of $\sigma^{-1}\nu$ as
\begin{align}
    \ell(\sigma,\nu) = C(\sigma^{-1}\nu) = k - |\sigma^{-1}\nu|,
\end{align}
and provides a metric on $S_k$.
Accordingly this metric fulfills the triangle inequality
\begin{align}\label{eq:triangle}
    \ell(\sigma,\rho)+\ell(\rho,\nu) \leq \ell(\sigma,\nu),
\end{align}
with equality if and only if $\rho$ lies on a geodesic from $\sigma$ to $\nu$. In terms of the number of cycles the triangle inequality reads
\begin{align}
    |\sigma^{-1}\rho|+|\rho^{-1}\nu| \leq k + |\sigma^{-1}\nu|.
\end{align}
Of particular interest are the geodesics from the identity $\circ$ to the cyclic shift $\smallsquare$, which are of length $\ell(\circ,\smallsquare)=C(\smallsquare)=k-1$, and the permutations $\sigma$ lying on such geodesics.
The set of such permutations is denoted as the interval $[\circ,\smallsquare]$ and comes equipped with a partial order, denoted by $\subseteq$, where $\nu \subseteq \sigma$ if and only if $\nu$ and $\sigma$ lie on a common geodesics from $\circ$ to $\smallsquare$ and $\ell(\circ,\nu)=C(\nu)\leq l(\circ,\sigma) = C(\sigma)$.
The resulting partially ordered set $[\circ,\smallsquare]$ is isomorphic to the noncrossing lattice~\cite{biane1997some}.

\emph{The noncrossing partition lattice.---} 
This isomorphism is, as described above, simply given by interpreting the cycles of a permutation $\sigma \in [\circ,\smallsquare]$ as the blocks of a partition $\sigma$, which then turns out to be noncrossing.
The noncrossing partitions inherit a partial ordering, where a noncrossing partition $\sigma \subseteq \nu$ if all blocks of $\sigma$ are contained in the blocks of $\nu$. This order is the dual of the usual refinement order of partitions.

The resulting ordering between noncrossing partitions can be represented in the noncrossing partition lattice, here illustrated for $k=4$:
\begin{align}
    \figeq[0.65\columnwidth]{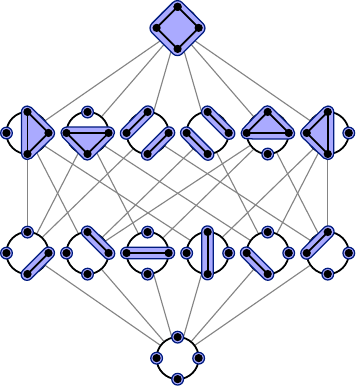} \, .
\end{align}
Noncrossing partitions $\sigma$ and $\nu$ are connected by a line if $\sigma \subseteq \nu$, and moving up a row in the lattice corresponds to increasing the rank of the noncrossing partitions by one.

\emph{Kreweras complement.---} The noncrossing partition lattice exhibits a symmetry that can be made explicit by defining the Kreweras complement~\cite{kreweras1972partitions}, which inverts the noncrossing partition lattice~\cite{biane1997some}. The Kreweras complement $\sigma^*$ of a noncrossing partition $\sigma$ is defined as $\sigma^{-1}\square$ and can be graphically represented as the largest noncrossing partition on the `dual' of the elements on the circle, e.g.
\begin{align}
     \figeq[0.28\columnwidth]{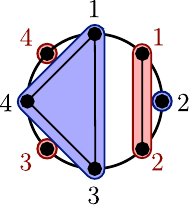}\,\,:\sigma&= (134)(2) \, \rightarrow\, \sigma^* = (12)(3)(4) \, ,\nonumber \\
     \figeq[0.28\columnwidth]{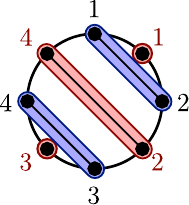}\,\,:\sigma&= (12)(34) \, \rightarrow\, \sigma^* = (1)(24)(3)\, . \nonumber
\end{align}
These satisfy $C(\sigma) + C(\sigma^*) = k-1$ or, equivalently, $|\sigma|+|\sigma^*|=k+1$.

\emph{Free cumulants.---} Mixed free cumulants are defined recursively through the combinatorics of noncrossing partitions as
\begin{align}\label{eq:def_mixedfreecumulants}
\varphi_k(a_1, a_2, \dots a_k) = \sum_{\sigma \in \textrm{NC}(k)} \kappa_{\sigma}(a_1, a_2, \dots a_k).
\end{align}
with $\kappa_1(a) = \varphi(a)$ and $\varphi_k(a_1, a_2, \dots a_k)=\varphi(a_1 a_2 \dots a_k)$. The sum is taken over all non–crossing partitions $\textrm{NC}(k)$ of the set $\{1,2, \dots k\}$ and $\kappa_{\sigma}$ factorizes according to the blocks of $\sigma = \{V_1, \dots V_{|\sigma|}\}$ as 
\begin{align}\label{eq:def_k}
&\kappa_{\sigma}(a_1, a_2 \dots a_k) = \prod_{V \in \sigma} \kappa_{V}[a_1, a_2 \dots a_k]
\end{align}
where each block $V = \{V(1), \dots V(s)\}$ of size $s=|V|$ defines a mixed free cumulant
\begin{align}
    \kappa_{V}[a_1, a_2 \dots a_k] = \kappa_{s}(a_{V(1)}, \dots ,a_{V(s)}). 
\end{align}
Mixed free cumulants vanish whenever two or more variables are freely independent, which can also be taken as the definition of free independence. Returning to the example of $a \in A$ and $b \in B$ freely independent, we have that
\begin{align}
    \kappa_s(\dots, a, \dots b ,\dots) = 0\,, \forall s\,.
\end{align}

\emph{M\"obius inversion.---} Inverting the definition of the mixed free cumulants~\eqref{eq:def_mixedfreecumulants} returns
\begin{align}
    \kappa_{\sigma}(a_1, a_2, \dots a_n) = \sum_{\nu \subseteq \sigma}\varphi_{\nu}(a_1, \dots, a_n)\,  \mu(\nu,\sigma)\,,
    \label{eq:mobius_inversion}
\end{align}
where $\varphi_{\nu}(a_1, \dots, a_n)$ is defined analogously to Eq.~\eqref{eq:def_k}. The function $\mu(\nu,\sigma)$ acts as the M\"obius function on the noncrossing partition lattice and has an explicit expression as
\begin{align}
\mu(\nu,\sigma) = \prod_{V \, \in \, \nu^{-1}\sigma} (-1)^{|V|-1} C_{|V|-1}\,.
\end{align}
For a fixed $\sigma$ and $\nu$ the M\"obius function satisfies
\begin{align}
    \sum_{\sigma \subseteq \rho \subseteq \nu}\mu(\nu,\rho)  = \delta_{\nu\sigma}\, ,
    \label{eq:sumrule_moebius}
\end{align}
which is, in fact, its defining property.
More precisely, the above property can be read as the Möbius function being the 
inverse of the zeta function of the noncrossing lattice defined by 
\begin{align}
    \zeta_{\nu \sigma} = \begin{cases}
        1 & \text{if } \nu \subseteq \sigma, \\
        0 & \text{otherwise}.
    \end{cases}
\end{align}

\section{Weingarten calculus and Haar-averaging}
\label{app:averaging}

In this Appendix we derive the thermodynamic limit of the $k$-OTOC leading to Eq.~\eqref{eq:OTOC_summation}. We start by briefly reviewing Haar averages for the unitary group, which allows for performing the average over the independent unitaries $V_t$ entering Eq.~\eqref{eq:kOTOC_single_finite_bath} and ultimately determines the averaged $k$-OTOC for finite system sizes.
We obtain the leading behavior in $d_E$, identifying the terms which survive in the thermodynamic limit $d_E \to \infty$. 
We then study fluctuations around this limiting averaged $k$-OTOC for single realizations and finite sizes and establish concentration bounds.
The latter imply that single realizations are close to the averaged result with a large probability and give rise to a notion of typicality, resulting in the stronger notion of almost sure convergence in the thermodynamic limit.

\subsection*{Haar Averages and Weingarten Calculus}

We start by reviewing the Weingarten calculus \cite{Col2003, ColSni2006} as a tool to compute the averages over the independent Haar random unitaries $V_t$.
From the independence of the $V_t$ the average factorizes and it suffices to compute the average $\bbE[V_t^{(k)}]$ of the $k$-folded gate in the replica picture. 
This average does not depend on the time step $t$ and we drop the index in the following.
Moreover, the average is the vectorization of the $k$-fold twirl channel of the unitary group, i.e. the orthonormal projection onto the space which is invariant under permutations of the $k$ replicas. 
In the vectorized (replica) picture, this space is spanned by all the permutation states $|\sigma)_{CE}$ for $\sigma \in S_k$ [Eq.~\eqref{eq:permutation_states}].
These states are linearly independent and form a basis for the permutation-invariant subspace, such that the orthogonal projection onto this space can be expressed as~\cite{Col2003, ColSni2006}
\begin{align}
    \bbE\, [ V^{(k)}] = \sum_{\nu, \sigma \in S_k}W(\nu, \sigma; d=d_Ed_C)\,\, {_{CE}|\nu)(\sigma|_{CE}} \, ,
    \label{eq:Haar_averages}
\end{align}
where $W$ denotes the Weingarten function~\cite{weingarten_asymptotic_1978}.
Viewed as a matrix with entries labeled by $\nu$ and $\sigma$, the latter is defined as the (pseudo-)inverse of the Gram matrix $G$ with entries $G(\nu, \sigma)={_{CE}(\nu|\sigma)_{CE}}$.
The Weingarten function $W(\nu, \sigma)$ depends only on the cycle structure of $\nu^{-1}\sigma$ and the dimension $d=d_Cd_E$ of the underlying Hilbert space.
It is a rational function of $d$ whose asymptotics for large dimension (large $d_E$ and fixed $d_C$ in our case) reads
\begin{align}
    W(\nu, \sigma; d) & = d^{-2k + |\nu^{-1}\sigma|}\prod_{V \in  \nu^{-1}{\sigma}}(-1)^{|V|-1}C_{|V|-1} \nonumber \\
    & \qquad \qquad \qquad \quad \times \left(1 +  \bigO\left(d_E^{-2}\right)\right).
    \label{eq:asymptotics_Wg}
\end{align}
Here the product runs over all cycles $V$ in $\nu^{-1}\sigma$, $|V|$ denotes the length of a cycle and $C_n$ denotes the $n$-th Catalan number.

\subsection*{Averaged $k$-OTOCs}

To obtain the averaged $k$-OTOC we now apply the above general result to obtain the averaged $k$-OTOC for a finite size $d_E$ of the environment.
We then identify the contributions which survive in the thermodynamic limit $d_E\to \infty$ and describe their combinatorial structure.
This is achieved by mapping the surviving contributions to so-called multichains in the noncrossing partition lattice.

To obtain these result we replace each of the $V_t^{(k)}$ by the averaged result $\bbE [ V^{(k)}]$ in the finite-size $k$-OTOC.
This substitution gives rise to a sum over $2t$ permutations $\nu_i$ and $\sigma_i$ for $i=1,\ldots,t$, minding that the $k$-OTOC at time $t$ does not depend on $V_t$ and where the boundary conditions from the initial and the final state enforce $\sigma_1=\circ$ and $\nu_t=\smallsquare$.
More precisely, the averaged $k$-OTOC reads
\begin{widetext}
\begin{align}\label{eq:kOTOC_summation_finite_dE}
     \bbE [ \kOTOCsingle(t; d_E)] = d_A^{-1} d_C^{k-1} \!\!\!\!\!\sum_{\substack{\ldots \sigma_i,  \nu_i,  \sigma_{i+1} \ldots \in S_k } }\left(\prod_{i=1}^{t-1}W(\nu_i,\sigma_{i+1};d_C)\right)(b_{\sq}|\mathcal{M}_{\nu_t \sigma_t} \dots \, \mathcal{M}_{\nu_2\sigma_2} \mathcal{M}_{\nu_1 \sigma_1}|a_\circ)\,
    d_E^{\Delta \ell(\sigma_1,\ldots,\nu_t)}\left(1 + \bigO(d_E^{-2})\right)\, .
\end{align}
\end{widetext}
Here the sum runs over all permutations $\nu_1,\ldots,\nu_{t-1},\sigma_2,\ldots,\sigma_t\in S_k$, with $\sigma_1=\circ$ and $\nu_t=\sq$ fixed.
The terms $W(\nu_i,\sigma_{i+1}; d_C)$, denoted by $W_{\nu_i \sigma_{i+1}}$ in the main text, are the leading part of the Weingarten functions and depend on $d_C$ as given by
\begin{align}
    W_{\nu\sigma} &= d_C^{-k+|\nu^{-1}\sigma|} \mu(\nu,\sigma),
\end{align}
where $\mu(\nu, \sigma)$ is a shorthand notation for
\begin{align}
     \mu(\nu,\sigma) = \prod_{V \in  \nu^{-1}{\sigma}}(-1)^{|V|-1}C_{|V|-1}
     \label{eq:mobius}
\end{align}
as it appears in Eq.~\eqref{eq:asymptotics_Wg}.
Moreover, the operators
\begin{align}
\mathcal{M}_{\nu \sigma} = \figeq[0.17\columnwidth]{fig_Mnusigma}
\end{align}
are non-expanding linear maps acting on $\Hil_A^{\otimes 2k}$.
In case of $\nu=\sigma$ they are unitarily equivalent (by permuting the replicas according to $\nu$) to $k$ products of the unital quantum channel $\mathcal{M}$, i.e., trace-preserving completely positive maps which preserve the identity, acting on operators on $\Hil_A^{\otimes k}$.
Explicitly, these maps are given by 
\begin{align}
\mathcal{M}_{\nu \sigma} =  d_C^{-k}\,\left( \bbId_A^{\otimes 2k} \otimes {_C}( \nu| \right) \left(U \otimes U^*\right)^{\otimes k} \left( \bbId_A^{\otimes 2k} \otimes |\sigma )_C \right) \, .
\end{align}
Note that being non-expanding is guaranteed by the introduction of the prefactor $d_C^{-k}$, which is compensated for by a corresponding rescaling of the Weingarten functions in Eq.~\eqref{eq:Weingarten_C} such that $W_{\nu\nu}=1$.

Ultimately, the crucial term for taking the thermodynamic limit is the factor containing $d_E$, in which we collect all the terms involving the dimension of the effective environment and in which the error term stems from the subleading corrections from the Weingarten function~\eqref{eq:asymptotics_Wg}.
The exponent $\Delta \ell$ depends on the sequence of permutations $\sigma_1,\ldots,\sigma_t,\nu_1,\ldots,\nu_t \in S_k$ and explicitly reads
\begin{align}
    \Delta \ell(\sigma_1,\ldots,\nu_t) & = -2k(t-1) - 1 \nonumber\\
    &+ \sum_{i=1}^{t}|\nu_i^{-1}\sigma_i|  + \sum_{i=1}^{t-1}|\sigma_i^{-1}\nu_{i+1}| \, ,
\end{align}
where the first sum originates from the Weingarten functions~\eqref{eq:asymptotics_Wg}, entering at each time step, the second sum is due to the overlaps between permutation states, Eq.~\eqref{eq:overlaps_permutations}, between subsequent time steps, and an additional $-1$ comes from the normalization of the trace.
The exponent $\Delta \ell$ has a geometric interpretation in the Cayley graph of the symmetric group $S_k$ as detailed in Appendix~\ref{app:FP}.
It can be seen as a difference in length, $\Delta \ell = \ell_0 - \ell_1$, between the length of a geodesic from the identity $\circ$ to the cyclic shift $\smallsquare$ given by $\ell_0 = \ell(\circ,\smallsquare)=k-1$ and the length $\ell_1$ of the shortest path that connects $\circ$ with $\sq$ while visiting all permutations $\nu_1,\sigma_1,\ldots,\sigma_{t-1},\nu_t$ in the prescribed order.
The latter is given by $\ell_1 = k(2t-1) - \sum_i|\nu_i^{-1}\sigma_i| - \sum_i|\sigma_i^{-1}\nu_{i+1}|$.

This geometric observation allows for identifying the leading contribution to the averaged $k$-OTOC as follows:
Since $\Delta \ell$ is non-positive it can be at most 0, which is satisfied when the permutations $\nu_1,\sigma_1,\ldots,\sigma_{t-1},\nu_t$ all lie on the same geodesic from $\circ$ to $\smallsquare$ with a non-decreasing distance to $\circ$.
Note, however, that subsequent permutations along the path are allowed to be equal.
This constraint forces all the permutations to correspond to  the noncrossing partitions of Appendix~\ref{app:FP} by identifying cycles of the permutations as blocks of a partition.
With this correspondence the function $\mu(\nu,\sigma)$ from Eq.~\eqref{eq:mobius} can be identified as the M{\"o}bius function of the noncrossing lattice.

Using the above description in terms of noncrossing partitions, the leading contributions to the $k$-OTOC are given in terms of nondecreasing paths on the noncrossing lattice starting at the smallest partition $\circ$ (the identity permutation) and ending at the largest partition $\sq$ (the cyclic shift).
Such a path is called a multi-chain and explicitly is given by the nondecreasing sequence 
\begin{align}
    \Sigma = \left( \sigma_1 \subseteq \nu_1 \subseteq \sigma_2 \subseteq \nu_2 \dots \subseteq \sigma_t \subseteq \nu_t \right) 
\end{align}
subject to the boundary conditions $\sigma_1 = \circ$ and $\nu_t=\smallsquare$.
All contributions to the $k$-OTOC not coming from such multichains are subleading and are at least suppressed as $d_E^{-2}$.
The latter follows from the difference of the lengths of two paths on the Cayley graph with same starting and end point being even.
Combining these argumenus, only the multichains contribute in the thermodynamic limit $d_E\to \infty$ and we obtain
\begin{align}
    \bbE [\kOTOCsingle(t; d_E) ]= \kOTOC(t) + \bigO(d_E^{-2})
    \label{eq:kOTOC_finite_size_corrections}
\end{align}
where, as stated in the main text [Eq.~\eqref{eq:OTOC_summation}], and repeated here for convenience
\begin{widetext}
\begin{align}\label{eq:OTOC_summation_app}
    C_{ab}^{(k)}(t) = d_A^{-1} d_C^{k-1} \sum_{\substack{\ldots \subseteq \sigma_i \subseteq \nu_i \subseteq \sigma_{i+1} \subseteq \ldots } }\left(\prod_{i=1}^{t-1}W_{\nu_i\sigma_{i+1}}\right) (b_{\sq}|\mathcal{M}_{\nu_t \sigma_t} \dots \, \mathcal{M}_{\nu_2\sigma_2} \mathcal{M}_{\nu_1 \sigma_1}|a_\circ) 
\end{align}
\end{widetext}
is the averaged $k$-OTOC in the thermodynamic limit.
Using the abbreviations
\begin{align}
    \mathcal{M}_\Sigma = \mathcal{M}_{\nu_t \sigma_t} \dots \, \mathcal{M}_{\nu_2\sigma_2} \mathcal{M}_{\nu_1 \sigma_1}
\end{align}
and 
\begin{align}
    W_\Sigma = \prod_{i=1}^{t-1}W_{\nu_i\sigma_{i+1}},
\end{align}
the above equation can be written in more compact form as
\begin{align}
     \label{eq:OTOC_summation_compact}
    C_{ab}^{(k)}(t) = d_A^{-1} d_C^{k-1} \sum_{\Sigma} W_\Sigma (b_{\sq}|\mathcal{M}_{\Sigma}|a_\circ) 
\end{align}
Restricting the first and the last noncrossing partion to be $\circ$ and $\smallsquare$, respectively, results in a total number $\propto t^{k-1}$ multichains that need to be considerd.

\subsection{Concentration Bounds and Typicality}

In the following we discuss fluctuations of the finite size $k$-OTOC $\kOTOCsingle(t, d_E)$ around its average and around $\kOTOC(t)$ .
We show that fluctuations are strongly suppressed in $d_E$ and single realizations are close to the averaged result $\kOTOC(t)$ with high probability by deriving concentration bounds for the distribution of the $k$-OTOC.
These bounds turn out to be strong enough to guarantee the convergence of the finite size $k$-OTOCs to $\kOTOC(t)$ given by Eq.~\eqref{eq:OTOC_summation} not only in the averaged case, i.e., in distribution, but also for typical realizations, i.e., in the almost sure sense.

To establish these results we start by considering the second moment of the finite size $k$-OTOC given by
\begin{align}
    \bbE \left[\kOTOCsingle(t; d_E)^2\right] = \frac{1}{D}\bbE \,\left(\tr\left[A_1(t)B_1\cdots A_k(t)B_k\right]^2\right) \, .
\end{align}
This average can again be computed by means of the general expression Eq.~\eqref{eq:Haar_averages} after replacing the number of replicas $k$ by $2k$ and adjusting the boundary conditions accordingly.
More precisely, the initial state $|a_\circ)$ remains the same up to interpreting $\circ$ as the identity permutation on $2k$ replicas and $a_i = a_{i+k}$, whereas the final state becomes  $(b_{\sq \otimes \sq}|  \otimes {_{CE}}(\sq|^{\otimes 2}$ with $\sq$ still being interpreted as the cyclic shift on $k$ elements and $b_i = b_{i+k}$.
The final state can be thought of as being associated with the permutation $\smallsquare \otimes \smallsquare = (1,2,\ldots,k)(k+1,k+2,\ldots,2k)\in S_{2k}$ of $2k$ elements.
We then proceed as in the computation of the average. 
Replacing $V_t^{(2k)}$ by $\bbE [ V^{(2k)}]$ we again obtain a sum over all $2t$-tuples of permutations (on $2k$ replicas) with the first entry of the tuple being $\circ$, but the last entry now being $\smallsquare \otimes \smallsquare$. 
These tuples have a similar geometric interpretation as paths on the Cayley graph of the symmetric group connecting $\circ$ with $\smallsquare\otimes \smallsquare$ and visiting all intermediate permuations in the prescribed order.
However, there are now two types of paths. 
The first type consists of paths which contain no edge in the graph corresponding to transpositions $(i,j)$ with $i \leq k$ and $j>k$ or vice versa.
These paths can be thought of ``products" of paths of permutations on the first $k$ replicas with paths of permutations on the last $k$ replicas.
As these paths of the first type do not couple the first and the last $k$ replicas, summing over all such paths contributes to the second moment with $\kOTOC(d; d_E)^2$, i.e., the average squared.
Moreover, they contain the ``products" of the geodesics which determine the thermodynamic limit.
The second type of paths from $\circ$ to $\smallsquare\otimes \smallsquare$ consists of paths which contain an edge corresponding to a transposition $(i, j)$ of the above form. 
Multiplication by such a transposition creates a cycle with elements from both the first and the last $k$-replicas for one of the intermediate permutations along the path.
The final permutation $\smallsquare\otimes \smallsquare$, however, does not contain such a cycle and hence there need to be at least another edge along the path corresponding to the transposition $(i',j')$ of the above form to split up the problematic cycle.
Consequently, each such path is at least two steps longer then the geodesics of the ``product" form and hence the contribution from summing all paths of the second type gives rise to a $\bigO(d_E^{-2})$ contribution to the second moment only.

We obtain for the variance 
\begin{align}
    \label{eq:variance_kOTOC_finite}
    \bbE \left[\kOTOCsingle(t; d_E)^2\right] - \bbE \left[\kOTOCsingle(t; d_E)\right]^2 = \bigO(d_E^{-2}),
\end{align}
and a similar estimate for the thermodynamic limit follows by using Eq.~\eqref{eq:kOTOC_finite_size_corrections} as
\begin{align}
    \label{eq:variance_kOTOC_infinite}
    \bbE \left[\kOTOCsingle(t; d_E)^2\right] - \kOTOC(t)^2 = \bigO(d_E^{-2}) .
\end{align}
This estimate of fluctuations around the average allows for deriving concentration bounds.
Explicitly, Eq.~\eqref{eq:variance_kOTOC_infinite} implies that there are constants $K_{t,k}$ depending on $t$ and $k$, but independent from $d_E$, such that the right hand side is upper bounded by $K_{t,k}d_E^{-2}$.
By fixing arbitrary $t_0$ and taking the maximum of $K_{t,k}$ we get a uniform bound for all $t\leq t_0$ of the form $K_{t_0,k}/d_E^{-2}$ for an appropriate constant $K_{t_0,k}$.
Using Markov's inequality we obtain for every $\epsilon>0$ and for all $t\leq t_0$
\begin{align}
\label{eq:concentration_bounds}
\mathbb{P}\left(|\kOTOCsingle(t, d_E) - \kOTOC(t) |> \epsilon \right) \leq \frac{K_{t_0,k}}{\epsilon^2 d_E^{2}}    \, ,
\end{align}
with the probability measure being the product Haar measure on $t_0$ independent copies of the unitary group.
For instance, by choosing $\epsilon=d_E^{-1/2}$, the above bound implies that the probability to find deviations of the finite size $k$-OTOC from the thermodynamic result for a single realization, i.e., $|\kOTOCsingle(t, d_E) - \kOTOC(t)|$, larger than $d_E^{-1/2}$ is on the order $\bigO(d_E^{-1})$ and hence vanishes in the thermodynamic limit.
In other words, typical realizations will be close to $\kOTOC(t)$ with high probability for all times $t\leq t_0$ for arbitrary chosen $t_0$ and $d_E$ large enough.

The above concentration bounds can also be used to promote the convergence of the averaged $k$-OTOC in the thermodynamic limit to almost sure convergence.
To this end, for fixed time $t \leq t_0$ we consider the sequence of $k$-OTOCs $\kOTOC(t; d_E)$ indexed by the dimension $d_E$ of the environment as random variables on the probability space given as the infinite product of $t_0$ copies of the unitary group of dimension $d_E$ equipped with the corresponding product of Haar measures.
These random variables are independent for different $d_E$ and the probabilities in Eq.~\eqref{eq:concentration_bounds} are summable in $d_E$. By the second Borell-Cantelli Lemma for almost all $d_E$ one has $|\kOTOCsingle(t, d_E) - \kOTOC(t) |\leq \epsilon $ and hence $\kOTOCsingle(t, d_E)$ converges almost surely to $\kOTOC(t)$ as $d_E\to \infty$ \cite{nica2006lectures}.

Intuitively, the above concentration bounds and the stronger form of convergence can be interpreted as the dynamics of $k$-OTOCs for typical realizations of the minimal circuit model closely following $\kOTOC(t)$ as given in Eq.~\eqref{eq:OTOC_summation} even without averaging.

\section{Properties of the quantum channel}
\label{app:channel}
In this Appendix we briefly review some properties of the single-particle quantum channel $\mathcal{M}$ and derive a bound on the operator entanglement that guarantees mixing dynamics. 
The action of $\mathcal{M}$ is defined as
\begin{align}\label{eq:app:def_M}
    \mathcal{M}(a) = \frac{1}{d_C} \mathrm{Tr}_C \left[U(a \otimes \mathbb{1}_C) U^{\dagger}\right],
\end{align}
where $a$ is an operator acting on $A$ and $U$ is a unitary matrix acting on $A$ and $C$. This expression can be graphically represented in the folded representation as
\begin{align}
    \mathcal{M} = \figeq[0.12\columnwidth]{diag_M_DU_1}\,\,,
\end{align}
with the folded operator acting on a single replica and hence $\circ=(1)$.

From unitarity it follows that $\mathcal{M}$ is trace preserving, i.e. $\mathrm{Tr}[\mathcal{M}(a)] = \mathrm{Tr}(a)$. The complete positivity follows from interpreting Eq.~\eqref{eq:app:def_M} as a Stinespring dilation, such that $\mathcal{M}$ realizes a unital quantum channel. This operator is necessarily contracting, such that all eigenvalues lie either on or inside the unit circle. By taking the complex conjugate of Eq.~\eqref{eq:app:def_M}, it follows that if $\mathcal{M}(a_\lambda) = \lambda\, a_\lambda$ then $\mathcal{M}(a_\lambda^{\dagger}) = \lambda^* a_\lambda^{\dagger}$. The eigenvalues of $\mathcal{M}$ are either real, with Hermitian eigenoperators (up to an overall phase), or appear in complex conjugate pairs, with eigenoperators related through Hermitian conjugation.

From unitality the quantum channel is guaranteed to have a trivial leading eigenvalue $1$ with the identity matrix as eigenoperator, $\mathcal{M}(\mathbb{1}_A) = \mathbb{1}_A$. As mentioned in the main text, it is possible to obtain a bound on the largest nontrivial eigenvalue $\lambda$ from the operator entanglement. The proof of this bound is analogous to the proof from Ref.~\cite{aravinda_dual-unitary_2021}, which presented a bound on the entangling power for dual-unitary gates that guarantees the corresponding quantum channels for light-cone dynamics to be mixing (following Ref.~\cite{bertini_exact_2019}). A similar proof appeared in Ref.~\cite{aravinda_ergodic_2024} for unitary gates with equal Hilbert space dimensions, i.e. $d_A=d_C$.

The operator Schmidt decomposition of $U$ is given by
\begin{align}
U = \sum_{j=1}^{d^2} \sqrt{\gamma_j}\, X_j \otimes Y_j,
\end{align}
with $d = \textrm{min}(d_A,d_C)$. Here $X_j$ and $Y_j$ are single-site operators acting on $A$ and $C$, respectively, orthonormalized such that $\mathrm{Tr}(X_j^{\dagger}X_k) = \mathrm{Tr}(Y_j^{\dagger}Y_k) = \delta_{jk}$. The linear entropy of the gate $U$, or operator entropy in short, is defined as~~\cite{rather_creating_2020,aravinda_dual-unitary_2021,rather_construction_2022,aravinda_ergodic_2024}
\begin{align}\label{app:eq:operatorentropy}
E(U) &= 1 - \frac{1}{(d_A d_C)^2}\sum_{j=1}^{d^2} \gamma_j^2 = 1-\frac{1}{d_A^2}\,\,\figeq[0.12\columnwidth]{fig_operatorent}\,\,,
\end{align}
here expressed in term of the folded gate acting on two replicas such that $\circ=(1)(2)$ and $\smallsquare=(12)$.
The operator entropy is bounded as $0 \leq E(U) \leq 1-1/d^2$.

In order to relate the operator entropy to the eigenvalues of $\mathcal{M}$, we first define $\widetilde{\mathcal{M}} = \mathcal{M} - |\circ)(\circ|$, where $|\circ)$ is the normalized vectorization of the identity, i.e. $(i,j|\circ)= \delta_{i,j}/\sqrt{d_A}$. This operator satisfies $\widetilde{\mathcal{M}}|\circ)=0$ and otherwise has the same nontrivial eigenvalues and eigenoperators as $\mathcal{M}$. We can rewrite Eq.~\eqref{app:eq:operatorentropy} as
\begin{align}
    E(U) = 1-\frac{1}{d_A^2}||\mathcal{M}||^2_2 = 1-\frac{1}{d_A^2}\left(1+||\widetilde{\mathcal{M}}||_2^2\right),
\end{align}
where in the first expression we have introduced the Frobenius norm of $\mathcal{M}$, which is in turn related to the Frobenius norm of $\widetilde{\mathcal{M}}$. The latter bounds the largest nontrivial eigenvalue $\lambda$, such that we find
\begin{align}
    |\lambda| \leq ||\widetilde{\mathcal{M}}||_2 = \sqrt{d_A^2[1-E(U)]-1}\,.
\end{align}
This bound becomes nontrivial when $E(U) > E^* = 1-2/d_A^2$, fixing $|\lambda|<1$ and hence indicating ergodic and mixing dynamics. 
Note that this bound can only be satisfied when $d_A \leq d_C$, since otherwise $E^*$ exceeds the maximally allowed value $1-1/d_C^2$.
For $d_A=d_C$ a maximal operator entanglement implies that $U$ is dual-unitary~\cite{rather_creating_2020}. Dual-unitarity~\cite{bertini_exact_2019,gopalakrishnan_unitary_2019} then directly fixes
\begin{align}
    \figeq[0.12\columnwidth]{diag_M_DU_1}\,\,=\frac{1}{d_A}\,\,\figeq[0.025\columnwidth]{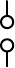}\,\,.
\end{align}
in which case $\mathcal{M}= |\circ)(\circ|$ is a projector and $|\lambda|=0$.

\section{Eigenstate Construction}
\label{app:eigenstates}

In this Appendix we construct the eigenstates of the Markovian matrix $\mathcal{T}$ [Eq.~\eqref{eq:Markov_T}]. 
We first derive the leading left and right eigenstates, before performing a biorthogonalization that can be used to construct the projector on this leading eigenspace.
We then discuss the construction of subleading eigenstates by dressing the leading eigenstates and show how they can be biorthogonalized.

\subsection*{Leading eigenstates}

\emph{Left eigenstates.---} From the triangular structure of $\mathcal{T}$, it follows that the left eigenstates can be directly obtained from the left eigenstates of the diagonal elements $\mathcal{M}_{\rho\rho}$. These operators satisfy $(\rho|\mathcal{M}_{\rho\rho}  = (\rho|$ with $(\rho|$ the permutation states defined in Eq.~\eqref{eq:permutation_states}. We can use these states to construct left eigenstates of $\mathcal{T}$, denoted as $((L_{\rho}|$, which we index by a noncrossing partition $\rho$ and write with a double bracket to highlight that these live in the joint Hilbert space of replicas and the auxiliary space of noncrossing partitions. These eigenstates have a single nonzero component in the auxiliary space and are given by
\begin{align}
    ((L_\rho| = (\rho|_{\rho}\,.
\end{align}
Note that here  $\rho$ both labels the permutation state $(\rho|$ and indicates the auxiliary space. From the triangular structure we only need to check the diagonal component, resulting in
\begin{align}
    (\rho|\mathcal{T}_{\rho \rho} = (\rho|\mathcal{M}_{\rho \rho} = (\rho| \,.
\end{align}

\emph{Right eigenstates.---} The right eigenstates can also be constructed, although they have a more involved structure. We can again use that the diagonal elements satisfy $\mathcal{M}_{\nu\nu} |\nu) = |\nu)$. Moreover, it can be directly checked that more generally
\begin{align}\label{eq:M_nusigma_sigma}
    \mathcal{M}_{\nu \sigma}|\sigma) = d_C^{-k+|\nu^{-1}\sigma|} |\sigma)\,.
\end{align}
This identity can be used to construct right eigenstates $|R_\sigma))$ of $\mathcal{T}$, which we again index by a noncrossing partition $\sigma$ and write with a double bracket, as
\begin{align}\label{eq:right_eigenstate_0}
    |R_\sigma)) = \sum_{\sigma \subseteq \rho}  d_C^{|\rho|}\, |\sigma)_{\rho}\,.
\end{align}
Note that in the right-hand side $|\sigma)$ indicates the permutation state, whereas the subscript $\rho$ is the index in the auxiliary space. The summation runs over all noncrossing partitions $\rho$ that lie above $\sigma$ in the noncrossing partition lattice.

It is a direct check that the states \eqref{eq:right_eigenstate_0} are right eigenstates of $\mathcal{T}$. We can calculate the component $\nu$ of $\mathcal{T} |R_\sigma))$ in the auxiliary space, where we separate the cases $\sigma \not\subseteq \nu$ and $\sigma \subseteq \nu$. For the former it follows that
\begin{align}
    \sum_{\sigma \subseteq \rho} \mathcal{T}_{\nu \rho}\, d_C^{|\rho|}\, |\sigma) = 0 \quad \textrm{if} \quad \sigma \not\subseteq \nu, 
\end{align}
since the summation runs over all components $\rho$ of $|R_\sigma))$, which are only nonzero if $\sigma \subseteq \rho$, but $\mathcal{T}_{\nu \rho}$ is only nonzero if $\rho \subseteq \nu$, and fixing both to be nonzero implies $\sigma \subseteq \rho \subseteq \nu$ and hence $\sigma \subseteq \nu$.
For the latter we have
\begin{align}
    &\sum_{\sigma \subseteq \rho}  \mathcal{T}_{\nu \rho}\, d_C^{|\rho|}\,  |\sigma) \nonumber\\
    &=\sum_{\sigma \subseteq \rho \subseteq \pi \subseteq \nu}\, \mu(\rho,\pi)\,d_C^{-k+|\rho^{-1}\pi|+|\rho|} \,\mathcal{M}_{\nu \pi} |\sigma) \,.
\end{align}
Here we can first use that for a geodesic $ \rho \subseteq \pi$ the triangle inequality becomes $ |\pi^{-1}\rho|+|\rho^{-1}| = k+|\pi^{-1}| = k+|\pi|$, to write
\begin{align}\label{eq:derivation_right_eigenstates}
    \sum_{\sigma \subseteq \rho  \subseteq \pi \subseteq \nu} & \, \mu(\rho,\pi)\,d_C^{| \pi|} \,\mathcal{M}_{\nu \pi} |\sigma) = \sum_{\sigma \subseteq \pi \subseteq \nu} \delta_{\sigma,\pi}\, d_C^{| \pi|} \,\mathcal{M}_{\nu \pi} |\sigma) \nonumber\\
    &\quad=d_C^{|\sigma|} \,\mathcal{M}_{\nu \sigma}|\sigma) = d_C^{-k+|\sigma|+|\sigma^{-1}\nu|}|\sigma) \nonumber\\
    &\quad = d_C^{|\nu|}|\sigma)
\end{align}
where in the first equality we used that M\"obius functions satisfy $\sum_{\sigma \subseteq \rho \subseteq \pi} \mu(\rho,\pi) = \delta_{\sigma, \pi}$ for fixed $\sigma$ and $\pi$ and a summation over $\rho$, see Eq.~\eqref{eq:sumrule_moebius}. In the second to last equality we applied Eq.~\eqref{eq:M_nusigma_sigma}, and in the final equality we again used that the triangle equality is saturated.
Taking these results together, we recover that the states from Eq.~\eqref{eq:right_eigenstate_0} are right eigenstates of the Markovian matrix $\mathcal{T}$ with eigenvalue 1.

\emph{Biorthogonalization.---} The previous constructions return $C_k$ left and right eigenstates with eigenvalue $1$, exhausting the leading eigenspace in the case of ergodic dynamics. However, these eigenstates are not biorthogonal. This biorthogonalization can be performed using the properties of the M\"obius functions. For reasons that are discussed in the main text, we write the resulting left eigenstates as $((\kappa_{\sigma}|$ and the right eigenstates as $|\varphi_{\sigma}))$,
\begin{align}
    |\varphi_{\sigma})) &= d_A^{|\sigma|-k}|R_{\sigma})) =  d_A^{|\sigma|-k}\sum_{\sigma \subseteq \rho}  d_C^{|\rho|}\, |\sigma)_{\rho}\,,\\
    ((\kappa_{\nu}| &= \sum_{\rho \subseteq \nu} (d_A d_C)^{-|\rho|}\,\mu(\nu,\rho) ((L_{\rho}|\, \nonumber\\
    &=\sum_{\rho \subseteq \nu} (d_A d_C)^{-|\rho|}\,\mu(\nu,\rho) (\rho|_{\rho}\,.
\end{align}
Using similar arguments as in our construction of the right eigenstates, it can be directly checked that these states satisfy
\begin{align}
    ((\kappa_{\nu}| \varphi_{\sigma})) = \delta_{\nu\sigma}\,.
\end{align}
To show this biorthogonality, we first note that
\begin{align}
    ((L_{\rho}|R_{\sigma})) = \begin{cases}
        d_A^{|\sigma^{-1}\rho|} d_C^{|\rho|} \quad &\textrm{if} \quad \sigma \subseteq \rho \\
        0 \quad &\textrm{if} \quad \sigma \not\subseteq \rho 
    \end{cases}
\end{align}
We directly observe that $((\kappa_{\nu}| \varphi_{\sigma})) =0$ if $\sigma \not\subseteq \nu$. If $\sigma \subseteq \nu$, the M\"obius functions enforce orthogonality since
\begin{align}
     ((\kappa_{\nu}| \varphi_{\sigma}))  &= d_A^{|\sigma|-k}\sum_{\sigma \subseteq \rho \subseteq \nu} (d_A d_C)^{-|\rho|}\, \mu(\rho,\nu)((L_{\rho}|R_{\sigma})) \nonumber\\
     &=\sum_{\sigma \subseteq \rho \subseteq \nu} d_A^{|\sigma|+|\sigma^{-1}\rho|-|\rho|-k} \mu(\rho,\nu) \nonumber\\
     &= \sum_{\sigma \subseteq \rho \subseteq \nu} \mu(\rho,\nu)= \delta_{\sigma\nu}\,,
\end{align}
using the same arguments as in Eq.~\eqref{eq:derivation_right_eigenstates}. 

\emph{Overlaps.---} We now consider the overlap of these eigenstates with the boundary states $|\psi_a))$ and $((\psi_b|$ [Eq.~\eqref{eq:Markov_boundaries}], whose definition we repeat here for convenience, 
\begin{align}
    ((\psi_b| &= (d_A d_C)^{-1}\,(b_{\sq}|_{\sq}\,, \\
    |\psi_a)) &= \sum_{\rho \in \textrm{NC}(k)}d_C^{|\rho|}\,|a_{\circ})_{\rho}\,.
\end{align}
The relevant overlap of the right eigenstates can be directly calculated as
\begin{align}
    ((\psi_b|\varphi_{\sigma})) &= d_A^{|\sigma|-k-1} (b_{\sq}|\smallsquare) = \varphi_{\sigma^*}(b_1,\dots,b_k), 
\end{align}
where we used that $(b_{\square}|\sigma)=d_A^{-|\sigma^*|}\varphi_{\sigma^*}(b_1,\dots,b_k)$ and $|\sigma^*| = k+1-|\sigma|$. For the left eigenstates we similarly find
\begin{align}
    ((\kappa_{\sigma}|\psi_a)) &= \sum_{\nu \subseteq \sigma} \mu(\sigma,\nu) d_A^{-|\nu|}  (\nu|a_{\circ}) \\
    &= \sum_{\nu \subseteq \sigma}\mu(\sigma,\nu) \varphi_{\nu}(a_1,\dots,a_k) \\
    & =\kappa_{\sigma}(a_1,\dots,a_k)\,,
\end{align}
now using that $\varphi_{\nu}(a_1,\dots,a_k) = d_A^{-|\nu|} (\nu|a_{\circ})$.

\subsection*{Subleading eigenstates}

We now consider eigenstates with eigenvalue $\lambda$ obtained by dressing these states with a right (left) eigenoperator $a=a_\lambda$ ($b=b_\lambda$) of the single-particle channel as
\begin{align}
    |\varphi^m_{\sigma})) = a_{\lambda,m}|\varphi_{\sigma})), \qquad
    ((\kappa^n_{\nu}| = ((\kappa_{\nu}| b_{\lambda,n},
\end{align}
where $m,n = 1 \dots k$ denote the contraction to be dressed and $a_{\lambda}$ ($b_{\lambda}$) acts on all permutations in the expansions of these states as by dressing the $m$'th contraction with this operator; see Eq.~\eqref{eq:permutation_states} and the surrounding discussion. In this appendix we will drop the subscript $\lambda$ for ease of notation.
For the left eigenstates the proof that these are eigenstates is fully analogous to the proof for the leading eigenstates. For the right eigenstates we only need the additional identity
\begin{align}\label{eq:subleading_Mnusigma}
    \mathcal{M}_{\nu \sigma}\, a_m|\sigma) = \lambda \, d_C^{-k+|\nu^{-1}\sigma|} a_m |\sigma),
\end{align}
which can be directly verified. 

\emph{Biorthogonalization.---} The overlaps can be calculated in a similar way as before, to return
\begin{align}\label{eq:overlaps_subleading_0}
    ((\kappa^n_{\nu}| \varphi^m_{\sigma})) = \sum_{\sigma \subseteq \rho \subseteq \nu} d_A^{-|\sigma^{-1}\rho|} \mu(\nu,\rho) (\rho|b_na_m|\sigma)\,.
\end{align}
The contraction $(\rho|b_na_m|\sigma)$ vanishes unless $m$ and $n$ are in the same cycle of $\sigma^{-1}\rho$, and hence of $\sigma^{-1}\nu$, since otherwise the contraction will contain factors $\tr(a)\tr(b) = 0$, whereas if these are in the same cycle the contraction contains a factor $\tr(ab) = 1$.

Furthermore, if $m$ and $n$ are in the same cycle of $\sigma^{-1}\nu$, there exists a unique permutation $\tilde\nu(m,n)$ with $\sigma \subseteq \tilde\nu(m,n) \subseteq \nu$, for which there are no permutations below $\tilde\nu(m,n)$ in which $m$ and $n$ are in the same cycle of $\sigma^{-1}\tilde\nu(m,n)$. For all permutations $\tilde\nu(m,n) \subseteq \rho \subset \nu$, $m$ and $n$ are then necessarily in the same cycle.
The overlap is fully determined by this permutation $\tilde\nu(m,n)$, since Eq.~\eqref{eq:overlaps_subleading_0} reduces to
\begin{align}\label{eq:overlaps_subleading_1}
    ((\kappa^n_{\nu}| \varphi^m_{\sigma})) = \!\!\!\! \sum_{\tilde\nu(m,n) \subseteq \rho \subseteq \nu}  \mu(\nu,\rho) = \delta_{\nu,\tilde\nu(m,n)}\,.
\end{align}
In other words, these states have unit overlap if and only if $m$ and $n$ are in the same cycle in $\sigma^{-1}\nu$ and in different cycles in all $\sigma^{-1}\rho$ with $\sigma \subseteq \rho \subset \nu$. In all other cases these states are orthogonal.

For $m=n$ this overlap directly returns $\delta_{\nu\sigma}$. For $m \neq n$ but in the same cycle of $\nu$, there is a single additional permutation for which this overlap is nonvanishing.
This permutation is uniquely determined by $\nu$ and $m \neq n$, motivating our notation $\tilde{\nu}(m,n)$.
This property can be proven by explicitly constructing a permutation that satisfies the above properties. For ease of notation we drop the explicit dependence on $m$ and $n$ and write $\tilde{\nu}(m,n) =\tilde{\nu}$.

First, we note that $m$ and $n$ need to be part of the same cycle of $\nu$, otherwise for all $\tilde{\nu} \subseteq \nu$ we have that $m$ and $n$ are part of different cycles in $\tilde{\nu}^{-1}\nu$.
Second, we note that all cycles in $\tilde{\nu}$ that do not contain $m$ or $n$ should be identical to cycles in $\nu$ that also do not contain $m$ or $n$. 
If this were not the case, we could merge two such cycles to create a new permutation $\rho$ that satisfies $\tilde{\nu} \subset \rho \subset \nu$ and for which $\rho^{-1}\nu$ has $m$ and $n$ in the same cycle (since this cycle would not be impacted by merging the other cycles), violating our assumption there are no permutations $\rho$ above $\tilde\nu$ for which $m$ and $n$ are in the same cycle of $\rho^{-1}\nu$. 
We hence find that $\tilde{\nu}$ is identical to $\nu$, except that the single cycle of $\nu$ that contains $m$ and $n$ is split into two cycles, one of which contains $m$ and the other of which contains $n$.
We take this cycle to have $\ell$ elements and denote it as $\nu_c = (i_1 i_2 \dots i_m \dots i_n \dots i_{\ell})$. This cycle can be identified with the cyclic permutation on $\ell$ elements. 
We hence want to find a way of splitting this permutation in two as $\tilde{\nu}_c = (\dots i_m \dots)(\dots i_n \dots)$, in such a way that $i_m$ and $i_n$ are in the same cycle in $\tilde{\nu}_c^{-1}\nu_c$ but in different cycles of $\rho_c^{-1}\nu_c$ for every $\tilde{\nu}_c \subset \rho_c \subseteq \nu_c$.
We note that multiplication with the inverse cyclic permutation is equivalent to constructing the Kreweras complement, such that we hence require that $i_m$ and $i_n$ are in the same cycle of the Kreweras complement $\tilde{\nu}_c^*$ of $\tilde{\nu}_c$, but in different cycles for the Kreweras complement of $\rho_c^*$ for $\tilde{\nu}_c \subset \rho_c$. The Kreweras complement reverses the ordering, such that we need to have $i_m$ and $i_n$ in different cycles for every $\rho_c^* \subset \tilde{\nu}_c^*$. This demand uniquely fixes $\tilde{\nu}^*_c$ as the permutation in which $i_m$ and $i_n$ form a transposition (a cycle of length two) and all other indices form singletons (cycles of length one). This construction then uniquely determines $\tilde{\nu}$ given $m,n$ and $\nu$.

We conclude this proof with some examples. For $\ell=2$ we can write $\nu_c = (i_1 i_2)$ with e.g. $i_m = i_1$ and $i_n=i_2$. We have that $\tilde{\nu}_c^* = (i_1 i_2)$ and the Kreweras complement of $(1)(2)$ is $(12)$, such that $\tilde{\nu}_c=(i_1)(i_2)$.  
For $\ell=3$ we can have ${\nu}_c=(i_1 i_2 i_3)$, and fixing $i_m = i_1$ and $i_n=i_2$, we have that $\tilde{\nu}_c^*=(i_1i_2)(i_3)$. Using that the Kreweras complement of $(13)(2)$ is $(12)(3)$, we have that $\tilde{\nu}_c = (i_1i_3)(i_2)$.
For $\ell=4$ and $\nu_c = (i_1 i_2 i_3 i_4)$, we can have e.g. $i_m=i_1$ and $i_n = i_2$, for which $\tilde{\nu}_c = (i_1 i_3 i_4)(i_2)$, or $i_m = i_1$ and $i_n=i_3$, for which $\tilde\nu_c = (i_1i_4)(i_2 i_3)$.

\emph{Singletons.---} To illustrate the construction of biorthogonalized eigenstates, we first choose $\nu$ such that $n$ is a cycle of length one (a singleton).
In this case, the condition for the overlap to be nonvanishing simplifies. Since $\sigma \subseteq \rho$, every singleton in $\rho$ needs to be a singleton in $\sigma$. As such, $m$ and $n$ can only be in the same cycle if $m=n$. This additionally implies that the lowest noncrossing partition in which $m$ and $n$ are in the same cycle is $\sigma$, i.e. $\tilde{\nu}(m,n) = \sigma$, such that the overlap reduces to
\begin{align}
    ((\kappa^n_{\nu}| \varphi^m_{\sigma})) = \delta_{mn} \delta_{\nu \sigma}\,.
\end{align}
We recover the expected biorthogonality. The free cumulant states in which $n$ is a singleton within $\nu$ are already properly biorthogonalized w.r.t. the moment states.

\emph{Transpositions.---} We now consider $\nu$ such that $n$ is part of a cycle of length two (a transposition). Denoting $\nu = (\dots)(n \tilde{n})(\dots)$ in which $n$ forms a cycle of length two with $\tilde{n}$, the overlap \eqref{eq:overlaps_subleading_1} is nonvanishing only if either $m=n$ or $m=\tilde{n}$. For $m=n$ we can directly use the previous result to find orthogonality, since $((\kappa^n_{\nu}| \varphi^n_{\sigma})) = \delta_{\nu \sigma}$ since $a$ and $b$ are always guaranteed to be in the same cycle. 

For $m = \tilde{n}$ we however have additionally nonvanishing overlaps.
Since $\sigma \subseteq \nu$ there are now only two options: Either $n$ and $m$ form a transposition within $\sigma = (\dots)(n m)(\dots)$, or they are both singletons and $\sigma = (\dots)(n)(m)(\dots)$. The former again leads to the expected orthogonality, since we can directly repeat the previous argument. In the latter case the overlap is nonvanishing if and only if $\sigma = \tilde\nu(m,n)$, which here corresponds to $\nu$ with the transposition split into two singletons. 

If $n$ forms a transposition with $m$ in $\nu$, then the two nonvanishing overlaps are
\begin{align}
    ((\kappa^n_{\nu}| \varphi^n_{\nu})) = ((\kappa^n_{\nu}| \varphi^{\tilde{n}}_{\tilde\nu({m,n})})) = 1\,.
\end{align}
Here the recursive construction comes into play. Since the (left) cumulant states in which a singleton is dressed already satisfy the orthogonality, and the additional state with nonvanishing overlap is exactly a (right) state in which a singleton is dressed, we can define orthogonal left states for transpositions by simply subtracting the left singleton state from this transposition state. For partitions $\nu$ in which $n$ and $m$ form a transposition, biorthogonalized left eigenstates are given by
\begin{align}
 ((\tilde{\kappa}^n_{\nu}| &=  ((\kappa^n_{\nu}| - ((\kappa^{\tilde{n}}_{\tilde\nu({m,n})}|\,.
\end{align}
These states satisfy
\begin{align}
((\tilde{\kappa}^n_{\nu}|\varphi^m_{\sigma})) = \delta_{mn} \delta_{\nu \sigma}\,.
\end{align}

\emph{General cycles.---} The previous construction can be directly extended to the case where $n$ is part of a larger cycle of $\nu$. Properly biorthogonalized right eigenstates can be defined as
\begin{align}
((\tilde{\kappa}_{\nu}^n| = ((\kappa_{\rho}^n| - \sum_{m \neq n} ((\tilde{\kappa}^{\tilde{n}}_{\tilde\nu(m,n)}|,
\end{align}
where the summation runs over all indices $m$ that are in the same cycle of $\nu$ as $n$. The corresponding permutation $\tilde\nu(m,n)$ is defined above. Since $n$ will always belong to a shorter cycle in this permutation, the construction can be done recursively, where if $n$ is a singleton in $\nu$ we define $((\tilde\kappa_{\nu}^n| = ((\kappa_{\nu}^n|$. These states now define an orthonormal basis
\begin{align}
((\tilde\kappa_{\nu}^n| \varphi_{\sigma}^m)) = \delta_{mn} \delta_{\nu\sigma}\,.
\end{align}
To conclude, we note that the above construction directly extends to states in which multiple contractions are dressed with (possibly different) eigenoperators of $\mathcal{M}$. The only restriction is that no two contractions within the same cycle can be dressed, since otherwise the equivalent of Eq.~\eqref{eq:subleading_Mnusigma} no longer holds.
This constraint prevents construction of the full set of eigenstates that would be allowed from the eigenstates of the channels $\channel{\nu}{\sigma}$ and ultimately gives rise to the observed Jordan structure.

\section{Additional Spatial Structure}
\label{app:extended_bath}

In this Appendix we briefly discuss a possible extensions of the minimal quantum circuit that includes additional spatial structure beyond the simple three-site setup.
While such generalizations seem natural, we argue that these are already captured by the minimal circuit, indicating the wider applicability of the presented results.

We first note that, since we did not make any assumptions on the Hilbert space for $A$, e.g. we might take $A$ as a finite one-dimensional lattice and $\gateboundary$ as a local quantum circuit which couples only the boundary of the lattice $A$ to the bottleneck $C$.
While this choice would influence the short-time dynamics, the late-time behavior does not depend on the specific structure of $\gateboundary$ but only on the spectral properties of the associated channel $\mathcal{M}$.
As for generic choices of local gates that comprise the quantum circuit $\gateboundary$ the latter will still be ergodic and mixing ($\lambda < 1$), all presented results remain valid but might apply at slightly later times.
Intuitively, this shift in time can be understood from additional constraints on the dynamics from causality and might be interpreted as the time it takes for local observables in $A$ to spread into the effective bath. 
Nevertheless, the minimal, three-site circuit accurately captured the dynamics at late times irrespective of the details within $A$ and the concrete structure of $\gateboundary$. 

Another natural extension of the minimal setup is to allow for additional spatial structure in the effective bath while still preserving solvability of the model.
This can be achieved by replacing the unitaries $V_i$ by a random quantum circuit with a brickwork geometry. A natural choice consistent with causality is by choosing the environment dynamics as
\begin{align}\label{eq:fig_extendedbath}
        \figeq[0.38\columnwidth]{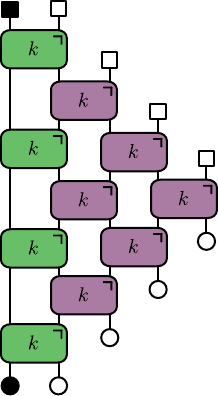}
\end{align}
We again consider Haar-random local gates, allowing for an exact solution in the limit of the local Hilbert space dimensions going to infinity.
Intuitively, the more involved environment should not change the dynamics as, in the minimal setting, a single Haar random unitary $V_t$ already perfectly scrambles any information leaking into the effective bath as $d_E \to \infty$.

This intuition can be made precise.
To this end, consider $E$ as a one-dimensional lattice of length $L$ with all local Hilbert spaces of the same dimension $d_E$ and replace $V_t$ by a random two-layer brickwork circuit $V_t=V_t^{(e)}V_t^{(o)}$
with 
\begin{align}
    V_t^{(e)} & = W_t^{(C,1)} \otimes W_t^{(2,3)} \otimes \cdots \otimes W_t^{(L-1,L)} \\
    V_t^{(o)} & = \bbId_C \otimes W_t^{(1,2)}  \otimes \cdots \otimes W_t^{(L-2,L-1)} \otimes \bbId_L \, ,
\end{align}
where we have chosen $L$ to be odd and $\bbId_L$ denotes the identity on site $L$. This dynamics gives rise to a causal light cone for the OTOC, as in Eq.~\eqref{eq:fig_extendedbath}.
The $W_t^{(C,1)}$ are $d_Cd_E$ dimensional Haar random unitaries acting on the bottleneck $C$ and the first site of the environment, whereas the $W_t^{(x,x+1)}$ are $d_E^2$ dimensional Haar random unitaries acting on the neighboring sites $x$ and $x+1$.
All random unitaries are assumed to be independent.

In this setting the ensemble averaged $k$-OTOCs in the limit $d_E \to \infty$ can be computed by the same methods as in the minimal setup and we only report the result here.
In particular, the $k$-OTOC can again be written as a sum over $L$-tuples of multichains on the noncrossing lattice 
\begin{align}
    \label{eq:kOTOC_summation_extended}
    \kOTOC(t) = d_A^{-1}d_C^{k-1} \!\!\!\!\sum_{\Sigma_1,\ldots,\Sigma_L}\!\!\!\!W_{\Sigma} \left(\prod_{x=2}^L \mu_{\Sigma_x}\right),
    ( b_{\sq} | \mathcal{M}_{\Sigma}|a_\circ )
\end{align}
similar to Eq.~\eqref{eq:OTOC_summation}.
Due to the two-layer brickwork layout and not using the causality indicated by Eq.~\eqref{eq:fig_extendedbath} the multichains $\Sigma_x$ for $x\neq L$ now are of length $4t$ with the appropriate boundary conditions $\circ$ and $\smallsquare$.
We denote them by $\Sigma_x =  \left(\ldots \subseteq \sigma_i^{(x-1,x)} \subseteq \nu_i^{(x-1,x)}\subseteq \sigma_i^{(x,x+1)} \subseteq \nu_i^{(x,x+1)} \subseteq \dots \right)$, keeping track from which averaged Haar-random gate $W_i^{(x,x+1)}$ the noncrossing partitions originate.
This enforces constraints as due to the brickwork structure most partitions are part of multiple multichains.
Due to the boundary conditions the multichain $\Sigma_L=\left(\ldots \subseteq \sigma_i^{(L-1,L)} \subseteq \nu_i^{(L-1,L)}\subseteq \sigma_{i+1}^{(L-1,L)} \dots \right)$ is only of length $2t$.
The coupling to $C$ is described by $\Sigma = \left(\ldots \subseteq \sigma_i^{(C,1)} \subseteq \nu_i^{(C,1)}\subseteq \sigma_{i+1}^{(C,1)} \dots \right)$, which has no independent partitions as it is derived from $\Sigma_1$ and hence is not summed over.
All other multichains enter exclusively via their weights related to the Möbius function
\begin{align}
    \mu_{\Sigma_x}=\prod_{i=1}^{t-1}\mu\left(\nu_{i}^{(x-1,x)}, \sigma_{i} ^{(x-1,x)}\right) \, .
\end{align}

We now utilize the defining properties of the Möbius function [Eq.~\eqref{eq:sumrule_moebius}] to simplify the above expression.
To this end we fix $\Sigma_1,\ldots,\Sigma_{L-1}$ and first perform the sum over $\Sigma_L$ subject to the constraints imposed by $\Sigma_{L-1}$.
That is, we sum only over the $\nu_i^{(L-1,L)}\equiv \nu $ and $\sigma_i^{(L-1,L)}\equiv \sigma $ subject to the constraints $\sigma_i^{(L-2,L)}\subseteq \nu_i^{(L-1,L)} \subseteq \sigma_i^{(L-1,L)} \subseteq \nu_{i+1}^{(L-2,L)}$.
The corresponding sum yields 
\begin{align}
    \sum_{\nu \subseteq \sigma}\mu(\nu, \sigma) =  \sum_{\sigma}\delta_{\sigma_i^{(L-2,L)}, \sigma} = 1 
\end{align}
and, applying this result to all time steps, we find that performing the sum over $\Sigma_L$ in Eq.~\eqref{eq:kOTOC_summation_extended} simply returns a factor of 1.
In other words, the $k$-OTOCs remains unchanged under replacing $L$ by $L-1$ and, by iterating this argument, is hence completely independent of $L$.
The only remaining sum runs over those noncrossing partitions not fixed by the above argument which are exactly those appearing in $\Sigma$.
The $k$-OTOC for the extended effective bath comprising a random brickwork circuit is identical to the one obtained in the minimal setup.


\input{main.bbl}
\end{document}

%% file: main.bbl
%

%% file: main.bbl
\begin{thebibliography}{112}%
\makeatletter
\providecommand \@ifxundefined [1]{%
 \@ifx{#1\undefined}
}%
\providecommand \@ifnum [1]{%
 \ifnum #1\expandafter \@firstoftwo
 \else \expandafter \@secondoftwo
 \fi
}%
\providecommand \@ifx [1]{%
 \ifx #1\expandafter \@firstoftwo
 \else \expandafter \@secondoftwo
 \fi
}%
\providecommand \natexlab [1]{#1}%
\providecommand \enquote  [1]{``#1''}%
\providecommand \bibnamefont  [1]{#1}%
\providecommand \bibfnamefont [1]{#1}%
\providecommand \citenamefont [1]{#1}%
\providecommand \href@noop [0]{\@secondoftwo}%
\providecommand \href [0]{\begingroup \@sanitize@url \@href}%
\providecommand \@href[1]{\@@startlink{#1}\@@href}%
\providecommand \@@href[1]{\endgroup#1\@@endlink}%
\providecommand \@sanitize@url [0]{\catcode `\\12\catcode `\$12\catcode `\&12\catcode `\#12\catcode `\^12\catcode `\_12\catcode `\%12\relax}%
\providecommand \@@startlink[1]{}%
\providecommand \@@endlink[0]{}%
\providecommand \url  [0]{\begingroup\@sanitize@url \@url }%
\providecommand \@url [1]{\endgroup\@href {#1}{\urlprefix }}%
\providecommand \urlprefix  [0]{URL }%
\providecommand \Eprint [0]{\href }%
\providecommand \doibase [0]{https://doi.org/}%
\providecommand \selectlanguage [0]{\@gobble}%
\providecommand \bibinfo  [0]{\@secondoftwo}%
\providecommand \bibfield  [0]{\@secondoftwo}%
\providecommand \translation [1]{[#1]}%
\providecommand \BibitemOpen [0]{}%
\providecommand \bibitemStop [0]{}%
\providecommand \bibitemNoStop [0]{.\EOS\space}%
\providecommand \EOS [0]{\spacefactor3000\relax}%
\providecommand \BibitemShut  [1]{\csname bibitem#1\endcsname}%
\let\auto@bib@innerbib\@empty
\bibitem [{\citenamefont {Srednicki}(1999)}]{srednickiApproachThermalEquilibrium1999}%
  \BibitemOpen
  \bibfield  {author} {\bibinfo {author} {\bibfnamefont {M.}~\bibnamefont {Srednicki}},\ }\bibfield  {title} {\bibinfo {title} {The approach to thermal equilibrium in quantized chaotic systems},\ }\href {https://doi.org/10.1088/0305-4470/32/7/007} {\bibfield  {journal} {\bibinfo  {journal} {J. Phys. A}\ }\textbf {\bibinfo {volume} {32}},\ \bibinfo {pages} {1163} (\bibinfo {year} {1999})}\BibitemShut {NoStop}%
\bibitem [{\citenamefont {Deutsch}(2018)}]{deutschEigenstateThermalizationHypothesis2018}%
  \BibitemOpen
  \bibfield  {author} {\bibinfo {author} {\bibfnamefont {J.~M.}\ \bibnamefont {Deutsch}},\ }\bibfield  {title} {\bibinfo {title} {Eigenstate thermalization hypothesis},\ }\href {https://doi.org/10.1088/1361-6633/aac9f1} {\bibfield  {journal} {\bibinfo  {journal} {Rep. Prog. Phys.}\ }\textbf {\bibinfo {volume} {81}},\ \bibinfo {pages} {082001} (\bibinfo {year} {2018})}\BibitemShut {NoStop}%
\bibitem [{\citenamefont {D'Alessio}\ \emph {et~al.}(2016)\citenamefont {D'Alessio}, \citenamefont {Kafri}, \citenamefont {Polkovnikov},\ and\ \citenamefont {Rigol}}]{dalessio_quantum_2016}%
  \BibitemOpen
  \bibfield  {author} {\bibinfo {author} {\bibfnamefont {L.}~\bibnamefont {D'Alessio}}, \bibinfo {author} {\bibfnamefont {Y.}~\bibnamefont {Kafri}}, \bibinfo {author} {\bibfnamefont {A.}~\bibnamefont {Polkovnikov}},\ and\ \bibinfo {author} {\bibfnamefont {M.}~\bibnamefont {Rigol}},\ }\bibfield  {title} {\bibinfo {title} {From quantum chaos and eigenstate thermalization to statistical mechanics and thermodynamics},\ }\href {https://doi.org/10.1080/00018732.2016.1198134} {\bibfield  {journal} {\bibinfo  {journal} {Adv. Phys.}\ }\textbf {\bibinfo {volume} {65}},\ \bibinfo {pages} {239} (\bibinfo {year} {2016})}\BibitemShut {NoStop}%
\bibitem [{\citenamefont {Ba\~nuls}\ \emph {et~al.}(2009)\citenamefont {Ba\~nuls}, \citenamefont {Hastings}, \citenamefont {Verstraete},\ and\ \citenamefont {Cirac}}]{banuls2009matrix}%
  \BibitemOpen
  \bibfield  {author} {\bibinfo {author} {\bibfnamefont {M.~C.}\ \bibnamefont {Ba\~nuls}}, \bibinfo {author} {\bibfnamefont {M.~B.}\ \bibnamefont {Hastings}}, \bibinfo {author} {\bibfnamefont {F.}~\bibnamefont {Verstraete}},\ and\ \bibinfo {author} {\bibfnamefont {J.~I.}\ \bibnamefont {Cirac}},\ }\bibfield  {title} {\bibinfo {title} {Matrix product states for dynamical simulation of infinite chains},\ }\href {https://doi.org/10.1103/PhysRevLett.102.240603} {\bibfield  {journal} {\bibinfo  {journal} {Phys. Rev. Lett.}\ }\textbf {\bibinfo {volume} {102}},\ \bibinfo {pages} {240603} (\bibinfo {year} {2009})}\BibitemShut {NoStop}%
\bibitem [{\citenamefont {Lerose}\ \emph {et~al.}(2021)\citenamefont {Lerose}, \citenamefont {Sonner},\ and\ \citenamefont {Abanin}}]{lerose_influence_2021}%
  \BibitemOpen
  \bibfield  {author} {\bibinfo {author} {\bibfnamefont {A.}~\bibnamefont {Lerose}}, \bibinfo {author} {\bibfnamefont {M.}~\bibnamefont {Sonner}},\ and\ \bibinfo {author} {\bibfnamefont {D.~A.}\ \bibnamefont {Abanin}},\ }\bibfield  {title} {\bibinfo {title} {Influence {Matrix} {Approach} to {Many}-{Body} {Floquet} {Dynamics}},\ }\href {https://doi.org/10.1103/PhysRevX.11.021040} {\bibfield  {journal} {\bibinfo  {journal} {Phys. Rev. X}\ }\textbf {\bibinfo {volume} {11}},\ \bibinfo {pages} {021040} (\bibinfo {year} {2021})}\BibitemShut {NoStop}%
\bibitem [{\citenamefont {Sonner}\ \emph {et~al.}(2021)\citenamefont {Sonner}, \citenamefont {Lerose},\ and\ \citenamefont {Abanin}}]{sonner_influence_2021}%
  \BibitemOpen
  \bibfield  {author} {\bibinfo {author} {\bibfnamefont {M.}~\bibnamefont {Sonner}}, \bibinfo {author} {\bibfnamefont {A.}~\bibnamefont {Lerose}},\ and\ \bibinfo {author} {\bibfnamefont {D.~A.}\ \bibnamefont {Abanin}},\ }\bibfield  {title} {\bibinfo {title} {Influence functional of many-body systems: {Temporal} entanglement and matrix-product state representation},\ }\href {https://doi.org/https://doi.org/10.1016/j.aop.2021.168677} {\bibfield  {journal} {\bibinfo  {journal} {Ann. Phys.}\ }\textbf {\bibinfo {volume} {435}},\ \bibinfo {pages} {168677} (\bibinfo {year} {2021})}\BibitemShut {NoStop}%
\bibitem [{\citenamefont {Feynman}\ and\ \citenamefont {Vernon}(1963)}]{feynman1963the}%
  \BibitemOpen
  \bibfield  {author} {\bibinfo {author} {\bibfnamefont {R.}~\bibnamefont {Feynman}}\ and\ \bibinfo {author} {\bibfnamefont {F.}~\bibnamefont {Vernon}},\ }\bibfield  {title} {\bibinfo {title} {The theory of a general quantum system interacting with a linear dissipative system},\ }\href {https://doi.org/https://doi.org/10.1016/0003-4916(63)90068-X} {\bibfield  {journal} {\bibinfo  {journal} {Ann. Phys}\ }\textbf {\bibinfo {volume} {24}},\ \bibinfo {pages} {118} (\bibinfo {year} {1963})}\BibitemShut {NoStop}%
\bibitem [{\citenamefont {M\"uller-Hermes}\ \emph {et~al.}(2012)\citenamefont {M\"uller-Hermes}, \citenamefont {Ignacio~Cirac},\ and\ \citenamefont {Ba\~nuls}}]{muellerhermes2012tensor}%
  \BibitemOpen
  \bibfield  {author} {\bibinfo {author} {\bibfnamefont {A.}~\bibnamefont {M\"uller-Hermes}}, \bibinfo {author} {\bibfnamefont {J.}~\bibnamefont {Ignacio~Cirac}},\ and\ \bibinfo {author} {\bibfnamefont {M.~C.}\ \bibnamefont {Ba\~nuls}},\ }\bibfield  {title} {\bibinfo {title} {Tensor network techniques for the computation of dynamical observables in one-dimensional quantum spin systems},\ }\href {https://doi.org/10.1088/1367-2630/14/7/075003} {\bibfield  {journal} {\bibinfo  {journal} {New J. Phys.}\ }\textbf {\bibinfo {volume} {14}},\ \bibinfo {pages} {075003} (\bibinfo {year} {2012})}\BibitemShut {NoStop}%
\bibitem [{\citenamefont {Hastings}\ and\ \citenamefont {Mahajan}(2015)}]{hastings2015connecting}%
  \BibitemOpen
  \bibfield  {author} {\bibinfo {author} {\bibfnamefont {M.~B.}\ \bibnamefont {Hastings}}\ and\ \bibinfo {author} {\bibfnamefont {R.}~\bibnamefont {Mahajan}},\ }\bibfield  {title} {\bibinfo {title} {Connecting entanglement in time and space: Improving the folding algorithm},\ }\href {https://doi.org/10.1103/PhysRevA.91.032306} {\bibfield  {journal} {\bibinfo  {journal} {Phys. Rev. A}\ }\textbf {\bibinfo {volume} {91}},\ \bibinfo {pages} {032306} (\bibinfo {year} {2015})}\BibitemShut {NoStop}%
\bibitem [{\citenamefont {Jørgensen}\ and\ \citenamefont {Pollock}(2019)}]{jorgensen_exploiting_2019}%
  \BibitemOpen
  \bibfield  {author} {\bibinfo {author} {\bibfnamefont {M.~R.}\ \bibnamefont {Jørgensen}}\ and\ \bibinfo {author} {\bibfnamefont {F.~A.}\ \bibnamefont {Pollock}},\ }\bibfield  {title} {\bibinfo {title} {Exploiting the {Causal} {Tensor} {Network} {Structure} of {Quantum} {Processes} to {Efficiently} {Simulate} {Non}-{Markovian} {Path} {Integrals}},\ }\href {https://doi.org/10.1103/PhysRevLett.123.240602} {\bibfield  {journal} {\bibinfo  {journal} {Phys. Rev. Lett.}\ }\textbf {\bibinfo {volume} {123}},\ \bibinfo {pages} {240602} (\bibinfo {year} {2019})}\BibitemShut {NoStop}%
\bibitem [{\citenamefont {Cygorek}\ \emph {et~al.}(2022)\citenamefont {Cygorek}, \citenamefont {Cosacchi}, \citenamefont {Vagov}, \citenamefont {Axt}, \citenamefont {Lovett}, \citenamefont {Keeling},\ and\ \citenamefont {Gauger}}]{cygorek_simulation_2022}%
  \BibitemOpen
  \bibfield  {author} {\bibinfo {author} {\bibfnamefont {M.}~\bibnamefont {Cygorek}}, \bibinfo {author} {\bibfnamefont {M.}~\bibnamefont {Cosacchi}}, \bibinfo {author} {\bibfnamefont {A.}~\bibnamefont {Vagov}}, \bibinfo {author} {\bibfnamefont {V.~M.}\ \bibnamefont {Axt}}, \bibinfo {author} {\bibfnamefont {B.~W.}\ \bibnamefont {Lovett}}, \bibinfo {author} {\bibfnamefont {J.}~\bibnamefont {Keeling}},\ and\ \bibinfo {author} {\bibfnamefont {E.~M.}\ \bibnamefont {Gauger}},\ }\bibfield  {title} {\bibinfo {title} {Simulation of open quantum systems by automated compression of arbitrary environments},\ }\href {https://doi.org/10.1038/s41567-022-01544-9} {\bibfield  {journal} {\bibinfo  {journal} {Nat. Phys.}\ }\textbf {\bibinfo {volume} {18}},\ \bibinfo {pages} {662} (\bibinfo {year} {2022})}\BibitemShut {NoStop}%
\bibitem [{\citenamefont {Fr\'{\i}as-P\'erez}\ and\ \citenamefont {Ba\~nuls}(2022)}]{friasperez2022light}%
  \BibitemOpen
  \bibfield  {author} {\bibinfo {author} {\bibfnamefont {M.}~\bibnamefont {Fr\'{\i}as-P\'erez}}\ and\ \bibinfo {author} {\bibfnamefont {M.~C.}\ \bibnamefont {Ba\~nuls}},\ }\bibfield  {title} {\bibinfo {title} {Light cone tensor network and time evolution},\ }\href {https://doi.org/10.1103/PhysRevB.106.115117} {\bibfield  {journal} {\bibinfo  {journal} {Phys. Rev. B}\ }\textbf {\bibinfo {volume} {106}},\ \bibinfo {pages} {115117} (\bibinfo {year} {2022})}\BibitemShut {NoStop}%
\bibitem [{\citenamefont {Giudice}\ \emph {et~al.}(2022)\citenamefont {Giudice}, \citenamefont {Giudici}, \citenamefont {Sonner}, \citenamefont {Thoenniss}, \citenamefont {Lerose}, \citenamefont {Abanin},\ and\ \citenamefont {Piroli}}]{giudice_temporal_2022}%
  \BibitemOpen
  \bibfield  {author} {\bibinfo {author} {\bibfnamefont {G.}~\bibnamefont {Giudice}}, \bibinfo {author} {\bibfnamefont {G.}~\bibnamefont {Giudici}}, \bibinfo {author} {\bibfnamefont {M.}~\bibnamefont {Sonner}}, \bibinfo {author} {\bibfnamefont {J.}~\bibnamefont {Thoenniss}}, \bibinfo {author} {\bibfnamefont {A.}~\bibnamefont {Lerose}}, \bibinfo {author} {\bibfnamefont {D.~A.}\ \bibnamefont {Abanin}},\ and\ \bibinfo {author} {\bibfnamefont {L.}~\bibnamefont {Piroli}},\ }\bibfield  {title} {\bibinfo {title} {Temporal entanglement, quasiparticles, and the role of interactions},\ }\href {https://doi.org/10.1103/PhysRevLett.128.220401} {\bibfield  {journal} {\bibinfo  {journal} {Phys. Rev. Lett.}\ }\textbf {\bibinfo {volume} {128}},\ \bibinfo {pages} {220401} (\bibinfo {year} {2022})}\BibitemShut {NoStop}%
\bibitem [{\citenamefont {Thoenniss}\ \emph {et~al.}(2023{\natexlab{a}})\citenamefont {Thoenniss}, \citenamefont {Sonner}, \citenamefont {Lerose},\ and\ \citenamefont {Abanin}}]{thoenniss_efficient_2023}%
  \BibitemOpen
  \bibfield  {author} {\bibinfo {author} {\bibfnamefont {J.}~\bibnamefont {Thoenniss}}, \bibinfo {author} {\bibfnamefont {M.}~\bibnamefont {Sonner}}, \bibinfo {author} {\bibfnamefont {A.}~\bibnamefont {Lerose}},\ and\ \bibinfo {author} {\bibfnamefont {D.~A.}\ \bibnamefont {Abanin}},\ }\bibfield  {title} {\bibinfo {title} {Efficient method for quantum impurity problems out of equilibrium},\ }\href {https://doi.org/10.1103/PhysRevB.107.L201115} {\bibfield  {journal} {\bibinfo  {journal} {Phys. Rev. B}\ }\textbf {\bibinfo {volume} {107}},\ \bibinfo {pages} {L201115} (\bibinfo {year} {2023}{\natexlab{a}})}\BibitemShut {NoStop}%
\bibitem [{\citenamefont {Foligno}\ \emph {et~al.}(2023)\citenamefont {Foligno}, \citenamefont {Zhou},\ and\ \citenamefont {Bertini}}]{foligno2023temporal}%
  \BibitemOpen
  \bibfield  {author} {\bibinfo {author} {\bibfnamefont {A.}~\bibnamefont {Foligno}}, \bibinfo {author} {\bibfnamefont {T.}~\bibnamefont {Zhou}},\ and\ \bibinfo {author} {\bibfnamefont {B.}~\bibnamefont {Bertini}},\ }\bibfield  {title} {\bibinfo {title} {Temporal entanglement in chaotic quantum circuits},\ }\href {https://doi.org/10.1103/PhysRevX.13.041008} {\bibfield  {journal} {\bibinfo  {journal} {Phys. Rev. X}\ }\textbf {\bibinfo {volume} {13}},\ \bibinfo {pages} {041008} (\bibinfo {year} {2023})}\BibitemShut {NoStop}%
\bibitem [{\citenamefont {Thoenniss}\ \emph {et~al.}(2023{\natexlab{b}})\citenamefont {Thoenniss}, \citenamefont {Lerose},\ and\ \citenamefont {Abanin}}]{thoenniss_nonequilibrium_2023}%
  \BibitemOpen
  \bibfield  {author} {\bibinfo {author} {\bibfnamefont {J.}~\bibnamefont {Thoenniss}}, \bibinfo {author} {\bibfnamefont {A.}~\bibnamefont {Lerose}},\ and\ \bibinfo {author} {\bibfnamefont {D.~A.}\ \bibnamefont {Abanin}},\ }\bibfield  {title} {\bibinfo {title} {Nonequilibrium quantum impurity problems via matrix-product states in the temporal domain},\ }\href {https://doi.org/10.1103/PhysRevB.107.195101} {\bibfield  {journal} {\bibinfo  {journal} {Phys. Rev. B}\ }\textbf {\bibinfo {volume} {107}},\ \bibinfo {pages} {195101} (\bibinfo {year} {2023}{\natexlab{b}})}\BibitemShut {NoStop}%
\bibitem [{\citenamefont {Ng}\ \emph {et~al.}(2023)\citenamefont {Ng}, \citenamefont {Park}, \citenamefont {Millis}, \citenamefont {Chan},\ and\ \citenamefont {Reichman}}]{ng_real-time_2023}%
  \BibitemOpen
  \bibfield  {author} {\bibinfo {author} {\bibfnamefont {N.}~\bibnamefont {Ng}}, \bibinfo {author} {\bibfnamefont {G.}~\bibnamefont {Park}}, \bibinfo {author} {\bibfnamefont {A.~J.}\ \bibnamefont {Millis}}, \bibinfo {author} {\bibfnamefont {G.~K.-L.}\ \bibnamefont {Chan}},\ and\ \bibinfo {author} {\bibfnamefont {D.~R.}\ \bibnamefont {Reichman}},\ }\bibfield  {title} {\bibinfo {title} {Real-time evolution of {Anderson} impurity models via tensor network influence functionals},\ }\href {https://doi.org/10.1103/PhysRevB.107.125103} {\bibfield  {journal} {\bibinfo  {journal} {Phys. Rev. B}\ }\textbf {\bibinfo {volume} {107}},\ \bibinfo {pages} {125103} (\bibinfo {year} {2023})}\BibitemShut {NoStop}%
\bibitem [{\citenamefont {Chen}\ \emph {et~al.}(2024)\citenamefont {Chen}, \citenamefont {Xu},\ and\ \citenamefont {Guo}}]{chen_grassmann_2024}%
  \BibitemOpen
  \bibfield  {author} {\bibinfo {author} {\bibfnamefont {R.}~\bibnamefont {Chen}}, \bibinfo {author} {\bibfnamefont {X.}~\bibnamefont {Xu}},\ and\ \bibinfo {author} {\bibfnamefont {C.}~\bibnamefont {Guo}},\ }\bibfield  {title} {\bibinfo {title} {Grassmann time-evolving matrix product operators for quantum impurity models},\ }\href {https://doi.org/10.1103/PhysRevB.109.045140} {\bibfield  {journal} {\bibinfo  {journal} {Phys. Rev. B}\ }\textbf {\bibinfo {volume} {109}},\ \bibinfo {pages} {045140} (\bibinfo {year} {2024})}\BibitemShut {NoStop}%
\bibitem [{\citenamefont {Nayak}\ \emph {et~al.}(2025)\citenamefont {Nayak}, \citenamefont {Thoenniss}, \citenamefont {Sonner}, \citenamefont {Abanin},\ and\ \citenamefont {Werner}}]{nayak_steady-state_2025}%
  \BibitemOpen
  \bibfield  {author} {\bibinfo {author} {\bibfnamefont {M.}~\bibnamefont {Nayak}}, \bibinfo {author} {\bibfnamefont {J.}~\bibnamefont {Thoenniss}}, \bibinfo {author} {\bibfnamefont {M.}~\bibnamefont {Sonner}}, \bibinfo {author} {\bibfnamefont {D.~A.}\ \bibnamefont {Abanin}},\ and\ \bibinfo {author} {\bibfnamefont {P.}~\bibnamefont {Werner}},\ }\bibfield  {title} {\bibinfo {title} {Steady-state dynamical mean field theory based on influence functional matrix product states},\ }\href {http://arxiv.org/abs/2503.02848} {\bibfield  {journal} {\bibinfo  {journal} {arXiv:2503.02848}\ } (\bibinfo {year} {2025})}\BibitemShut {NoStop}%
\bibitem [{\citenamefont {Klobas}\ and\ \citenamefont {Bertini}(2021)}]{klobas2021entanglement}%
  \BibitemOpen
  \bibfield  {author} {\bibinfo {author} {\bibfnamefont {K.}~\bibnamefont {Klobas}}\ and\ \bibinfo {author} {\bibfnamefont {B.}~\bibnamefont {Bertini}},\ }\bibfield  {title} {\bibinfo {title} {{Entanglement dynamics in Rule 54: Exact results and quasiparticle picture}},\ }\href {https://doi.org/10.21468/SciPostPhys.11.6.107} {\bibfield  {journal} {\bibinfo  {journal} {SciPost Phys.}\ }\textbf {\bibinfo {volume} {11}},\ \bibinfo {pages} {107} (\bibinfo {year} {2021})}\BibitemShut {NoStop}%
\bibitem [{\citenamefont {Bertini}\ \emph {et~al.}(2022{\natexlab{a}})\citenamefont {Bertini}, \citenamefont {Klobas}, \citenamefont {Alba}, \citenamefont {Lagnese},\ and\ \citenamefont {Calabrese}}]{bertini2022growth}%
  \BibitemOpen
  \bibfield  {author} {\bibinfo {author} {\bibfnamefont {B.}~\bibnamefont {Bertini}}, \bibinfo {author} {\bibfnamefont {K.}~\bibnamefont {Klobas}}, \bibinfo {author} {\bibfnamefont {V.}~\bibnamefont {Alba}}, \bibinfo {author} {\bibfnamefont {G.}~\bibnamefont {Lagnese}},\ and\ \bibinfo {author} {\bibfnamefont {P.}~\bibnamefont {Calabrese}},\ }\bibfield  {title} {\bibinfo {title} {{Growth of R\'enyi Entropies in Interacting Integrable Models and the Breakdown of the Quasiparticle Picture}},\ }\href {https://doi.org/10.1103/PhysRevX.12.031016} {\bibfield  {journal} {\bibinfo  {journal} {Phys. Rev. X}\ }\textbf {\bibinfo {volume} {12}},\ \bibinfo {pages} {031016} (\bibinfo {year} {2022}{\natexlab{a}})}\BibitemShut {NoStop}%
\bibitem [{\citenamefont {Bertini}\ \emph {et~al.}(2022{\natexlab{b}})\citenamefont {Bertini}, \citenamefont {Klobas},\ and\ \citenamefont {Lu}}]{bertini2022entanglement}%
  \BibitemOpen
  \bibfield  {author} {\bibinfo {author} {\bibfnamefont {B.}~\bibnamefont {Bertini}}, \bibinfo {author} {\bibfnamefont {K.}~\bibnamefont {Klobas}},\ and\ \bibinfo {author} {\bibfnamefont {T.-C.}\ \bibnamefont {Lu}},\ }\bibfield  {title} {\bibinfo {title} {Entanglement negativity and mutual information after a quantum quench: Exact link from space-time duality},\ }\href {https://doi.org/10.1103/PhysRevLett.129.140503} {\bibfield  {journal} {\bibinfo  {journal} {Phys. Rev. Lett.}\ }\textbf {\bibinfo {volume} {129}},\ \bibinfo {pages} {140503} (\bibinfo {year} {2022}{\natexlab{b}})}\BibitemShut {NoStop}%
\bibitem [{\citenamefont {Rampp}\ \emph {et~al.}(2024)\citenamefont {Rampp}, \citenamefont {Rather},\ and\ \citenamefont {Claeys}}]{rampp_entanglement_2023}%
  \BibitemOpen
  \bibfield  {author} {\bibinfo {author} {\bibfnamefont {M.~A.}\ \bibnamefont {Rampp}}, \bibinfo {author} {\bibfnamefont {S.~A.}\ \bibnamefont {Rather}},\ and\ \bibinfo {author} {\bibfnamefont {P.~W.}\ \bibnamefont {Claeys}},\ }\bibfield  {title} {\bibinfo {title} {Entanglement membrane in exactly solvable lattice models},\ }\href {https://doi.org/10.1103/PhysRevResearch.6.033271} {\bibfield  {journal} {\bibinfo  {journal} {Phys. Rev. Res.}\ }\textbf {\bibinfo {volume} {6}},\ \bibinfo {pages} {033271} (\bibinfo {year} {2024})}\BibitemShut {NoStop}%
\bibitem [{\citenamefont {Bertini}\ \emph {et~al.}(2023)\citenamefont {Bertini}, \citenamefont {Calabrese}, \citenamefont {Collura}, \citenamefont {Klobas},\ and\ \citenamefont {Rylands}}]{bertini2023nonequilibrium}%
  \BibitemOpen
  \bibfield  {author} {\bibinfo {author} {\bibfnamefont {B.}~\bibnamefont {Bertini}}, \bibinfo {author} {\bibfnamefont {P.}~\bibnamefont {Calabrese}}, \bibinfo {author} {\bibfnamefont {M.}~\bibnamefont {Collura}}, \bibinfo {author} {\bibfnamefont {K.}~\bibnamefont {Klobas}},\ and\ \bibinfo {author} {\bibfnamefont {C.}~\bibnamefont {Rylands}},\ }\bibfield  {title} {\bibinfo {title} {Nonequilibrium full counting statistics and symmetry-resolved entanglement from space-time duality},\ }\href {https://doi.org/10.1103/PhysRevLett.131.140401} {\bibfield  {journal} {\bibinfo  {journal} {Phys. Rev. Lett.}\ }\textbf {\bibinfo {volume} {131}},\ \bibinfo {pages} {140401} (\bibinfo {year} {2023})}\BibitemShut {NoStop}%
\bibitem [{\citenamefont {Yao}\ and\ \citenamefont {Claeys}(2024)}]{yao_temporal_2024}%
  \BibitemOpen
  \bibfield  {author} {\bibinfo {author} {\bibfnamefont {J.}~\bibnamefont {Yao}}\ and\ \bibinfo {author} {\bibfnamefont {P.~W.}\ \bibnamefont {Claeys}},\ }\bibfield  {title} {\bibinfo {title} {Temporal entanglement barriers in dual-unitary {Clifford} circuits with measurements},\ }\href {https://doi.org/10.1103/PhysRevResearch.6.043077} {\bibfield  {journal} {\bibinfo  {journal} {Phys. Rev. Res.}\ }\textbf {\bibinfo {volume} {6}},\ \bibinfo {pages} {043077} (\bibinfo {year} {2024})}\BibitemShut {NoStop}%
\bibitem [{\citenamefont {Bertini}\ \emph {et~al.}(2024)\citenamefont {Bertini}, \citenamefont {Klobas}, \citenamefont {Collura}, \citenamefont {Calabrese},\ and\ \citenamefont {Rylands}}]{bertini2024dynamics}%
  \BibitemOpen
  \bibfield  {author} {\bibinfo {author} {\bibfnamefont {B.}~\bibnamefont {Bertini}}, \bibinfo {author} {\bibfnamefont {K.}~\bibnamefont {Klobas}}, \bibinfo {author} {\bibfnamefont {M.}~\bibnamefont {Collura}}, \bibinfo {author} {\bibfnamefont {P.}~\bibnamefont {Calabrese}},\ and\ \bibinfo {author} {\bibfnamefont {C.}~\bibnamefont {Rylands}},\ }\bibfield  {title} {\bibinfo {title} {Dynamics of charge fluctuations from asymmetric initial states},\ }\href {https://doi.org/10.1103/PhysRevB.109.184312} {\bibfield  {journal} {\bibinfo  {journal} {Phys. Rev. B}\ }\textbf {\bibinfo {volume} {109}},\ \bibinfo {pages} {184312} (\bibinfo {year} {2024})}\BibitemShut {NoStop}%
\bibitem [{\citenamefont {Link}\ \emph {et~al.}(2024)\citenamefont {Link}, \citenamefont {Tu},\ and\ \citenamefont {Strunz}}]{link_open_2024}%
  \BibitemOpen
  \bibfield  {author} {\bibinfo {author} {\bibfnamefont {V.}~\bibnamefont {Link}}, \bibinfo {author} {\bibfnamefont {H.-H.}\ \bibnamefont {Tu}},\ and\ \bibinfo {author} {\bibfnamefont {W.~T.}\ \bibnamefont {Strunz}},\ }\bibfield  {title} {\bibinfo {title} {Open {Quantum} {System} {Dynamics} from {Infinite} {Tensor} {Network} {Contraction}},\ }\href {https://doi.org/10.1103/PhysRevLett.132.200403} {\bibfield  {journal} {\bibinfo  {journal} {Phys. Rev. Lett.}\ }\textbf {\bibinfo {volume} {132}},\ \bibinfo {pages} {200403} (\bibinfo {year} {2024})}\BibitemShut {NoStop}%
\bibitem [{\citenamefont {Sonner}\ \emph {et~al.}(2025)\citenamefont {Sonner}, \citenamefont {Link},\ and\ \citenamefont {Abanin}}]{sonner_semi-group_2025}%
  \BibitemOpen
  \bibfield  {author} {\bibinfo {author} {\bibfnamefont {M.}~\bibnamefont {Sonner}}, \bibinfo {author} {\bibfnamefont {V.}~\bibnamefont {Link}},\ and\ \bibinfo {author} {\bibfnamefont {D.~A.}\ \bibnamefont {Abanin}},\ }\bibfield  {title} {\bibinfo {title} {Semi-group influence matrices for non-equilibrium quantum impurity models},\ }\href {http://arxiv.org/abs/2502.00109} {\bibfield  {journal} {\bibinfo  {journal} {arXiv:2502.00109}\ } (\bibinfo {year} {2025})}\BibitemShut {NoStop}%
\bibitem [{\citenamefont {Cotler}\ \emph {et~al.}(2023)\citenamefont {Cotler}, \citenamefont {Mark}, \citenamefont {Huang}, \citenamefont {Hernández}, \citenamefont {Choi}, \citenamefont {Shaw}, \citenamefont {Endres},\ and\ \citenamefont {Choi}}]{cotler_emergent_2023}%
  \BibitemOpen
  \bibfield  {author} {\bibinfo {author} {\bibfnamefont {J.~S.}\ \bibnamefont {Cotler}}, \bibinfo {author} {\bibfnamefont {D.~K.}\ \bibnamefont {Mark}}, \bibinfo {author} {\bibfnamefont {H.-Y.}\ \bibnamefont {Huang}}, \bibinfo {author} {\bibfnamefont {F.}~\bibnamefont {Hernández}}, \bibinfo {author} {\bibfnamefont {J.}~\bibnamefont {Choi}}, \bibinfo {author} {\bibfnamefont {A.~L.}\ \bibnamefont {Shaw}}, \bibinfo {author} {\bibfnamefont {M.}~\bibnamefont {Endres}},\ and\ \bibinfo {author} {\bibfnamefont {S.}~\bibnamefont {Choi}},\ }\bibfield  {title} {\bibinfo {title} {Emergent {Quantum} {State} {Designs} from {Individual} {Many}-{Body} {Wave} {Functions}},\ }\href {https://doi.org/10.1103/PRXQuantum.4.010311} {\bibfield  {journal} {\bibinfo  {journal} {PRX Quantum}\ }\textbf {\bibinfo {volume} {4}},\ \bibinfo {pages} {010311} (\bibinfo {year} {2023})}\BibitemShut {NoStop}%
\bibitem [{\citenamefont {Choi}\ \emph {et~al.}(2023)\citenamefont {Choi}, \citenamefont {Shaw}, \citenamefont {Madjarov}, \citenamefont {Xie}, \citenamefont {Finkelstein}, \citenamefont {Covey}, \citenamefont {Cotler}, \citenamefont {Mark}, \citenamefont {Huang}, \citenamefont {Kale}, \citenamefont {Pichler}, \citenamefont {Brandão}, \citenamefont {Choi},\ and\ \citenamefont {Endres}}]{choi_preparing_2023}%
  \BibitemOpen
  \bibfield  {author} {\bibinfo {author} {\bibfnamefont {J.}~\bibnamefont {Choi}}, \bibinfo {author} {\bibfnamefont {A.~L.}\ \bibnamefont {Shaw}}, \bibinfo {author} {\bibfnamefont {I.~S.}\ \bibnamefont {Madjarov}}, \bibinfo {author} {\bibfnamefont {X.}~\bibnamefont {Xie}}, \bibinfo {author} {\bibfnamefont {R.}~\bibnamefont {Finkelstein}}, \bibinfo {author} {\bibfnamefont {J.~P.}\ \bibnamefont {Covey}}, \bibinfo {author} {\bibfnamefont {J.~S.}\ \bibnamefont {Cotler}}, \bibinfo {author} {\bibfnamefont {D.~K.}\ \bibnamefont {Mark}}, \bibinfo {author} {\bibfnamefont {H.-Y.}\ \bibnamefont {Huang}}, \bibinfo {author} {\bibfnamefont {A.}~\bibnamefont {Kale}}, \bibinfo {author} {\bibfnamefont {H.}~\bibnamefont {Pichler}}, \bibinfo {author} {\bibfnamefont {F.~G. S.~L.}\ \bibnamefont {Brandão}}, \bibinfo {author} {\bibfnamefont {S.}~\bibnamefont {Choi}},\ and\ \bibinfo {author} {\bibfnamefont {M.}~\bibnamefont {Endres}},\ }\bibfield  {title} {\bibinfo {title} {Preparing random states and benchmarking with many-body
  quantum chaos},\ }\href {https://doi.org/10.1038/s41586-022-05442-1} {\bibfield  {journal} {\bibinfo  {journal} {Nature}\ }\textbf {\bibinfo {volume} {613}},\ \bibinfo {pages} {468} (\bibinfo {year} {2023})}\BibitemShut {NoStop}%
\bibitem [{\citenamefont {Ippoliti}\ and\ \citenamefont {Ho}(2022)}]{ippoliti_solvable_2022}%
  \BibitemOpen
  \bibfield  {author} {\bibinfo {author} {\bibfnamefont {M.}~\bibnamefont {Ippoliti}}\ and\ \bibinfo {author} {\bibfnamefont {W.~W.}\ \bibnamefont {Ho}},\ }\bibfield  {title} {\bibinfo {title} {Solvable model of deep thermalization with distinct design times},\ }\href {https://doi.org/10.22331/q-2022-12-29-886} {\bibfield  {journal} {\bibinfo  {journal} {Quantum}\ }\textbf {\bibinfo {volume} {6}},\ \bibinfo {pages} {886} (\bibinfo {year} {2022})}\BibitemShut {NoStop}%
\bibitem [{\citenamefont {Ho}\ and\ \citenamefont {Choi}(2022)}]{ho_exact_2022}%
  \BibitemOpen
  \bibfield  {author} {\bibinfo {author} {\bibfnamefont {W.~W.}\ \bibnamefont {Ho}}\ and\ \bibinfo {author} {\bibfnamefont {S.}~\bibnamefont {Choi}},\ }\bibfield  {title} {\bibinfo {title} {Exact {Emergent} {Quantum} {State} {Designs} from {Quantum} {Chaotic} {Dynamics}},\ }\href {https://doi.org/10.1103/PhysRevLett.128.060601} {\bibfield  {journal} {\bibinfo  {journal} {Phys. Rev. Lett.}\ }\textbf {\bibinfo {volume} {128}},\ \bibinfo {pages} {060601} (\bibinfo {year} {2022})}\BibitemShut {NoStop}%
\bibitem [{\citenamefont {Claeys}\ and\ \citenamefont {Lamacraft}(2022)}]{claeys_emergent_2022}%
  \BibitemOpen
  \bibfield  {author} {\bibinfo {author} {\bibfnamefont {P.~W.}\ \bibnamefont {Claeys}}\ and\ \bibinfo {author} {\bibfnamefont {A.}~\bibnamefont {Lamacraft}},\ }\bibfield  {title} {\bibinfo {title} {Emergent quantum state designs and biunitarity in dual-unitary circuit dynamics},\ }\href {https://doi.org/10.22331/q-2022-06-15-738} {\bibfield  {journal} {\bibinfo  {journal} {Quantum}\ }\textbf {\bibinfo {volume} {6}},\ \bibinfo {pages} {738} (\bibinfo {year} {2022})}\BibitemShut {NoStop}%
\bibitem [{\citenamefont {Wilming}\ and\ \citenamefont {Roth}(2022)}]{wilming_high-temperature_2022}%
  \BibitemOpen
  \bibfield  {author} {\bibinfo {author} {\bibfnamefont {H.}~\bibnamefont {Wilming}}\ and\ \bibinfo {author} {\bibfnamefont {I.}~\bibnamefont {Roth}},\ }\bibfield  {title} {\bibinfo {title} {High-temperature thermalization implies the emergence of quantum state designs},\ }\href {http://arxiv.org/abs/2202.01669} {\bibfield  {journal} {\bibinfo  {journal} {arXiv:2202.01669}\ } (\bibinfo {year} {2022})}\BibitemShut {NoStop}%
\bibitem [{\citenamefont {Ippoliti}\ and\ \citenamefont {Ho}(2023)}]{ippoliti_dynamical_2023}%
  \BibitemOpen
  \bibfield  {author} {\bibinfo {author} {\bibfnamefont {M.}~\bibnamefont {Ippoliti}}\ and\ \bibinfo {author} {\bibfnamefont {W.~W.}\ \bibnamefont {Ho}},\ }\bibfield  {title} {\bibinfo {title} {Dynamical {Purification} and the {Emergence} of {Quantum} {State} {Designs} from the {Projected} {Ensemble}},\ }\href {https://doi.org/10.1103/PRXQuantum.4.030322} {\bibfield  {journal} {\bibinfo  {journal} {PRX Quantum}\ }\textbf {\bibinfo {volume} {4}},\ \bibinfo {pages} {030322} (\bibinfo {year} {2023})}\BibitemShut {NoStop}%
\bibitem [{\citenamefont {Bhore}\ \emph {et~al.}(2023)\citenamefont {Bhore}, \citenamefont {Desaules},\ and\ \citenamefont {Papić}}]{bhore_deep_2023}%
  \BibitemOpen
  \bibfield  {author} {\bibinfo {author} {\bibfnamefont {T.}~\bibnamefont {Bhore}}, \bibinfo {author} {\bibfnamefont {J.-Y.}\ \bibnamefont {Desaules}},\ and\ \bibinfo {author} {\bibfnamefont {Z.}~\bibnamefont {Papić}},\ }\bibfield  {title} {\bibinfo {title} {Deep thermalization in constrained quantum systems},\ }\href {https://doi.org/10.1103/PhysRevB.108.104317} {\bibfield  {journal} {\bibinfo  {journal} {Phys. Rev. B}\ }\textbf {\bibinfo {volume} {108}},\ \bibinfo {pages} {104317} (\bibinfo {year} {2023})}\BibitemShut {NoStop}%
\bibitem [{\citenamefont {Liu}\ \emph {et~al.}(2024)\citenamefont {Liu}, \citenamefont {Huang},\ and\ \citenamefont {Ho}}]{liu_deep_2024}%
  \BibitemOpen
  \bibfield  {author} {\bibinfo {author} {\bibfnamefont {C.}~\bibnamefont {Liu}}, \bibinfo {author} {\bibfnamefont {Q.~C.}\ \bibnamefont {Huang}},\ and\ \bibinfo {author} {\bibfnamefont {W.~W.}\ \bibnamefont {Ho}},\ }\bibfield  {title} {\bibinfo {title} {Deep thermalization in {Gaussian} continuous-variable quantum systems},\ }\href {https://doi.org/10.1103/PhysRevLett.133.260401} {\bibfield  {journal} {\bibinfo  {journal} {Phys. Rev. Lett.}\ }\textbf {\bibinfo {volume} {133}},\ \bibinfo {pages} {260401} (\bibinfo {year} {2024})}\BibitemShut {NoStop}%
\bibitem [{\citenamefont {Chang}\ \emph {et~al.}(2025)\citenamefont {Chang}, \citenamefont {Shrotriya}, \citenamefont {Ho},\ and\ \citenamefont {Ippoliti}}]{chang_deep_2025}%
  \BibitemOpen
  \bibfield  {author} {\bibinfo {author} {\bibfnamefont {R.-A.}\ \bibnamefont {Chang}}, \bibinfo {author} {\bibfnamefont {H.}~\bibnamefont {Shrotriya}}, \bibinfo {author} {\bibfnamefont {W.~W.}\ \bibnamefont {Ho}},\ and\ \bibinfo {author} {\bibfnamefont {M.}~\bibnamefont {Ippoliti}},\ }\bibfield  {title} {\bibinfo {title} {Deep thermalization under charge-conserving quantum dynamics},\ }\href {http://arxiv.org/abs/2408.15325} {\bibfield  {journal} {\bibinfo  {journal} {arXiv:2408.15325}\ } (\bibinfo {year} {2025})}\BibitemShut {NoStop}%
\bibitem [{\citenamefont {Lucas}\ \emph {et~al.}(2023)\citenamefont {Lucas}, \citenamefont {Piroli}, \citenamefont {De~Nardis},\ and\ \citenamefont {De~Luca}}]{lucas_generalized_2023}%
  \BibitemOpen
  \bibfield  {author} {\bibinfo {author} {\bibfnamefont {M.}~\bibnamefont {Lucas}}, \bibinfo {author} {\bibfnamefont {L.}~\bibnamefont {Piroli}}, \bibinfo {author} {\bibfnamefont {J.}~\bibnamefont {De~Nardis}},\ and\ \bibinfo {author} {\bibfnamefont {A.}~\bibnamefont {De~Luca}},\ }\bibfield  {title} {\bibinfo {title} {Generalized deep thermalization for free fermions},\ }\href {https://doi.org/10.1103/PhysRevA.107.032215} {\bibfield  {journal} {\bibinfo  {journal} {Phys. Rev. A}\ }\textbf {\bibinfo {volume} {107}},\ \bibinfo {pages} {032215} (\bibinfo {year} {2023})}\BibitemShut {NoStop}%
\bibitem [{\citenamefont {Mark}\ \emph {et~al.}(2024)\citenamefont {Mark}, \citenamefont {Surace}, \citenamefont {Elben}, \citenamefont {Shaw}, \citenamefont {Choi}, \citenamefont {Refael}, \citenamefont {Endres},\ and\ \citenamefont {Choi}}]{mark_maximum_2024}%
  \BibitemOpen
  \bibfield  {author} {\bibinfo {author} {\bibfnamefont {D.~K.}\ \bibnamefont {Mark}}, \bibinfo {author} {\bibfnamefont {F.}~\bibnamefont {Surace}}, \bibinfo {author} {\bibfnamefont {A.}~\bibnamefont {Elben}}, \bibinfo {author} {\bibfnamefont {A.~L.}\ \bibnamefont {Shaw}}, \bibinfo {author} {\bibfnamefont {J.}~\bibnamefont {Choi}}, \bibinfo {author} {\bibfnamefont {G.}~\bibnamefont {Refael}}, \bibinfo {author} {\bibfnamefont {M.}~\bibnamefont {Endres}},\ and\ \bibinfo {author} {\bibfnamefont {S.}~\bibnamefont {Choi}},\ }\bibfield  {title} {\bibinfo {title} {Maximum {Entropy} {Principle} in {Deep} {Thermalization} and in {Hilbert}-{Space} {Ergodicity}},\ }\href {https://doi.org/10.1103/PhysRevX.14.041051} {\bibfield  {journal} {\bibinfo  {journal} {Phys. Rev. X}\ }\textbf {\bibinfo {volume} {14}},\ \bibinfo {pages} {041051} (\bibinfo {year} {2024})}\BibitemShut {NoStop}%
\bibitem [{\citenamefont {Manna}\ \emph {et~al.}(2025)\citenamefont {Manna}, \citenamefont {Roy},\ and\ \citenamefont {Sreejith}}]{manna_projected_2025}%
  \BibitemOpen
  \bibfield  {author} {\bibinfo {author} {\bibfnamefont {S.}~\bibnamefont {Manna}}, \bibinfo {author} {\bibfnamefont {S.}~\bibnamefont {Roy}},\ and\ \bibinfo {author} {\bibfnamefont {G.~J.}\ \bibnamefont {Sreejith}},\ }\bibfield  {title} {\bibinfo {title} {Projected ensemble in a system with conserved charges with local support},\ }\href {http://arxiv.org/abs/2501.01823} {\bibfield  {journal} {\bibinfo  {journal} {arXiv:2501.01823}\ } (\bibinfo {year} {2025})}\BibitemShut {NoStop}%
\bibitem [{\citenamefont {Chan}\ and\ \citenamefont {De~Luca}(2024)}]{chan_projected_2024}%
  \BibitemOpen
  \bibfield  {author} {\bibinfo {author} {\bibfnamefont {A.}~\bibnamefont {Chan}}\ and\ \bibinfo {author} {\bibfnamefont {A.}~\bibnamefont {De~Luca}},\ }\bibfield  {title} {\bibinfo {title} {Projected state ensemble of a generic model of many-body quantum chaos},\ }\href {https://doi.org/10.1088/1751-8121/ad7211} {\bibfield  {journal} {\bibinfo  {journal} {J. Phys. A: Math. Theor.}\ }\textbf {\bibinfo {volume} {57}},\ \bibinfo {pages} {405001} (\bibinfo {year} {2024})}\BibitemShut {NoStop}%
\bibitem [{\citenamefont {Varikuti}\ and\ \citenamefont {Bandyopadhyay}(2024)}]{varikuti_unraveling_2024}%
  \BibitemOpen
  \bibfield  {author} {\bibinfo {author} {\bibfnamefont {N.~D.}\ \bibnamefont {Varikuti}}\ and\ \bibinfo {author} {\bibfnamefont {S.}~\bibnamefont {Bandyopadhyay}},\ }\bibfield  {title} {\bibinfo {title} {Unraveling the emergence of quantum state designs in systems with symmetry},\ }\href {https://doi.org/10.22331/q-2024-08-29-1456} {\bibfield  {journal} {\bibinfo  {journal} {Quantum}\ }\textbf {\bibinfo {volume} {8}},\ \bibinfo {pages} {1456} (\bibinfo {year} {2024})}\BibitemShut {NoStop}%
\bibitem [{\citenamefont {Shrotriya}\ and\ \citenamefont {Ho}(2025)}]{shrotriya_nonlocality_2025}%
  \BibitemOpen
  \bibfield  {author} {\bibinfo {author} {\bibfnamefont {H.}~\bibnamefont {Shrotriya}}\ and\ \bibinfo {author} {\bibfnamefont {W.~W.}\ \bibnamefont {Ho}},\ }\bibfield  {title} {\bibinfo {title} {Nonlocality of deep thermalization},\ }\href {https://doi.org/10.21468/SciPostPhys.18.3.107} {\bibfield  {journal} {\bibinfo  {journal} {SciPost Phys.}\ }\textbf {\bibinfo {volume} {18}},\ \bibinfo {pages} {107} (\bibinfo {year} {2025})}\BibitemShut {NoStop}%
\bibitem [{\citenamefont {Mok}\ \emph {et~al.}(2025)\citenamefont {Mok}, \citenamefont {Haug}, \citenamefont {Shaw}, \citenamefont {Endres},\ and\ \citenamefont {Preskill}}]{mok_optimal_2025}%
  \BibitemOpen
  \bibfield  {author} {\bibinfo {author} {\bibfnamefont {W.-K.}\ \bibnamefont {Mok}}, \bibinfo {author} {\bibfnamefont {T.}~\bibnamefont {Haug}}, \bibinfo {author} {\bibfnamefont {A.~L.}\ \bibnamefont {Shaw}}, \bibinfo {author} {\bibfnamefont {M.}~\bibnamefont {Endres}},\ and\ \bibinfo {author} {\bibfnamefont {J.}~\bibnamefont {Preskill}},\ }\bibfield  {title} {\bibinfo {title} {Optimal {Conversion} from {Classical} to {Quantum} {Randomness} via {Quantum} {Chaos}},\ }\href {http://arxiv.org/abs/2410.05181} {\bibfield  {journal} {\bibinfo  {journal} {arXiv:2410.05181}\ } (\bibinfo {year} {2025})}\BibitemShut {NoStop}%
\bibitem [{\citenamefont {Yu}\ \emph {et~al.}(2025)\citenamefont {Yu}, \citenamefont {Ho},\ and\ \citenamefont {Kos}}]{yu2025mixed}%
  \BibitemOpen
  \bibfield  {author} {\bibinfo {author} {\bibfnamefont {X.-H.}\ \bibnamefont {Yu}}, \bibinfo {author} {\bibfnamefont {W.~W.}\ \bibnamefont {Ho}},\ and\ \bibinfo {author} {\bibfnamefont {P.}~\bibnamefont {Kos}},\ }\bibfield  {title} {\bibinfo {title} {Mixed state deep thermalization},\ }\href {https://doi.org/10.48550/arXiv.2505.07795} {\bibfield  {journal} {\bibinfo  {journal} {arXiv:2505.07795}\ } (\bibinfo {year} {2025})}\BibitemShut {NoStop}%
\bibitem [{\citenamefont {Lami}\ \emph {et~al.}(2025)\citenamefont {Lami}, \citenamefont {De~Luca}, \citenamefont {Turkeshi},\ and\ \citenamefont {De~Nardis}}]{lami2025quantum}%
  \BibitemOpen
  \bibfield  {author} {\bibinfo {author} {\bibfnamefont {G.}~\bibnamefont {Lami}}, \bibinfo {author} {\bibfnamefont {A.}~\bibnamefont {De~Luca}}, \bibinfo {author} {\bibfnamefont {X.}~\bibnamefont {Turkeshi}},\ and\ \bibinfo {author} {\bibfnamefont {J.}~\bibnamefont {De~Nardis}},\ }\bibfield  {title} {\bibinfo {title} {Quantum state design and emergent confinement mechanism in measured tensor network states},\ }\bibfield  {journal} {\bibinfo  {journal} {arXiv:2504.16995}\ }\href {https://doi.org/https://doi.org/10.48550/arXiv.2504.16995} {https://doi.org/10.48550/arXiv.2504.16995} (\bibinfo {year} {2025})\BibitemShut {NoStop}%
\bibitem [{\citenamefont {Maldacena}\ \emph {et~al.}(2016)\citenamefont {Maldacena}, \citenamefont {Shenker},\ and\ \citenamefont {Stanford}}]{maldacena2016bound}%
  \BibitemOpen
  \bibfield  {author} {\bibinfo {author} {\bibfnamefont {J.}~\bibnamefont {Maldacena}}, \bibinfo {author} {\bibfnamefont {S.~H.}\ \bibnamefont {Shenker}},\ and\ \bibinfo {author} {\bibfnamefont {D.}~\bibnamefont {Stanford}},\ }\bibfield  {title} {\bibinfo {title} {A bound on chaos},\ }\href {https://doi.org/https://doi.org/10.1007/jhep08(2016)106} {\bibfield  {journal} {\bibinfo  {journal} {J. High Energy Phys.}\ }\textbf {\bibinfo {volume} {2016}}\bibinfo  {number} { (8)}}\BibitemShut {NoStop}%
\bibitem [{\citenamefont {Swingle}(2018)}]{swingle_unscrambling_2018}%
  \BibitemOpen
\bibfield  {number} {  }\bibfield  {author} {\bibinfo {author} {\bibfnamefont {B.}~\bibnamefont {Swingle}},\ }\bibfield  {title} {\bibinfo {title} {Unscrambling the physics of out-of-time-order correlators},\ }\href {https://doi.org/10.1038/s41567-018-0295-5} {\bibfield  {journal} {\bibinfo  {journal} {Nat. Phys.}\ }\textbf {\bibinfo {volume} {14}},\ \bibinfo {pages} {988} (\bibinfo {year} {2018})}\BibitemShut {NoStop}%
\bibitem [{\citenamefont {Fava}\ \emph {et~al.}(2021)\citenamefont {Fava}, \citenamefont {Biswas}, \citenamefont {Gopalakrishnan}, \citenamefont {Vasseur},\ and\ \citenamefont {Parameswaran}}]{fava2021hydrodynamic}%
  \BibitemOpen
  \bibfield  {author} {\bibinfo {author} {\bibfnamefont {M.}~\bibnamefont {Fava}}, \bibinfo {author} {\bibfnamefont {S.}~\bibnamefont {Biswas}}, \bibinfo {author} {\bibfnamefont {S.}~\bibnamefont {Gopalakrishnan}}, \bibinfo {author} {\bibfnamefont {R.}~\bibnamefont {Vasseur}},\ and\ \bibinfo {author} {\bibfnamefont {S.}~\bibnamefont {Parameswaran}},\ }\bibfield  {title} {\bibinfo {title} {Hydrodynamic nonlinear response of interacting integrable systems},\ }\href {https://doi.org/10.1073/pnas.2106945118} {\bibfield  {journal} {\bibinfo  {journal} {Proc. Natl. Acad. Sci.}\ }\textbf {\bibinfo {volume} {118}},\ \bibinfo {pages} {e2106945118} (\bibinfo {year} {2021})}\BibitemShut {NoStop}%
\bibitem [{\citenamefont {Doyon}\ and\ \citenamefont {Myers}(2020)}]{doyon2020fluctuations}%
  \BibitemOpen
  \bibfield  {author} {\bibinfo {author} {\bibfnamefont {B.}~\bibnamefont {Doyon}}\ and\ \bibinfo {author} {\bibfnamefont {J.}~\bibnamefont {Myers}},\ }\bibfield  {title} {\bibinfo {title} {Fluctuations in ballistic transport from euler hydrodynamics},\ }in\ \href {https://doi.org/10.1007/s00023-019-00860-w} {\emph {\bibinfo {booktitle} {Annales Henri Poincar{\'e}}}},\ Vol.~\bibinfo {volume} {21}\ (\bibinfo {organization} {Springer},\ \bibinfo {year} {2020})\ pp.\ \bibinfo {pages} {255--302}\BibitemShut {NoStop}%
\bibitem [{\citenamefont {Myers}\ \emph {et~al.}(2020)\citenamefont {Myers}, \citenamefont {Bhaseen}, \citenamefont {Harris},\ and\ \citenamefont {Doyon}}]{myers2020transport}%
  \BibitemOpen
  \bibfield  {author} {\bibinfo {author} {\bibfnamefont {J.}~\bibnamefont {Myers}}, \bibinfo {author} {\bibfnamefont {J.}~\bibnamefont {Bhaseen}}, \bibinfo {author} {\bibfnamefont {R.~J.}\ \bibnamefont {Harris}},\ and\ \bibinfo {author} {\bibfnamefont {B.}~\bibnamefont {Doyon}},\ }\bibfield  {title} {\bibinfo {title} {Transport fluctuations in integrable models out of equilibrium},\ }\href {10.21468/SciPostPhys.8.1.007} {\bibfield  {journal} {\bibinfo  {journal} {SciPost Phys.}\ }\textbf {\bibinfo {volume} {8}},\ \bibinfo {pages} {007} (\bibinfo {year} {2020})}\BibitemShut {NoStop}%
\bibitem [{\citenamefont {Foini}\ and\ \citenamefont {Kurchan}(2019)}]{foiniEigenstateThermalizationHypothesis2019a}%
  \BibitemOpen
  \bibfield  {author} {\bibinfo {author} {\bibfnamefont {L.}~\bibnamefont {Foini}}\ and\ \bibinfo {author} {\bibfnamefont {J.}~\bibnamefont {Kurchan}},\ }\bibfield  {title} {\bibinfo {title} {Eigenstate thermalization hypothesis and out of time order correlators},\ }\href {https://doi.org/10.1103/PhysRevE.99.042139} {\bibfield  {journal} {\bibinfo  {journal} {Phys. Rev. E}\ }\textbf {\bibinfo {volume} {99}},\ \bibinfo {pages} {042139} (\bibinfo {year} {2019})}\BibitemShut {NoStop}%
\bibitem [{\citenamefont {Brenes}\ \emph {et~al.}(2021)\citenamefont {Brenes}, \citenamefont {Pappalardi}, \citenamefont {Mitchison}, \citenamefont {Goold},\ and\ \citenamefont {Silva}}]{brenesOutoftimeorderCorrelationsFine2021c}%
  \BibitemOpen
  \bibfield  {author} {\bibinfo {author} {\bibfnamefont {M.}~\bibnamefont {Brenes}}, \bibinfo {author} {\bibfnamefont {S.}~\bibnamefont {Pappalardi}}, \bibinfo {author} {\bibfnamefont {M.~T.}\ \bibnamefont {Mitchison}}, \bibinfo {author} {\bibfnamefont {J.}~\bibnamefont {Goold}},\ and\ \bibinfo {author} {\bibfnamefont {A.}~\bibnamefont {Silva}},\ }\bibfield  {title} {\bibinfo {title} {Out-of-time-order correlations and the fine structure of eigenstate thermalization},\ }\href {https://doi.org/10.1103/PhysRevE.104.034120} {\bibfield  {journal} {\bibinfo  {journal} {Phys. Rev. E}\ }\textbf {\bibinfo {volume} {104}},\ \bibinfo {pages} {034120} (\bibinfo {year} {2021})}\BibitemShut {NoStop}%
\bibitem [{\citenamefont {Pappalardi}\ \emph {et~al.}(2022)\citenamefont {Pappalardi}, \citenamefont {Foini},\ and\ \citenamefont {Kurchan}}]{pappalardiEigenstateThermalizationHypothesis2022}%
  \BibitemOpen
  \bibfield  {author} {\bibinfo {author} {\bibfnamefont {S.}~\bibnamefont {Pappalardi}}, \bibinfo {author} {\bibfnamefont {L.}~\bibnamefont {Foini}},\ and\ \bibinfo {author} {\bibfnamefont {J.}~\bibnamefont {Kurchan}},\ }\bibfield  {title} {\bibinfo {title} {Eigenstate {{Thermalization Hypothesis}} and {{Free Probability}}},\ }\href {https://doi.org/10.1103/PhysRevLett.129.170603} {\bibfield  {journal} {\bibinfo  {journal} {Phys. Rev. Lett.}\ }\textbf {\bibinfo {volume} {129}},\ \bibinfo {pages} {170603} (\bibinfo {year} {2022})}\BibitemShut {NoStop}%
\bibitem [{\citenamefont {Jindal}\ and\ \citenamefont {Hosur}(2024)}]{jindalGeneralizedFreeCumulants2024}%
  \BibitemOpen
  \bibfield  {author} {\bibinfo {author} {\bibfnamefont {S.}~\bibnamefont {Jindal}}\ and\ \bibinfo {author} {\bibfnamefont {P.}~\bibnamefont {Hosur}},\ }\bibfield  {title} {\bibinfo {title} {Generalized {{Free Cumulants}} for {{Quantum Chaotic Systems}}},\ }\href {https://doi.org/10.1007/JHEP09(2024)066} {\bibfield  {journal} {\bibinfo  {journal} {arXiv:2401.13829}\ } (\bibinfo {year} {2024})}\BibitemShut {NoStop}%
\bibitem [{\citenamefont {Vallini}\ and\ \citenamefont {Pappalardi}(2024)}]{valliniLongtimeFreenessKicked2024}%
  \BibitemOpen
  \bibfield  {author} {\bibinfo {author} {\bibfnamefont {E.}~\bibnamefont {Vallini}}\ and\ \bibinfo {author} {\bibfnamefont {S.}~\bibnamefont {Pappalardi}},\ }\bibfield  {title} {\bibinfo {title} {Long-time {{Freeness}} in the {{Kicked Top}}},\ }\href {https://doi.org/10.48550/arXiv.2411.12050} {\bibfield  {journal} {\bibinfo  {journal} {arXiv:2411.12050}\ } (\bibinfo {year} {2024})}\BibitemShut {NoStop}%
\bibitem [{\citenamefont {Pappalardi}\ \emph {et~al.}(2025)\citenamefont {Pappalardi}, \citenamefont {Fritzsch},\ and\ \citenamefont {Prosen}}]{pappalardi_full_2025}%
  \BibitemOpen
  \bibfield  {author} {\bibinfo {author} {\bibfnamefont {S.}~\bibnamefont {Pappalardi}}, \bibinfo {author} {\bibfnamefont {F.}~\bibnamefont {Fritzsch}},\ and\ \bibinfo {author} {\bibfnamefont {T.}~\bibnamefont {Prosen}},\ }\bibfield  {title} {\bibinfo {title} {Full {Eigenstate} {Thermalization} via {Free} {Cumulants} in {Quantum} {Lattice} {Systems}},\ }\href {https://doi.org/10.1103/PhysRevLett.134.140404} {\bibfield  {journal} {\bibinfo  {journal} {Phys. Rev. Lett.}\ }\textbf {\bibinfo {volume} {134}},\ \bibinfo {pages} {140404} (\bibinfo {year} {2025})}\BibitemShut {NoStop}%
\bibitem [{\citenamefont {Fava}\ \emph {et~al.}(2025)\citenamefont {Fava}, \citenamefont {Kurchan},\ and\ \citenamefont {Pappalardi}}]{fava_designs_2025}%
  \BibitemOpen
  \bibfield  {author} {\bibinfo {author} {\bibfnamefont {M.}~\bibnamefont {Fava}}, \bibinfo {author} {\bibfnamefont {J.}~\bibnamefont {Kurchan}},\ and\ \bibinfo {author} {\bibfnamefont {S.}~\bibnamefont {Pappalardi}},\ }\bibfield  {title} {\bibinfo {title} {Designs via {Free} {Probability}},\ }\href {https://doi.org/10.1103/PhysRevX.15.011031} {\bibfield  {journal} {\bibinfo  {journal} {Phys. Rev. X}\ }\textbf {\bibinfo {volume} {15}},\ \bibinfo {pages} {011031} (\bibinfo {year} {2025})}\BibitemShut {NoStop}%
\bibitem [{\citenamefont {Alves}\ \emph {et~al.}(2025)\citenamefont {Alves}, \citenamefont {Fritzsch},\ and\ \citenamefont {Claeys}}]{alves_probes_2025}%
  \BibitemOpen
  \bibfield  {author} {\bibinfo {author} {\bibfnamefont {G.~O.}\ \bibnamefont {Alves}}, \bibinfo {author} {\bibfnamefont {F.}~\bibnamefont {Fritzsch}},\ and\ \bibinfo {author} {\bibfnamefont {P.~W.}\ \bibnamefont {Claeys}},\ }\bibfield  {title} {\bibinfo {title} {Probes of {Full} {Eigenstate} {Thermalization} in {Ergodicity}-{Breaking} {Quantum} {Circuits}},\ }\href {http://arxiv.org/abs/2504.08517} {\bibfield  {journal} {\bibinfo  {journal} {arXiv:2504.08517}\ } (\bibinfo {year} {2025})}\BibitemShut {NoStop}%
\bibitem [{\citenamefont {Fritzsch}\ \emph {et~al.}(2025)\citenamefont {Fritzsch}, \citenamefont {Prosen},\ and\ \citenamefont {Pappalardi}}]{fritzschMicrocanonicalFreeCumulants2024}%
  \BibitemOpen
  \bibfield  {author} {\bibinfo {author} {\bibfnamefont {F.}~\bibnamefont {Fritzsch}}, \bibinfo {author} {\bibfnamefont {T.}~\bibnamefont {Prosen}},\ and\ \bibinfo {author} {\bibfnamefont {S.}~\bibnamefont {Pappalardi}},\ }\bibfield  {title} {\bibinfo {title} {Microcanonical free cumulants in lattice systems},\ }\href {https://doi.org/10.1103/PhysRevB.111.054303} {\bibfield  {journal} {\bibinfo  {journal} {Phys. Rev. B}\ }\textbf {\bibinfo {volume} {111}},\ \bibinfo {pages} {054303} (\bibinfo {year} {2025})}\BibitemShut {NoStop}%
\bibitem [{\citenamefont {Chan}\ \emph {et~al.}(2019)\citenamefont {Chan}, \citenamefont {De~Luca},\ and\ \citenamefont {Chalker}}]{chan2019eigenstate}%
  \BibitemOpen
  \bibfield  {author} {\bibinfo {author} {\bibfnamefont {A.}~\bibnamefont {Chan}}, \bibinfo {author} {\bibfnamefont {A.}~\bibnamefont {De~Luca}},\ and\ \bibinfo {author} {\bibfnamefont {J.}~\bibnamefont {Chalker}},\ }\bibfield  {title} {\bibinfo {title} {Eigenstate correlations, thermalization, and the butterfly effect},\ }\href {https://doi.org/https://doi.org/10.1103/PhysRevLett.122.220601} {\bibfield  {journal} {\bibinfo  {journal} {Phys. Rev. Lett.}\ }\textbf {\bibinfo {volume} {122}},\ \bibinfo {pages} {220601} (\bibinfo {year} {2019})}\BibitemShut {NoStop}%
\bibitem [{\citenamefont {Hahn}\ \emph {et~al.}(2024)\citenamefont {Hahn}, \citenamefont {Luitz},\ and\ \citenamefont {Chalker}}]{hahn2024eigenstate}%
  \BibitemOpen
  \bibfield  {author} {\bibinfo {author} {\bibfnamefont {D.}~\bibnamefont {Hahn}}, \bibinfo {author} {\bibfnamefont {D.~J.}\ \bibnamefont {Luitz}},\ and\ \bibinfo {author} {\bibfnamefont {J.}~\bibnamefont {Chalker}},\ }\bibfield  {title} {\bibinfo {title} {Eigenstate correlations, the eigenstate thermalization hypothesis, and quantum information dynamics in chaotic many-body quantum systems},\ }\href {https://doi.org/https://doi.org/10.1103/PhysRevX.14.031029} {\bibfield  {journal} {\bibinfo  {journal} {Phys. Rev. X}\ }\textbf {\bibinfo {volume} {14}},\ \bibinfo {pages} {031029} (\bibinfo {year} {2024})}\BibitemShut {NoStop}%
\bibitem [{\citenamefont {Jafferis}\ \emph {et~al.}(2023{\natexlab{a}})\citenamefont {Jafferis}, \citenamefont {Kolchmeyer}, \citenamefont {Mukhametzhanov},\ and\ \citenamefont {Sonner}}]{jafferis2023matrix}%
  \BibitemOpen
  \bibfield  {author} {\bibinfo {author} {\bibfnamefont {D.~L.}\ \bibnamefont {Jafferis}}, \bibinfo {author} {\bibfnamefont {D.~K.}\ \bibnamefont {Kolchmeyer}}, \bibinfo {author} {\bibfnamefont {B.}~\bibnamefont {Mukhametzhanov}},\ and\ \bibinfo {author} {\bibfnamefont {J.}~\bibnamefont {Sonner}},\ }\bibfield  {title} {\bibinfo {title} {Matrix models for eigenstate thermalization},\ }\href {https://doi.org/https://doi.org/10.1103/PhysRevX.13.031033} {\bibfield  {journal} {\bibinfo  {journal} {Phys.~Rev.~X}\ }\textbf {\bibinfo {volume} {13}},\ \bibinfo {pages} {031033} (\bibinfo {year} {2023}{\natexlab{a}})}\BibitemShut {NoStop}%
\bibitem [{\citenamefont {Jafferis}\ \emph {et~al.}(2023{\natexlab{b}})\citenamefont {Jafferis}, \citenamefont {Kolchmeyer}, \citenamefont {Mukhametzhanov},\ and\ \citenamefont {Sonner}}]{jafferis2023jackiw}%
  \BibitemOpen
  \bibfield  {author} {\bibinfo {author} {\bibfnamefont {D.~L.}\ \bibnamefont {Jafferis}}, \bibinfo {author} {\bibfnamefont {D.~K.}\ \bibnamefont {Kolchmeyer}}, \bibinfo {author} {\bibfnamefont {B.}~\bibnamefont {Mukhametzhanov}},\ and\ \bibinfo {author} {\bibfnamefont {J.}~\bibnamefont {Sonner}},\ }\bibfield  {title} {\bibinfo {title} {Jackiw-teitelboim gravity with matter, generalized eigenstate thermalization hypothesis, and random matrices},\ }\href {https://doi.org/https://doi.org/10.1103/PhysRevD.108.066015} {\bibfield  {journal} {\bibinfo  {journal} {Phys.~Rev.~D}\ }\textbf {\bibinfo {volume} {108}},\ \bibinfo {pages} {066015} (\bibinfo {year} {2023}{\natexlab{b}})}\BibitemShut {NoStop}%
\bibitem [{\citenamefont {Voiculescu}(1991)}]{voiculescuFreeNoncommutativeRandom1991}%
  \BibitemOpen
  \bibfield  {author} {\bibinfo {author} {\bibfnamefont {D.}~\bibnamefont {Voiculescu}},\ }\bibfield  {title} {\bibinfo {title} {Free {{Noncommutative Random Variables}}, {{Random Matrices}} and the {II}$_1$ {Factors} of {{Free Groups}}},\ }in\ \href {https://doi.org/10.1142/9789814360203_0031} {\emph {\bibinfo {booktitle} {Quantum {{Probability}} and {{Related Topics}}}}},\ Vol.\ \bibinfo {volume} {Volume 6}\ (\bibinfo  {publisher} {World Scientific},\ \bibinfo {year} {1991})\ pp.\ \bibinfo {pages} {473--487}\BibitemShut {NoStop}%
\bibitem [{\citenamefont {Speicher}(2003)}]{speicherFreeProbabilityTheory2003d}%
  \BibitemOpen
  \bibfield  {author} {\bibinfo {author} {\bibfnamefont {R.}~\bibnamefont {Speicher}},\ }\bibfield  {title} {\bibinfo {title} {Free {{Probability Theory}} and {{Random Matrices}}},\ }in\ \href {https://doi.org/10.1007/3-540-44890-X_3} {\emph {\bibinfo {booktitle} {Asymptotic {{Combinatorics}} with {{Applications}} to {{Mathematical Physics}}}}},\ Vol.\ \bibinfo {volume} {1815}\ (\bibinfo  {publisher} {Springer Berlin Heidelberg},\ \bibinfo {address} {Berlin, Heidelberg},\ \bibinfo {year} {2003})\ pp.\ \bibinfo {pages} {53--73}\BibitemShut {NoStop}%
\bibitem [{\citenamefont {Mingo}\ and\ \citenamefont {Speicher}(2017)}]{mingoFreeProbabilityRandom2017}%
  \BibitemOpen
  \bibfield  {author} {\bibinfo {author} {\bibfnamefont {J.~A.}\ \bibnamefont {Mingo}}\ and\ \bibinfo {author} {\bibfnamefont {R.}~\bibnamefont {Speicher}},\ }\href {https://doi.org/10.1007/978-1-4939-6942-5} {\emph {\bibinfo {title} {Free {{Probability}} and {{Random Matrices}}}}},\ \bibinfo {series} {Fields {{Institute Monographs}}}, Vol.~\bibinfo {volume} {35}\ (\bibinfo  {publisher} {Springer New York},\ \bibinfo {address} {New York, NY},\ \bibinfo {year} {2017})\BibitemShut {NoStop}%
\bibitem [{\citenamefont {Nica}\ and\ \citenamefont {Speicher}(2006)}]{nica2006lectures}%
  \BibitemOpen
  \bibfield  {author} {\bibinfo {author} {\bibfnamefont {A.}~\bibnamefont {Nica}}\ and\ \bibinfo {author} {\bibfnamefont {R.}~\bibnamefont {Speicher}},\ }\href {https://rolandspeicher.com/wp-content/uploads/2020/06/nica-speicher-book.pdf} {\emph {\bibinfo {title} {Lectures on the combinatorics of free probability}}},\ Vol.~\bibinfo {volume} {13}\ (\bibinfo  {publisher} {Cambridge University Press},\ \bibinfo {year} {2006})\BibitemShut {NoStop}%
\bibitem [{\citenamefont {Novak}\ and\ \citenamefont {LaCroix}(2012)}]{novakThreeLecturesFree2012}%
  \BibitemOpen
  \bibfield  {author} {\bibinfo {author} {\bibfnamefont {J.}~\bibnamefont {Novak}}\ and\ \bibinfo {author} {\bibfnamefont {M.}~\bibnamefont {LaCroix}},\ }\bibfield  {title} {\bibinfo {title} {Three lectures on free probability},\ }\href {https://doi.org/10.48550/arXiv.1205.2097} {\bibfield  {journal} {\bibinfo  {journal} {arXiv:1205.2097}\ } (\bibinfo {year} {2012})}\BibitemShut {NoStop}%
\bibitem [{\citenamefont {Speicher}(2017)}]{speicherFreeProbabilityTheory2017b}%
  \BibitemOpen
  \bibfield  {author} {\bibinfo {author} {\bibfnamefont {R.}~\bibnamefont {Speicher}},\ }\bibfield  {title} {\bibinfo {title} {Free {{Probability Theory}}},\ }\href {https://doi.org/10.1365/s13291-016-0150-5} {\bibfield  {journal} {\bibinfo  {journal} {Jahresber. Deutsch. Math.-Verein.}\ }\textbf {\bibinfo {volume} {119}},\ \bibinfo {pages} {3} (\bibinfo {year} {2017})}\BibitemShut {NoStop}%
\bibitem [{\citenamefont {Hruza}\ and\ \citenamefont {Bernard}(2023)}]{hruza2023coherent}%
  \BibitemOpen
  \bibfield  {author} {\bibinfo {author} {\bibfnamefont {L.}~\bibnamefont {Hruza}}\ and\ \bibinfo {author} {\bibfnamefont {D.}~\bibnamefont {Bernard}},\ }\bibfield  {title} {\bibinfo {title} {Coherent fluctuations in noisy mesoscopic systems, the open quantum ssep, and free probability},\ }\href {https://doi.org/https://doi.org/10.1103/PhysRevX.13.011045} {\bibfield  {journal} {\bibinfo  {journal} {Physical Review X}\ }\textbf {\bibinfo {volume} {13}},\ \bibinfo {pages} {011045} (\bibinfo {year} {2023})}\BibitemShut {NoStop}%
\bibitem [{\citenamefont {Chen}\ and\ \citenamefont {Kudler-Flam}(2025)}]{chen_free_2025}%
  \BibitemOpen
  \bibfield  {author} {\bibinfo {author} {\bibfnamefont {H.~J.}\ \bibnamefont {Chen}}\ and\ \bibinfo {author} {\bibfnamefont {J.}~\bibnamefont {Kudler-Flam}},\ }\bibfield  {title} {\bibinfo {title} {Free independence and the noncrossing partition lattice in dual-unitary quantum circuits},\ }\href {https://doi.org/10.1103/PhysRevB.111.014311} {\bibfield  {journal} {\bibinfo  {journal} {Phys. Rev. B}\ }\textbf {\bibinfo {volume} {111}},\ \bibinfo {pages} {014311} (\bibinfo {year} {2025})}\BibitemShut {NoStop}%
\bibitem [{\citenamefont {Camargo}\ \emph {et~al.}(2025)\citenamefont {Camargo}, \citenamefont {Fu}, \citenamefont {Jahnke}, \citenamefont {Pal},\ and\ \citenamefont {Kim}}]{camargo2025quantum}%
  \BibitemOpen
  \bibfield  {author} {\bibinfo {author} {\bibfnamefont {H.~A.}\ \bibnamefont {Camargo}}, \bibinfo {author} {\bibfnamefont {Y.}~\bibnamefont {Fu}}, \bibinfo {author} {\bibfnamefont {V.}~\bibnamefont {Jahnke}}, \bibinfo {author} {\bibfnamefont {K.}~\bibnamefont {Pal}},\ and\ \bibinfo {author} {\bibfnamefont {K.-Y.}\ \bibnamefont {Kim}},\ }\bibfield  {title} {\bibinfo {title} {Quantum signatures of chaos from free probability},\ }\href {https://arxiv.org/abs/2503.20338} {\bibfield  {journal} {\bibinfo  {journal} {arXiv:2503.20338}\ } (\bibinfo {year} {2025})}\BibitemShut {NoStop}%
\bibitem [{\citenamefont {Jahnke}\ \emph {et~al.}(2025)\citenamefont {Jahnke}, \citenamefont {Nandy}, \citenamefont {Pal}, \citenamefont {Camargo},\ and\ \citenamefont {Kim}}]{jahnke2025free}%
  \BibitemOpen
  \bibfield  {author} {\bibinfo {author} {\bibfnamefont {V.}~\bibnamefont {Jahnke}}, \bibinfo {author} {\bibfnamefont {P.}~\bibnamefont {Nandy}}, \bibinfo {author} {\bibfnamefont {K.}~\bibnamefont {Pal}}, \bibinfo {author} {\bibfnamefont {H.~A.}\ \bibnamefont {Camargo}},\ and\ \bibinfo {author} {\bibfnamefont {K.-Y.}\ \bibnamefont {Kim}},\ }\bibfield  {title} {\bibinfo {title} {Free probability approach to spectral and operator statistics in rosenzweig-porter random matrix ensembles},\ }\bibfield  {journal} {\bibinfo  {journal} {arXiv:2506.04520}\ }\href {https://doi.org/https://doi.org/10.48550/arXiv.2506.04520} {https://doi.org/10.48550/arXiv.2506.04520} (\bibinfo {year} {2025})\BibitemShut {NoStop}%
\bibitem [{\citenamefont {Roberts}\ and\ \citenamefont {Stanford}(2015)}]{roberts2015diagnosing}%
  \BibitemOpen
  \bibfield  {author} {\bibinfo {author} {\bibfnamefont {D.~A.}\ \bibnamefont {Roberts}}\ and\ \bibinfo {author} {\bibfnamefont {D.}~\bibnamefont {Stanford}},\ }\bibfield  {title} {\bibinfo {title} {Diagnosing chaos using four-point functions in two-dimensional conformal field theory},\ }\href {https://doi.org/https://doi.org/10.1103/PhysRevLett.115.131603} {\bibfield  {journal} {\bibinfo  {journal} {Phys. Rev. Lett}\ }\textbf {\bibinfo {volume} {115}},\ \bibinfo {pages} {131603} (\bibinfo {year} {2015})}\BibitemShut {NoStop}%
\bibitem [{\citenamefont {Roberts}\ \emph {et~al.}(2015)\citenamefont {Roberts}, \citenamefont {Stanford},\ and\ \citenamefont {Susskind}}]{roberts2015localized}%
  \BibitemOpen
  \bibfield  {author} {\bibinfo {author} {\bibfnamefont {D.~A.}\ \bibnamefont {Roberts}}, \bibinfo {author} {\bibfnamefont {D.}~\bibnamefont {Stanford}},\ and\ \bibinfo {author} {\bibfnamefont {L.}~\bibnamefont {Susskind}},\ }\bibfield  {title} {\bibinfo {title} {Localized shocks},\ }\href {https://doi.org/https://doi.org/10.1007/JHEP12(2014)046} {\bibfield  {journal} {\bibinfo  {journal} {JHEP}\ }\textbf {\bibinfo {volume} {2015}}\bibinfo  {number} { (3)},\ \bibinfo {pages} {1}}\BibitemShut {NoStop}%
\bibitem [{\citenamefont {Das}\ \emph {et~al.}(2024)\citenamefont {Das}, \citenamefont {Dutta},\ and\ \citenamefont {Maji}}]{das2024late}%
  \BibitemOpen
\bibfield  {number} {  }\bibfield  {author} {\bibinfo {author} {\bibfnamefont {R.~N.}\ \bibnamefont {Das}}, \bibinfo {author} {\bibfnamefont {S.}~\bibnamefont {Dutta}},\ and\ \bibinfo {author} {\bibfnamefont {A.}~\bibnamefont {Maji}},\ }\bibfield  {title} {\bibinfo {title} {Late time dynamics in susy saddle-dominated scrambling through higher-point otoc},\ }\href {https://doi.org/https://doi.org/10.1088/1402-4896/ad629d} {\bibfield  {journal} {\bibinfo  {journal} {Phys.~Scr.}\ }\textbf {\bibinfo {volume} {99}},\ \bibinfo {pages} {085246} (\bibinfo {year} {2024})}\BibitemShut {NoStop}%
\bibitem [{\citenamefont {Maldacena}\ and\ \citenamefont {Stanford}(2016)}]{maldacena2016remarks}%
  \BibitemOpen
  \bibfield  {author} {\bibinfo {author} {\bibfnamefont {J.}~\bibnamefont {Maldacena}}\ and\ \bibinfo {author} {\bibfnamefont {D.}~\bibnamefont {Stanford}},\ }\bibfield  {title} {\bibinfo {title} {Remarks on the sachdev-ye-kitaev model},\ }\href {https://doi.org/https://doi.org/10.1103/PhysRevD.94.106002} {\bibfield  {journal} {\bibinfo  {journal} {Phys. Rev. D}\ }\textbf {\bibinfo {volume} {94}},\ \bibinfo {pages} {106002} (\bibinfo {year} {2016})}\BibitemShut {NoStop}%
\bibitem [{\citenamefont {Nahum}\ \emph {et~al.}(2018)\citenamefont {Nahum}, \citenamefont {Vijay},\ and\ \citenamefont {Haah}}]{nahum_operator_2018}%
  \BibitemOpen
  \bibfield  {author} {\bibinfo {author} {\bibfnamefont {A.}~\bibnamefont {Nahum}}, \bibinfo {author} {\bibfnamefont {S.}~\bibnamefont {Vijay}},\ and\ \bibinfo {author} {\bibfnamefont {J.}~\bibnamefont {Haah}},\ }\bibfield  {title} {\bibinfo {title} {Operator {Spreading} in {Random} {Unitary} {Circuits}},\ }\href {https://doi.org/10.1103/PhysRevX.8.021014} {\bibfield  {journal} {\bibinfo  {journal} {Phys. Rev. X}\ }\textbf {\bibinfo {volume} {8}},\ \bibinfo {pages} {021014} (\bibinfo {year} {2018})}\BibitemShut {NoStop}%
\bibitem [{\citenamefont {Chan}\ \emph {et~al.}(2018)\citenamefont {Chan}, \citenamefont {De~Luca},\ and\ \citenamefont {Chalker}}]{chan2018solution}%
  \BibitemOpen
  \bibfield  {author} {\bibinfo {author} {\bibfnamefont {A.}~\bibnamefont {Chan}}, \bibinfo {author} {\bibfnamefont {A.}~\bibnamefont {De~Luca}},\ and\ \bibinfo {author} {\bibfnamefont {J.~T.}\ \bibnamefont {Chalker}},\ }\bibfield  {title} {\bibinfo {title} {Solution of a minimal model for many-body quantum chaos},\ }\href {https://doi.org/10.1103/PhysRevX.8.041019} {\bibfield  {journal} {\bibinfo  {journal} {Phys. Rev. X}\ }\textbf {\bibinfo {volume} {8}},\ \bibinfo {pages} {041019} (\bibinfo {year} {2018})}\BibitemShut {NoStop}%
\bibitem [{\citenamefont {Yoshimura}\ \emph {et~al.}(2025)\citenamefont {Yoshimura}, \citenamefont {Garratt},\ and\ \citenamefont {Chalker}}]{yoshimura_operator_2025}%
  \BibitemOpen
  \bibfield  {author} {\bibinfo {author} {\bibfnamefont {T.}~\bibnamefont {Yoshimura}}, \bibinfo {author} {\bibfnamefont {S.~J.}\ \bibnamefont {Garratt}},\ and\ \bibinfo {author} {\bibfnamefont {J.~T.}\ \bibnamefont {Chalker}},\ }\bibfield  {title} {\bibinfo {title} {Operator dynamics in {Floquet} many-body systems},\ }\href {https://doi.org/10.1103/PhysRevB.111.094316} {\bibfield  {journal} {\bibinfo  {journal} {Phys. Rev. B}\ }\textbf {\bibinfo {volume} {111}},\ \bibinfo {pages} {094316} (\bibinfo {year} {2025})}\BibitemShut {NoStop}%
\bibitem [{\citenamefont {Claeys}\ and\ \citenamefont {Lamacraft}(2020)}]{claeys_maximum_2020}%
  \BibitemOpen
  \bibfield  {author} {\bibinfo {author} {\bibfnamefont {P.~W.}\ \bibnamefont {Claeys}}\ and\ \bibinfo {author} {\bibfnamefont {A.}~\bibnamefont {Lamacraft}},\ }\bibfield  {title} {\bibinfo {title} {Maximum velocity quantum circuits},\ }\href {https://doi.org/10.1103/PhysRevResearch.2.033032} {\bibfield  {journal} {\bibinfo  {journal} {Phys. Rev. Res.}\ }\textbf {\bibinfo {volume} {2}},\ \bibinfo {pages} {033032} (\bibinfo {year} {2020})}\BibitemShut {NoStop}%
\bibitem [{\citenamefont {Bertini}\ and\ \citenamefont {Piroli}(2020)}]{bertini2020scrambling}%
  \BibitemOpen
  \bibfield  {author} {\bibinfo {author} {\bibfnamefont {B.}~\bibnamefont {Bertini}}\ and\ \bibinfo {author} {\bibfnamefont {L.}~\bibnamefont {Piroli}},\ }\bibfield  {title} {\bibinfo {title} {Scrambling in random unitary circuits: Exact results},\ }\href {https://doi.org/10.1103/PhysRevB.102.064305} {\bibfield  {journal} {\bibinfo  {journal} {Phys. Rev. B}\ }\textbf {\bibinfo {volume} {102}},\ \bibinfo {pages} {064305} (\bibinfo {year} {2020})}\BibitemShut {NoStop}%
\bibitem [{Note1()}]{Note1}%
  \BibitemOpen
  \bibinfo {note} {Note that the order of limits matters, since for finite systems there is no expected convergence in time due to revivals.}\BibitemShut {Stop}%
\bibitem [{\citenamefont {Bertini}\ \emph {et~al.}(2025)\citenamefont {Bertini}, \citenamefont {Claeys},\ and\ \citenamefont {Prosen}}]{bertini_exactly_2025}%
  \BibitemOpen
  \bibfield  {author} {\bibinfo {author} {\bibfnamefont {B.}~\bibnamefont {Bertini}}, \bibinfo {author} {\bibfnamefont {P.~W.}\ \bibnamefont {Claeys}},\ and\ \bibinfo {author} {\bibfnamefont {T.}~\bibnamefont {Prosen}},\ }\bibfield  {title} {\bibinfo {title} {Exactly solvable many-body dynamics from space-time duality},\ }\href {http://arxiv.org/abs/2505.11489} {\bibfield  {journal} {\bibinfo  {journal} {arXiv:2505.11489}\ } (\bibinfo {year} {2025})}\BibitemShut {NoStop}%
\bibitem [{Note2()}]{Note2}%
  \BibitemOpen
  \bibinfo {note} {Due to unitarity of $V_t$ or, equivalently, the strict causality in the brickwork circuit, the above $k$-OTOC does not depend on $V_t$.}\BibitemShut {Stop}%
\bibitem [{\citenamefont {Nielsen}\ and\ \citenamefont {Chuang}(2010)}]{nielsen2010quantum}%
  \BibitemOpen
  \bibfield  {author} {\bibinfo {author} {\bibfnamefont {M.}~\bibnamefont {Nielsen}}\ and\ \bibinfo {author} {\bibfnamefont {I.}~\bibnamefont {Chuang}},\ }\href {https://books.google.co.uk/books?id=-s4DEy7o-a0C} {\emph {\bibinfo {title} {Quantum Computation and Quantum Information}}}\ (\bibinfo  {publisher} {Cambridge University Press},\ \bibinfo {year} {2010})\BibitemShut {NoStop}%
\bibitem [{\citenamefont {Bertini}\ \emph {et~al.}(2019)\citenamefont {Bertini}, \citenamefont {Kos},\ and\ \citenamefont {Prosen}}]{bertini_exact_2019}%
  \BibitemOpen
  \bibfield  {author} {\bibinfo {author} {\bibfnamefont {B.}~\bibnamefont {Bertini}}, \bibinfo {author} {\bibfnamefont {P.}~\bibnamefont {Kos}},\ and\ \bibinfo {author} {\bibfnamefont {T.}~\bibnamefont {Prosen}},\ }\bibfield  {title} {\bibinfo {title} {Exact {Correlation} {Functions} for {Dual}-{Unitary} {Lattice} {Models} in 1 + 1 {Dimensions}},\ }\href {https://doi.org/10.1103/PhysRevLett.123.210601} {\bibfield  {journal} {\bibinfo  {journal} {Phys. Rev. Lett.}\ }\textbf {\bibinfo {volume} {123}},\ \bibinfo {pages} {210601} (\bibinfo {year} {2019})}\BibitemShut {NoStop}%
\bibitem [{\citenamefont {Rather}\ \emph {et~al.}(2020)\citenamefont {Rather}, \citenamefont {Aravinda},\ and\ \citenamefont {Lakshminarayan}}]{rather_creating_2020}%
  \BibitemOpen
  \bibfield  {author} {\bibinfo {author} {\bibfnamefont {S.~A.}\ \bibnamefont {Rather}}, \bibinfo {author} {\bibfnamefont {S.}~\bibnamefont {Aravinda}},\ and\ \bibinfo {author} {\bibfnamefont {A.}~\bibnamefont {Lakshminarayan}},\ }\bibfield  {title} {\bibinfo {title} {Creating {Ensembles} of {Dual} {Unitary} and {Maximally} {Entangling} {Quantum} {Evolutions}},\ }\href {https://doi.org/10.1103/PhysRevLett.125.070501} {\bibfield  {journal} {\bibinfo  {journal} {Phys. Rev. Lett.}\ }\textbf {\bibinfo {volume} {125}},\ \bibinfo {pages} {070501} (\bibinfo {year} {2020})}\BibitemShut {NoStop}%
\bibitem [{\citenamefont {Aravinda}\ \emph {et~al.}(2021)\citenamefont {Aravinda}, \citenamefont {Rather},\ and\ \citenamefont {Lakshminarayan}}]{aravinda_dual-unitary_2021}%
  \BibitemOpen
  \bibfield  {author} {\bibinfo {author} {\bibfnamefont {S.}~\bibnamefont {Aravinda}}, \bibinfo {author} {\bibfnamefont {S.~A.}\ \bibnamefont {Rather}},\ and\ \bibinfo {author} {\bibfnamefont {A.}~\bibnamefont {Lakshminarayan}},\ }\bibfield  {title} {\bibinfo {title} {From dual-unitary to quantum {Bernoulli} circuits: {Role} of the entangling power in constructing a quantum ergodic hierarchy},\ }\href {https://doi.org/10.1103/PhysRevResearch.3.043034} {\bibfield  {journal} {\bibinfo  {journal} {Phys. Rev. Research}\ }\textbf {\bibinfo {volume} {3}},\ \bibinfo {pages} {043034} (\bibinfo {year} {2021})}\BibitemShut {NoStop}%
\bibitem [{\citenamefont {Rather}\ \emph {et~al.}(2022)\citenamefont {Rather}, \citenamefont {Aravinda},\ and\ \citenamefont {Lakshminarayan}}]{rather_construction_2022}%
  \BibitemOpen
  \bibfield  {author} {\bibinfo {author} {\bibfnamefont {S.~A.}\ \bibnamefont {Rather}}, \bibinfo {author} {\bibfnamefont {S.}~\bibnamefont {Aravinda}},\ and\ \bibinfo {author} {\bibfnamefont {A.}~\bibnamefont {Lakshminarayan}},\ }\bibfield  {title} {\bibinfo {title} {Construction and {Local} {Equivalence} of {Dual}-{Unitary} {Operators}: {From} {Dynamical} {Maps} to {Quantum} {Combinatorial} {Designs}},\ }\href {https://doi.org/10.1103/PRXQuantum.3.040331} {\bibfield  {journal} {\bibinfo  {journal} {PRX Quantum}\ }\textbf {\bibinfo {volume} {3}},\ \bibinfo {pages} {040331} (\bibinfo {year} {2022})}\BibitemShut {NoStop}%
\bibitem [{\citenamefont {Aravinda}\ \emph {et~al.}(2024)\citenamefont {Aravinda}, \citenamefont {Banerjee},\ and\ \citenamefont {Modak}}]{aravinda_ergodic_2024}%
  \BibitemOpen
  \bibfield  {author} {\bibinfo {author} {\bibfnamefont {S.}~\bibnamefont {Aravinda}}, \bibinfo {author} {\bibfnamefont {S.}~\bibnamefont {Banerjee}},\ and\ \bibinfo {author} {\bibfnamefont {R.}~\bibnamefont {Modak}},\ }\bibfield  {title} {\bibinfo {title} {Ergodic and mixing quantum channels: {From} two-qubit to many-body quantum systems},\ }\href {https://doi.org/10.1103/PhysRevA.110.042607} {\bibfield  {journal} {\bibinfo  {journal} {Phys. Rev. A}\ }\textbf {\bibinfo {volume} {110}},\ \bibinfo {pages} {042607} (\bibinfo {year} {2024})}\BibitemShut {NoStop}%
\bibitem [{\citenamefont {Gopalakrishnan}\ and\ \citenamefont {Lamacraft}(2019)}]{gopalakrishnan_unitary_2019}%
  \BibitemOpen
  \bibfield  {author} {\bibinfo {author} {\bibfnamefont {S.}~\bibnamefont {Gopalakrishnan}}\ and\ \bibinfo {author} {\bibfnamefont {A.}~\bibnamefont {Lamacraft}},\ }\bibfield  {title} {\bibinfo {title} {Unitary circuits of finite depth and infinite width from quantum channels},\ }\href {https://doi.org/10.1103/PhysRevB.100.064309} {\bibfield  {journal} {\bibinfo  {journal} {Phys. Rev. B}\ }\textbf {\bibinfo {volume} {100}},\ \bibinfo {pages} {064309} (\bibinfo {year} {2019})}\BibitemShut {NoStop}%
\bibitem [{\citenamefont {Rampp}\ \emph {et~al.}(2023)\citenamefont {Rampp}, \citenamefont {Moessner},\ and\ \citenamefont {Claeys}}]{rampp_dual_2023}%
  \BibitemOpen
  \bibfield  {author} {\bibinfo {author} {\bibfnamefont {M.~A.}\ \bibnamefont {Rampp}}, \bibinfo {author} {\bibfnamefont {R.}~\bibnamefont {Moessner}},\ and\ \bibinfo {author} {\bibfnamefont {P.~W.}\ \bibnamefont {Claeys}},\ }\bibfield  {title} {\bibinfo {title} {From {Dual} {Unitarity} to {Generic} {Quantum} {Operator} {Spreading}},\ }\href {https://doi.org/10.1103/PhysRevLett.130.130402} {\bibfield  {journal} {\bibinfo  {journal} {Phys. Rev. Lett.}\ }\textbf {\bibinfo {volume} {130}},\ \bibinfo {pages} {130402} (\bibinfo {year} {2023})}\BibitemShut {NoStop}%
\bibitem [{\citenamefont {Zhou}\ and\ \citenamefont {Harrow}(2022)}]{zhou_maximal_2022}%
  \BibitemOpen
  \bibfield  {author} {\bibinfo {author} {\bibfnamefont {T.}~\bibnamefont {Zhou}}\ and\ \bibinfo {author} {\bibfnamefont {A.~W.}\ \bibnamefont {Harrow}},\ }\bibfield  {title} {\bibinfo {title} {Maximal entanglement velocity implies dual unitarity},\ }\href {https://doi.org/10.1103/PhysRevB.106.L201104} {\bibfield  {journal} {\bibinfo  {journal} {Phys. Rev. B}\ }\textbf {\bibinfo {volume} {106}},\ \bibinfo {pages} {L201104} (\bibinfo {year} {2022})}\BibitemShut {NoStop}%
\bibitem [{\citenamefont {Kos}\ \emph {et~al.}(2021)\citenamefont {Kos}, \citenamefont {Bertini},\ and\ \citenamefont {Prosen}}]{kos_correlations_2021}%
  \BibitemOpen
  \bibfield  {author} {\bibinfo {author} {\bibfnamefont {P.}~\bibnamefont {Kos}}, \bibinfo {author} {\bibfnamefont {B.}~\bibnamefont {Bertini}},\ and\ \bibinfo {author} {\bibfnamefont {T.}~\bibnamefont {Prosen}},\ }\bibfield  {title} {\bibinfo {title} {Correlations in {Perturbed} {Dual}-{Unitary} {Circuits}: {Efficient} {Path}-{Integral} {Formula}},\ }\href {https://doi.org/10.1103/PhysRevX.11.011022} {\bibfield  {journal} {\bibinfo  {journal} {Phys. Rev. X}\ }\textbf {\bibinfo {volume} {11}},\ \bibinfo {pages} {011022} (\bibinfo {year} {2021})}\BibitemShut {NoStop}%
\bibitem [{\citenamefont {Riddell}\ \emph {et~al.}(2024)\citenamefont {Riddell}, \citenamefont {von Keyserlingk}, \citenamefont {Prosen},\ and\ \citenamefont {Bertini}}]{riddell2024structural}%
  \BibitemOpen
  \bibfield  {author} {\bibinfo {author} {\bibfnamefont {J.}~\bibnamefont {Riddell}}, \bibinfo {author} {\bibfnamefont {C.}~\bibnamefont {von Keyserlingk}}, \bibinfo {author} {\bibfnamefont {T.}~\bibnamefont {Prosen}},\ and\ \bibinfo {author} {\bibfnamefont {B.}~\bibnamefont {Bertini}},\ }\bibfield  {title} {\bibinfo {title} {Structural stability hypothesis of dual unitary quantum chaos},\ }\href {https://doi.org/10.1103/PhysRevResearch.6.033226} {\bibfield  {journal} {\bibinfo  {journal} {Phys. Rev. Res.}\ }\textbf {\bibinfo {volume} {6}},\ \bibinfo {pages} {033226} (\bibinfo {year} {2024})}\BibitemShut {NoStop}%
\bibitem [{\citenamefont {von Keyserlingk}\ \emph {et~al.}(2018)\citenamefont {von Keyserlingk}, \citenamefont {Rakovszky}, \citenamefont {Pollmann},\ and\ \citenamefont {Sondhi}}]{von_keyserlingk_operator_2018}%
  \BibitemOpen
  \bibfield  {author} {\bibinfo {author} {\bibfnamefont {C.~W.}\ \bibnamefont {von Keyserlingk}}, \bibinfo {author} {\bibfnamefont {T.}~\bibnamefont {Rakovszky}}, \bibinfo {author} {\bibfnamefont {F.}~\bibnamefont {Pollmann}},\ and\ \bibinfo {author} {\bibfnamefont {S.~L.}\ \bibnamefont {Sondhi}},\ }\bibfield  {title} {\bibinfo {title} {Operator {Hydrodynamics}, {OTOCs}, and {Entanglement} {Growth} in {Systems} without {Conservation} {Laws}},\ }\href {https://doi.org/10.1103/PhysRevX.8.021013} {\bibfield  {journal} {\bibinfo  {journal} {Phys. Rev. X}\ }\textbf {\bibinfo {volume} {8}},\ \bibinfo {pages} {021013} (\bibinfo {year} {2018})}\BibitemShut {NoStop}%
\bibitem [{\citenamefont {Fisher}\ \emph {et~al.}(2023)\citenamefont {Fisher}, \citenamefont {Khemani}, \citenamefont {Nahum},\ and\ \citenamefont {Vijay}}]{fisher_random_2023}%
  \BibitemOpen
  \bibfield  {author} {\bibinfo {author} {\bibfnamefont {M.~P.}\ \bibnamefont {Fisher}}, \bibinfo {author} {\bibfnamefont {V.}~\bibnamefont {Khemani}}, \bibinfo {author} {\bibfnamefont {A.}~\bibnamefont {Nahum}},\ and\ \bibinfo {author} {\bibfnamefont {S.}~\bibnamefont {Vijay}},\ }\bibfield  {title} {\bibinfo {title} {Random {Quantum} {Circuits}},\ }\href {https://doi.org/10.1146/annurev-conmatphys-031720-030658} {\bibfield  {journal} {\bibinfo  {journal} {Annu. Rev. Conden. Ma. P.}\ }\textbf {\bibinfo {volume} {14}},\ \bibinfo {pages} {335–79} (\bibinfo {year} {2023})}\BibitemShut {NoStop}%
\bibitem [{\citenamefont {Bensa}\ and\ \citenamefont {Žnidarič}(2022)}]{bensa_two-step_2022}%
  \BibitemOpen
  \bibfield  {author} {\bibinfo {author} {\bibfnamefont {J.}~\bibnamefont {Bensa}}\ and\ \bibinfo {author} {\bibfnamefont {M.}~\bibnamefont {Žnidarič}},\ }\bibfield  {title} {\bibinfo {title} {Two-step phantom relaxation of out-of-time-ordered correlations in random circuits},\ }\href {https://doi.org/10.1103/PhysRevResearch.4.013228} {\bibfield  {journal} {\bibinfo  {journal} {Phys. Rev. Research}\ }\textbf {\bibinfo {volume} {4}},\ \bibinfo {pages} {013228} (\bibinfo {year} {2022})}\BibitemShut {NoStop}%
\bibitem [{\citenamefont {Žnidarič}(2023)}]{znidaric_two-step_2023}%
  \BibitemOpen
  \bibfield  {author} {\bibinfo {author} {\bibfnamefont {M.}~\bibnamefont {Žnidarič}},\ }\bibfield  {title} {\bibinfo {title} {Two-step relaxation in local many-body {Floquet} systems},\ }\href {https://doi.org/10.1088/1751-8121/acfc05} {\bibfield  {journal} {\bibinfo  {journal} {J. Phys. A: Math. Theor.}\ }\textbf {\bibinfo {volume} {56}},\ \bibinfo {pages} {434001} (\bibinfo {year} {2023})}\BibitemShut {NoStop}%
\bibitem [{\citenamefont {Jonay}\ and\ \citenamefont {Zhou}(2024)}]{jonay_physical_2024}%
  \BibitemOpen
  \bibfield  {author} {\bibinfo {author} {\bibfnamefont {C.}~\bibnamefont {Jonay}}\ and\ \bibinfo {author} {\bibfnamefont {T.}~\bibnamefont {Zhou}},\ }\bibfield  {title} {\bibinfo {title} {Physical theory of two-stage thermalization},\ }\href {https://doi.org/10.1103/PhysRevB.110.L020306} {\bibfield  {journal} {\bibinfo  {journal} {Phys. Rev. B}\ }\textbf {\bibinfo {volume} {110}},\ \bibinfo {pages} {L020306} (\bibinfo {year} {2024})}\BibitemShut {NoStop}%
\bibitem [{\citenamefont {Jonay}\ \emph {et~al.}(2025)\citenamefont {Jonay}, \citenamefont {Li},\ and\ \citenamefont {Zhou}}]{jonay_two-stage_2025}%
  \BibitemOpen
  \bibfield  {author} {\bibinfo {author} {\bibfnamefont {C.}~\bibnamefont {Jonay}}, \bibinfo {author} {\bibfnamefont {C.}~\bibnamefont {Li}},\ and\ \bibinfo {author} {\bibfnamefont {T.}~\bibnamefont {Zhou}},\ }\bibfield  {title} {\bibinfo {title} {Two-stage relaxation of operators through domain wall and magnon dynamics},\ }\href {https://doi.org/10.1103/PhysRevB.111.224304} {\bibfield  {journal} {\bibinfo  {journal} {Phys. Rev. B}\ }\textbf {\bibinfo {volume} {111}},\ \bibinfo {pages} {224304} (\bibinfo {year} {2025})}\BibitemShut {NoStop}%
\bibitem [{\citenamefont {Bertini}\ \emph {et~al.}(2020)\citenamefont {Bertini}, \citenamefont {Kos},\ and\ \citenamefont {Prosen}}]{bertini_operator_i_2020}%
  \BibitemOpen
  \bibfield  {author} {\bibinfo {author} {\bibfnamefont {B.}~\bibnamefont {Bertini}}, \bibinfo {author} {\bibfnamefont {P.}~\bibnamefont {Kos}},\ and\ \bibinfo {author} {\bibfnamefont {T.}~\bibnamefont {Prosen}},\ }\bibfield  {title} {\bibinfo {title} {Operator {Entanglement} in {Local} {Quantum} {Circuits} {I}: {Chaotic} {Dual}-{Unitary} {Circuits}},\ }\href {https://doi.org/10.21468/SciPostPhys.8.4.067} {\bibfield  {journal} {\bibinfo  {journal} {SciPost Phys.}\ }\textbf {\bibinfo {volume} {8}},\ \bibinfo {pages} {067} (\bibinfo {year} {2020})}\BibitemShut {NoStop}%
\bibitem [{\citenamefont {Huang}\ \emph {et~al.}(2023)\citenamefont {Huang}, \citenamefont {Li}, \citenamefont {Huse},\ and\ \citenamefont {Chan}}]{huang_out--time-order_2023}%
  \BibitemOpen
  \bibfield  {author} {\bibinfo {author} {\bibfnamefont {K.}~\bibnamefont {Huang}}, \bibinfo {author} {\bibfnamefont {X.}~\bibnamefont {Li}}, \bibinfo {author} {\bibfnamefont {D.~A.}\ \bibnamefont {Huse}},\ and\ \bibinfo {author} {\bibfnamefont {A.}~\bibnamefont {Chan}},\ }\bibfield  {title} {\bibinfo {title} {Out-of-time-order correlator, many-body quantum chaos, light-like generators, and singular values},\ }\href {https://arxiv.org/abs/2308.16179} {\bibfield  {journal} {\bibinfo  {journal} {arXiv:2308.16179}\ } (\bibinfo {year} {2023})}\BibitemShut {NoStop}%
\bibitem [{\citenamefont {Fritzsch}\ and\ \citenamefont {Claeys}(2025)}]{zenodo}%
  \BibitemOpen
  \bibfield  {author} {\bibinfo {author} {\bibfnamefont {F.}~\bibnamefont {Fritzsch}}\ and\ \bibinfo {author} {\bibfnamefont {P.~W.}\ \bibnamefont {Claeys}},\ }\href {https://doi.org/10.5281/zenodo.15682409} {\bibinfo {title} {Data and numerical code for "Free probability in a minimal quantum circuit model"}} (\bibinfo {year} {2025}),\ \bibinfo {note} {[Data Set], Zenodo}\BibitemShut {NoStop}%
\bibitem [{\citenamefont {Kreweras}(1972)}]{kreweras1972partitions}%
  \BibitemOpen
  \bibfield  {author} {\bibinfo {author} {\bibfnamefont {G.}~\bibnamefont {Kreweras}},\ }\bibfield  {title} {\bibinfo {title} {Sur les partitions non crois{\'e}es d'un cycle},\ }\href {https://doi.org/https://doi.org/10.1016/0012-365X(72)90041-6} {\bibfield  {journal} {\bibinfo  {journal} {Discrete Math.}\ }\textbf {\bibinfo {volume} {1}},\ \bibinfo {pages} {333} (\bibinfo {year} {1972})}\BibitemShut {NoStop}%
\bibitem [{\citenamefont {Biane}(1997)}]{biane1997some}%
  \BibitemOpen
  \bibfield  {author} {\bibinfo {author} {\bibfnamefont {P.}~\bibnamefont {Biane}},\ }\bibfield  {title} {\bibinfo {title} {Some properties of crossings and partitions},\ }\href {https://doi.org/https://doi.org/10.1016/S0012-365X(96)00139-2} {\bibfield  {journal} {\bibinfo  {journal} {Discrete Math.}\ }\textbf {\bibinfo {volume} {175}},\ \bibinfo {pages} {41} (\bibinfo {year} {1997})}\BibitemShut {NoStop}%
\bibitem [{\citenamefont {Collins}(2003)}]{Col2003}%
  \BibitemOpen
  \bibfield  {author} {\bibinfo {author} {\bibfnamefont {B.}~\bibnamefont {Collins}},\ }\bibfield  {title} {\bibinfo {title} {Moments and cumulants of polynomial random variables on unitary groups, the {I}tzykson-{Z}uber integral, and free probability},\ }\href {https://doi.org/10.1155/S107379280320917X} {\bibfield  {journal} {\bibinfo  {journal} {Int.~Math.~Res.}\ }\textbf {\bibinfo {volume} {2003}},\ \bibinfo {pages} {953} (\bibinfo {year} {2003})}\BibitemShut {NoStop}%
\bibitem [{\citenamefont {Collins}\ and\ \citenamefont {{\'S}niady}(2006)}]{ColSni2006}%
  \BibitemOpen
  \bibfield  {author} {\bibinfo {author} {\bibfnamefont {B.}~\bibnamefont {Collins}}\ and\ \bibinfo {author} {\bibfnamefont {P.}~\bibnamefont {{\'S}niady}},\ }\bibfield  {title} {\bibinfo {title} {{Integration with Respect to the Haar Measure on Unitary, Orthogonal and Symplectic Group}},\ }\href {https://doi.org/10.1007/s00220-006-1554-3} {\bibfield  {journal} {\bibinfo  {journal} {Commun.~Math.~Phys.}\ }\textbf {\bibinfo {volume} {264}},\ \bibinfo {pages} {773} (\bibinfo {year} {2006})}\BibitemShut {NoStop}%
\bibitem [{\citenamefont {Weingarten}(1978)}]{weingarten_asymptotic_1978}%
  \BibitemOpen
  \bibfield  {author} {\bibinfo {author} {\bibfnamefont {D.}~\bibnamefont {Weingarten}},\ }\bibfield  {title} {\bibinfo {title} {Asymptotic behavior of group integrals in the limit of infinite rank},\ }\href {https://doi.org/10.1063/1.523807} {\bibfield  {journal} {\bibinfo  {journal} {J. Math. Phys.}\ }\textbf {\bibinfo {volume} {19}},\ \bibinfo {pages} {999} (\bibinfo {year} {1978})}\BibitemShut {NoStop}%
\end{thebibliography}
